\documentclass[manuscript]{aastex61}

\def\teff{$T_\mathrm{eff}$}

\def\logg{$\log g$}
\def\feoh{[Fe/H]}

\newcommand{\kms}{km~s$^{-1}$}

\newcommand{\lam}{$\lambda$}
\newcommand{\NIR}{near-infrared~}
\newcommand{\um}{$\mu$m}

\shorttitle{IGRINS Spectral Library}
\shortauthors{Park et al.}

\begin{document}

\title{IGRINS Spectral Library}

\correspondingauthor{Jeong-Eun Lee}
\email{jeongeun.lee@khu.ac.kr}

\author{Sunkyung Park}
\affil{School of Space Research, Kyung Hee University \\
1732, Deogyeong-daero, Giheung-gu, Yongin-si, Gyeonggi-do, 17104, Republic of Korea }

\author{Jeong-Eun Lee}
\affil{School of Space Research, Kyung Hee University \\
1732, Deogyeong-daero, Giheung-gu, Yongin-si, Gyeonggi-do, 17104, Republic of Korea }

\author{Wonseok Kang}
\affiliation{National Youth Space Center \\
200, Deokheungyangjjok-gil, Dongil-myeon, Goheung-gun, Jeollanam-do, 59567, Republic of Korea}

\author{Sang-Gak Lee}
\affiliation{Seoul National University \\
1 Gwanak-ro, Gwanak-gu, Seoul 08826, Republic of Korea}

\author{Moo-Young Chun}
\affiliation{Korea Astronomy and Space Science Institute \\
776, Daedeok-daero, Yuseong-gu, Daejeon, 34055, Republic of Korea}

\author{Kang-Min Kim}
\affiliation{Korea Astronomy and Space Science Institute \\
776, Daedeok-daero, Yuseong-gu, Daejeon, 34055, Republic of Korea}

\author{In-Soo Yuk}
\affiliation{Korea Astronomy and Space Science Institute \\
776, Daedeok-daero, Yuseong-gu, Daejeon, 34055, Republic of Korea}

\author{Jae-Joon Lee}
\affiliation{Korea Astronomy and Space Science Institute \\
776, Daedeok-daero, Yuseong-gu, Daejeon, 34055, Republic of Korea}

\author{Gregory N. Mace}
\affiliation{Department of Astronomy, University of Texas at Austin \\
2515 Speedway, Austin, TX, USA}
 
\author{Hwihyun Kim}
\affiliation{Korea Astronomy and Space Science Institute \\
776, Daedeok-daero, Yuseong-gu, Daejeon, 34055, Republic of Korea}
\affiliation{Gemini Observatory, La Serena, Chile}

\author{Kyle F. Kaplan}
\affiliation{Department of Astronomy, University of Texas at Austin \\
2515 Speedway, Austin, TX, USA}

\author{Chan Park}
\affiliation{Korea Astronomy and Space Science Institute \\
776, Daedeok-daero, Yuseong-gu, Daejeon, 34055, Republic of Korea}
\author{Jae Sok Oh}
\affiliation{Korea Astronomy and Space Science Institute \\
776, Daedeok-daero, Yuseong-gu, Daejeon, 34055, Republic of Korea}
\author{Sungho Lee}
\affiliation{Korea Astronomy and Space Science Institute \\
776, Daedeok-daero, Yuseong-gu, Daejeon, 34055, Republic of Korea}

\author{Daniel T. Jaffe}
\affiliation{Department of Astronomy, University of Texas at Austin \\
2515 Speedway, Austin, TX, USA}

\begin{abstract}

We present a library of high-resolution (R $\equiv$ \lam/$\Delta$\lam ~$\sim$ 45,000) and high signal-to-noise ratio (S/N $\geq$~200) \NIR spectra for stars of a wide range of spectral types and luminosity classes.
The spectra were obtained with the Immersion GRating INfrared Spectrograph (IGRINS) covering the full range of the H (1.496-1.780~\um) and K (2.080-2.460~\um) atmospheric windows. 
The targets were primarily selected for being MK standard stars covering a wide range of effective temperatures and surface gravities with metallicities close to the Solar value. 
Currently, the library includes flux-calibrated and telluric-absorption-corrected spectra of 84 stars, with prospects for expansion to provide denser coverage of the parametric space.
Throughout the H and K atmospheric windows, we identified spectral lines that are sensitive to \teff\ or \logg\ and defined corresponding spectral indices.
We also provide their equivalent widths.
For those indices, we derive empirical relations between the measured equivalent widths and the stellar atmospheric parameters. 
Therefore, the derived empirical equations can be used to calculate \teff\ and \logg\ of a star without requiring stellar atmospheric models.
\end{abstract}


\section{Introduction} \label{sec:intro}
Spectral libraries are important for science efficiency and repeatability. 
By understanding standard stars very well, we can also study stellar populations and chemical abundances throughout the Galaxy.
While a number of stellar spectral libraries are available, many are at optical wavelengths and low- to moderate-resolution. 
The number of spectral libraries with high spectral resolution in the \NIR (NIR) is very limited.
NIR spectral libraries are useful for studying the cool phenomena of the universe: physics of cool stars, circumstellar objects such as planets and disks surrounding young stellar objects (YSOs), and the extended atmospheres of evolved stars. 
Compared to the red part of their optical spectra, the NIR spectra of cool stars are better suited for detailed spectroscopic study (especially in the J and H atmospheric windows, hereafter bands) as they contain fewer molecular bands. 
When determining equivalent widths (EWs) the broad molecular bands make the continuum level hard to determine and affect the measurement of narrow atomic lines.
Also, the smaller extinction in the NIR, relative to optical wavelengths, provides us access to dust obscured stars across the Galaxy.
Additionally, a NIR stellar library is useful for population synthesis of cool stars at the wavelengths where they are brightest.

Over the past four decades, a number of infrared spectral libraries have been developed (see Table 1 in Rayner et al. 2009). 
For example, \citet{rayner09} presented an intermediate-resolution (R $\equiv $ \lam/$\Delta$\lam ~$\sim$ 2000) NIR spectral library of 210 F, G, K, and M stars that covers  0.8-5~\um, and this library provided flux calibrated spectra across this wide wavelength range for the first time.
Moderate-resolution spectra are suitable for studying variations in strong spectral features, but many lines remain blended.
For example, Figure~\ref{comp_spec} shows how the important Na I features at 2.2~\um\ are contaminated by lines of Si and Sc when observed at lower resolution \citep[see Table 7 in][]{ramirez97, doppmann03}. 
High-resolution spectroscopy \textquotedblleft resolves\textquotedblright~this problem. 
In addition, \citet{lebzelter12} provided a library of high-resolution (R $\sim$ 100,000) NIR spectra for 25 stars observed with CRIRES at the VLT.
However, the small number of stars observed with CRIRES leaves the spectral type and luminosity class space coarsely sampled. 
The primary reason for the small sample in the CRIRES library is that the broad spectral coverage (1-5~\um) required about 
200 instrument settings to cover the full wavelength range, and about 70 settings to cover the entire H- and K-bands. 
Other large surveys like APOGEE \citep[][and references therein]{zamora15} will obtain hundreds of thousands of NIR spectra, but the planned spectral coverage is narrow and it will not provide a comprehensive library covering the entire parametric space.

Currently, a uniform library of high-resolution NIR spectra covering all spectral types and luminosity classes is not available.
Although low- and moderate-resolution spectra are appropriate for extragalactic studies, because of the faintness of the targets and their internal velocity dispersions of 100 to 250~\kms, there is an ongoing need for a high-resolution spectral libraries in the Galactic context.
One consequence of the absence of such libraries is that only strong and isolated lines have been used for studies.
High-resolution NIR spectra will reveal lines with small equivalent widths that can be more sensitive to stellar properties and to abundances of less common elements \citep{afsar16}.
At R=45,000, lines in most stars are well-resolved and the line shapes offer an additional window into the stellar properties.
High-resolution stellar spectra covering wide ranges of effective temperature and surface gravity will test atmospheric models with unprecedented accuracy and permit modeling to improve infrared atomic and molecular line data.

In this paper, we present a high-resolution (R $\sim$ 45,000) and high signal-to-noise ratio (S/N $\geq$~200) NIR spectral 
library covering a broad range of spectral parameters. The spectra were obtained with the Immersion GRating INfrared Spectrograph \citep[IGRINS,][]{yuk10, park14, mace16}.
IGRINS has a resolving power of R $\sim$ 45,000 and covers the full H (1.49-1.80~\um) and K (1.96-2.46~\um) bands, simultaneously. 
The data presented in this paper was obtained as part of the \textquotedblleft IGRINS spectral library (PI : Jeong-Eun Lee)\textquotedblright~Legacy program\footnote{Korean IGRINS Legacy Program - \\ \url{http://kgmt.kasi.re.kr/kgmtscience/kgmt_igrins/legacy/selected.html}}. 
We present spectra, derived equivalent widths, and empirical relations based on the spectral catalog.

\section{Target Selection} \label{sec:target selection}
The targets in our library were selected to be MK standard stars \citep{walborn73, garcia89, garrison94, montes97, montes98, munari99, marrese03, cushing05, rayner09} with well-defined spectral types and luminosity classes. 
We have adopted the following criteria for making our selection:
1) MK standard star, 2) low rotational velocity, 3) low chromospheric activity, 4) low variability, and 5) not a known binary/multiple system.
This paper presents spectra of 84 targets.
While this sample covers spectral types between O and M (O4-M6) with luminosity classes between I and V, they mostly have solar metallicity and are relatively bright stars.
Also, the sample is biased toward the late-type stars since they have more atomic and molecular lines in the NIR.
More than half of the targets (63 stars) are stars later than F0. 
Table~\ref{tbl_comp} lists the spectral composition of the IGRINS Spectral Library.
Other than one star (HD 216868, G5 with unknown luminosity class), all of our sample has well established spectral types and luminosity classes.

Individual target information is listed in Table~\ref{tbl_info} which contains object name, spectral type, Two Micron All Sky Survey 
\citep[2MASS;][]{skrutskie06} H- and K-band magnitudes, observation date, total exposure time, S/N at 2.2~\um, A0 standard star for telluric correction, heliocentric radial velocity, radial velocity, references for radial velocity, and references for target selection. 
The S/N for each star in Table~\ref{tbl_info} is the medians of the S/N values per resolution element for $\lambda$=2.21~$\sim$~2.24~\um.
About half of our targets (47 stars), have their stellar atmospheric parameters (\teff, \logg, and \feoh) taken from the literature.
Among these 47 stars, the stellar atmospheric parameters of 36 stars are from the ELODIE \citep{prugniel07}, MILES \citep{prugniel11} and CFLIB libraries \citep{wu11}. 
This is an inhomogeneous collection of the stellar atmospheric parameters, therefore, we tried to compile the data as uniformly as possible.
In the ELODIE library \citep{prugniel07}, stars were selected that have known stellar atmospheric parameters: \teff, \logg, and \feoh.
The majority of atmospheric parameters are from the catalogue by \citet{cayreldestrobel97} and \citet{cayreldestrobel01}. 
The MILES \citep{prugniel11} and CFLIB \citep{wu11} libraries redetermined atmospheric parameters by using the ULySS program \citep{koleva09}.
The uncertainties of these three libraries are compared in Table~\ref{tbl_uncertainty}.

For stars without \teff\ information from the literature, \teff\ was calculated from the \teff(B-V) relation \citep{flower96, torres10}.
The standard deviation of the differences between \teff(Ref.) and \teff(B-V) is about 200~K for common stars, and this value is adopted as the error of \teff(B-V).
Most of the calculated \teff(B-V) are in the 1$\sigma$ uncertainty and we therefore have adopted the \teff(B-V) for all of our sources. 
However, if the B-V value is greater than 1.5, the calculated \teff(B-V) shows greater scatter \citep{gray05} and objects above this cutoff were not included in our analysis.

\begin{figure}
\plotone{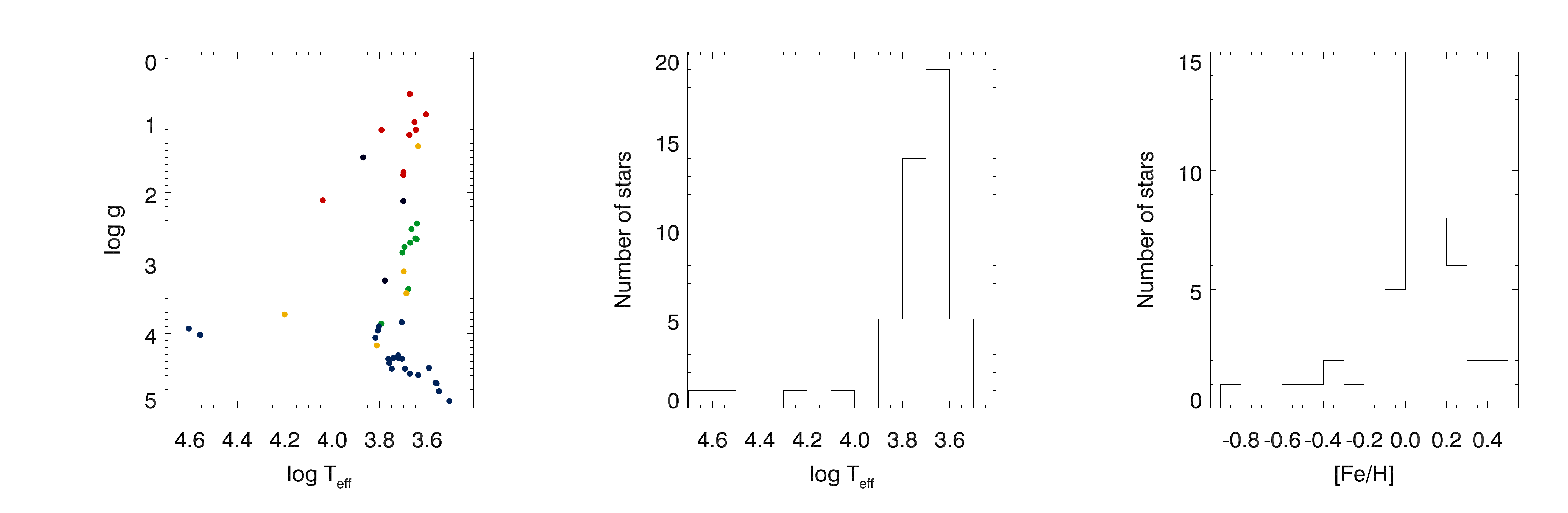} 
\caption{
Physical parameters for the targets in the IGRINS Spectral Library. 
In total, 47 stars have physical parameters in the literature. 
The detailed values and references are listed in Table~\ref{tbl_ew_param}. 
Left: The surface gravity as a function of effective temperature for our targets.
Red, black, green, orange, and blue symbols represent supergiants, bright giants, giants, subgiants, and dwarfs (main-sequence stars), respectively.
Middle: The effective temperature distribution of the targets, which reveals that a majority of the targets are late-type stars. 
Right: The metallicity distribution of the targets in the spectral library. 
The typical value of the metallicity (\feoh~$\sim$ 0.05) is approximately solar composition. 
\label{stellar_param}
}
\end{figure}

The distribution of our selected targets on the modeler's Hertzsprung-Russell Diagram, and the histograms of effective temperature and metallicity for the library are shown in Figure~\ref{stellar_param}, where stars with their known photospheric parameters (47 stars) are presented. 
The median and average value of \feoh\ is about 0.05 and 0.02, respectively, which is a typical value of the solar neighborhood. 

\section{Observations and Data reduction} \label{sec:observations and data reduction}
\subsection{Observations}\label{subsec:observations}
The spectra in this library were obtained with IGRINS on the 2.7~m Harlan J. Smith Telescope (HJST) at McDonald Observatory during multiple observing runs from 2014 to 2016.
IGRINS makes use of low-noise H2RG detectors, a silicon immersion grating, and volume-phase holographic (VPH) cross-dispersers to maximize throughput and efficiency. 
As a result, IGRINS provides the full spectral range of the H (1.49-1.80~\um) and K (1.96-2.46~\um) bands with R $\sim$ 45,000 (corresponding to a velocity resolution of $\Delta v $=($\Delta$\lam $/$\lam~$\times$ c)$~\sim$ 7 \kms) with a single exposure. 
IGRINS mounted at the Cassegrain focus of the HJST has the slit size of 1\arcsec$\times$15\arcsec\, and 3.5 pixel sampling across the slit.

Individual stars were observed several times at different positions on the slit to better subtract the sky background.
The majority of the stars, whose K-band magnitudes are fainter than 3~mag, were obtained in a series of ABBA nodding observations (nod-on-slit mode) along the slit at positions separated by 7\arcsec. 
Brighter targets were observed with the nod-off-slit mode where the A position is on the slit and the B position places the target off the slit and we observe adjacent sky. 
For the telluric correction, a nearby A0 star was observed immediately after or before the observation of each target and telluric standards are shared among targets of similar airmass.
The average difference in airmass between targets and their corresponding A0 telluric correction stars is 0.06.

\subsection{Data Reduction}\label{subsec:data reduction}
\begin{figure}
 \includegraphics[width=18cm, height=5cm]{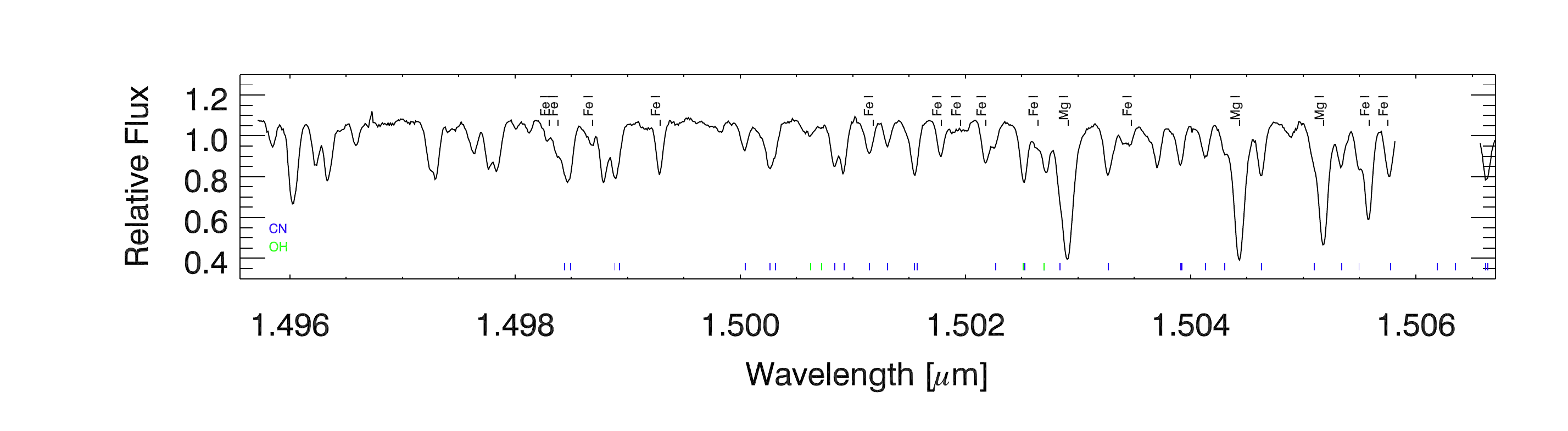}
 \includegraphics[width=18cm, height=5cm]{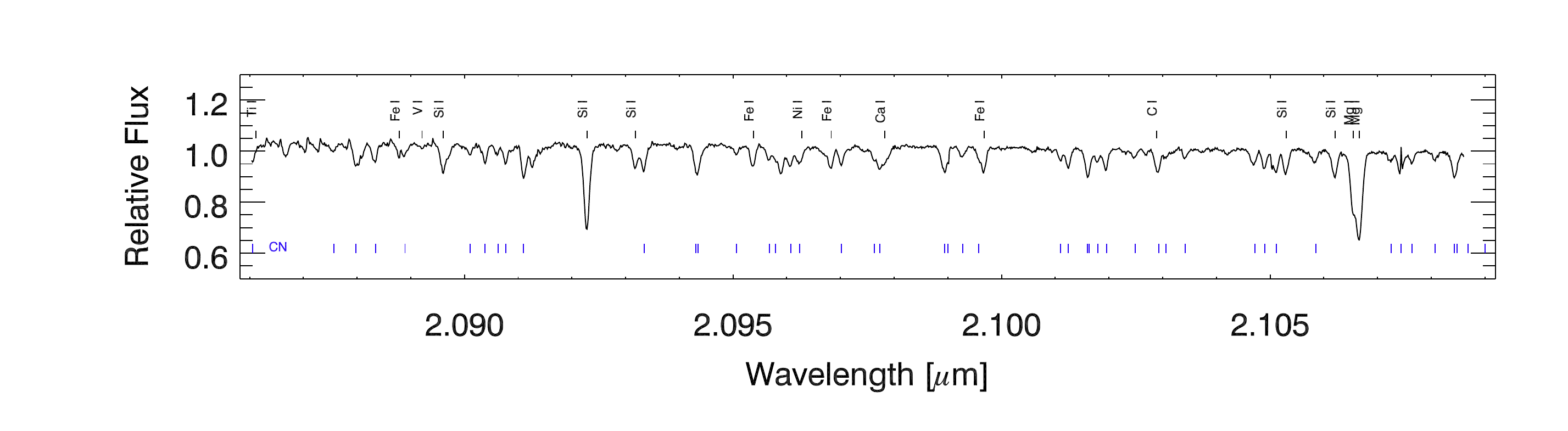}
\caption{
Representative H- and K-band spectra of HD 44391 (K0 Ib).
Line identification was made using the Arcturus line list from \citet{hinkle95}.
The full spectrum is presented in the Appendix.
\label{spec_HK}
}
\end{figure}

The spectra were reduced using the IGRINS pipeline version 2 \citep{lee17}. 
The raw data were flat-fielded to correct the pixel-to-pixel variations, and bad pixels were masked out.
The spectra were extracted from pair-subtracted images (A-B), which effectively removes sky background.
The 2-d spectra were first corrected for the distortion in the dispersion direction, then combined into a final 2-d spectrum and a modified version of the optimal extraction algorithms of \citet{horne86} was applied to produce 1-d spectra for each order.
The wavelength solutions were derived using OH emission lines from the spectra of blank sky. 
In the longer wavelength part of the K-band where there are fewer OH lines, other molecular lines are used. 
Typical velocity uncertainties are~$<$~0.2~\kms\ in the H-band spectrum and better than~$\sim$~1~\kms\ in the K-band spectrum.
Furthermore, refined wavelength solutions are generated using telluric absorption features.
The final wavelengths are given in vacuum.
A representative spectrum from the IGRINS Spectral Library for HD~44391~(K0~Ib) is shown in Figure~\ref{spec_HK}.

\begin{figure}
\epsscale{.80}
\plotone{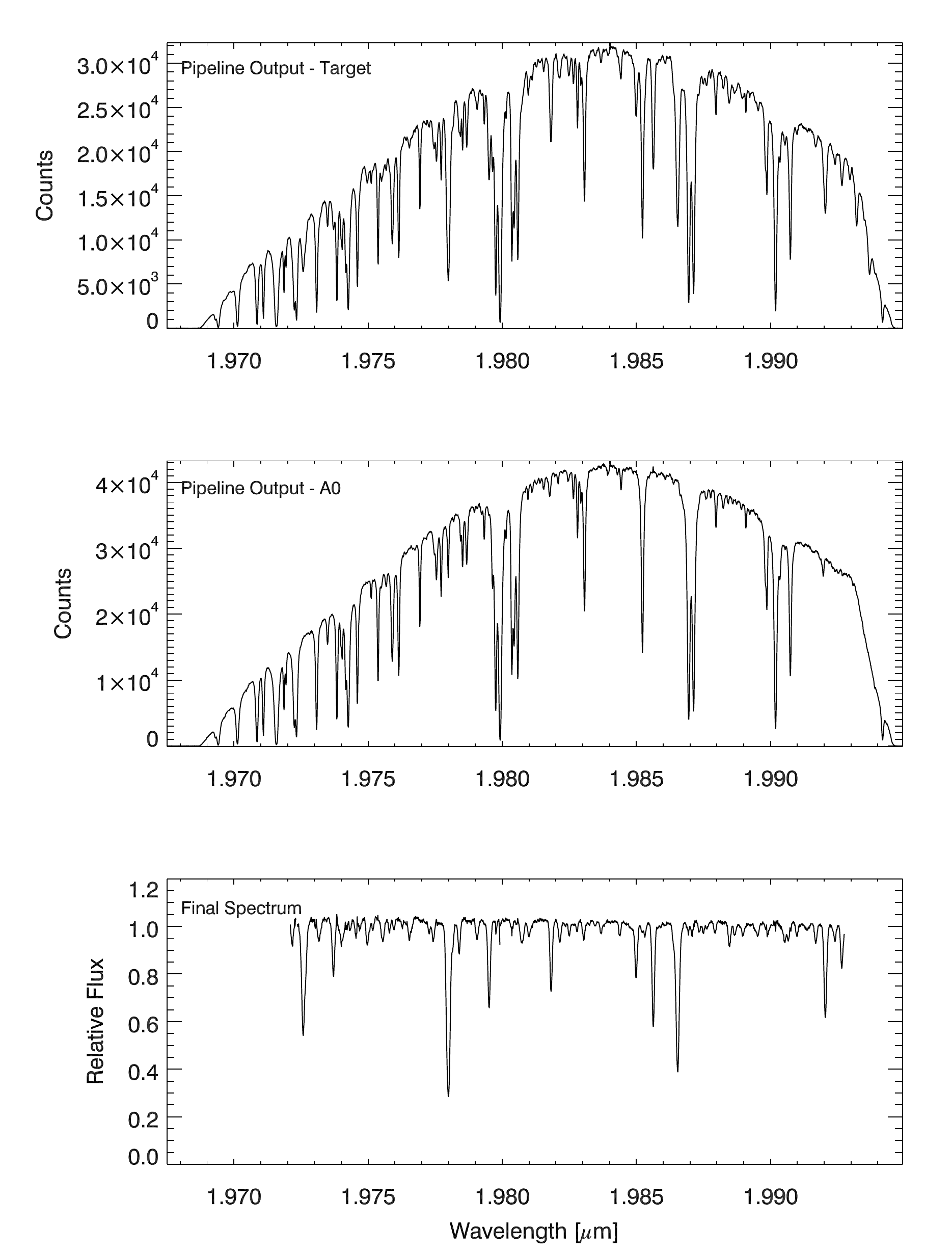}
\caption{Example of the telluric correction method.
The top and middle panels show the pipeline output spectrum of HD 44391 (K0 Ib) and HIP 31434 (A0~V), respectively.
The bottom panel shows the telluric corrected spectrum, which is created by dividing the target spectrum by the Vega-corrected, shifted and scaled A0 spectrum.
\label{telcorr}}
\end{figure}

The target spectrum is then corrected for the shape of the blaze function and the telluric absorption using the spectra of A0~V stars. 
Estimating the shape of the blaze function requires the intrinsic broad hydrogen lines of A0~V to first be divided out. 
The pipeline employs the Kurucz model of Vega\footnote{\url{http://kurucz.harvard.edu/stars/vega}} convolved to the spectral resolution of IGRINS. 
While this provides adequate results for most of our targets, some display residuals larger than our average flux uncertainties at the location of the hydrogen lines. 
Further adjustments to line profiles of the Kurucz model provided marginal improvements and these objects are marked with a dagger symbol ($\dagger$) in Table~\ref{tbl_info}. 
We note that Hydrogen lines are intrinsically broad and their persistence has a negligible effect on the analysis of the narrow features we are interested in.
After the removal of hydrogen lines, the data were corrected for telluric absorption lines in the individual orders, as shown in Figure~\ref{telcorr}. 
While the spectra of telluric standards are observed at a similar airmass to that of the target, changes in instrument orientation at the Cassegrain focus of the HJST introduce small amounts of shift in the wavelength solution \citep[typically much less than 1 pixel;][]{mace16}.
This shift is measured and compensated by refining the telluric spectrum wavelength solution to match the target wavelength solution using sky line locations in the raw spectra.
The change in airmass (although small) is further compensated by scaling the spectra.
The A0 spectra were scaled by the difference of airmass between the target and the A0 star ($A0^{(airmass_{Target}/airmass_{A0})}$).
Figure~\ref{telcorr} shows an example spectrum of a target, an A0 telluric standard, and the telluric corrected
target spectrum where not only are the telluric absorption features removed but where the division also corrects for variations in instrumental throughput and flattens the continuum.

\begin{figure}
\epsscale{0.2}

 \includegraphics[width=9cm, height=7cm]{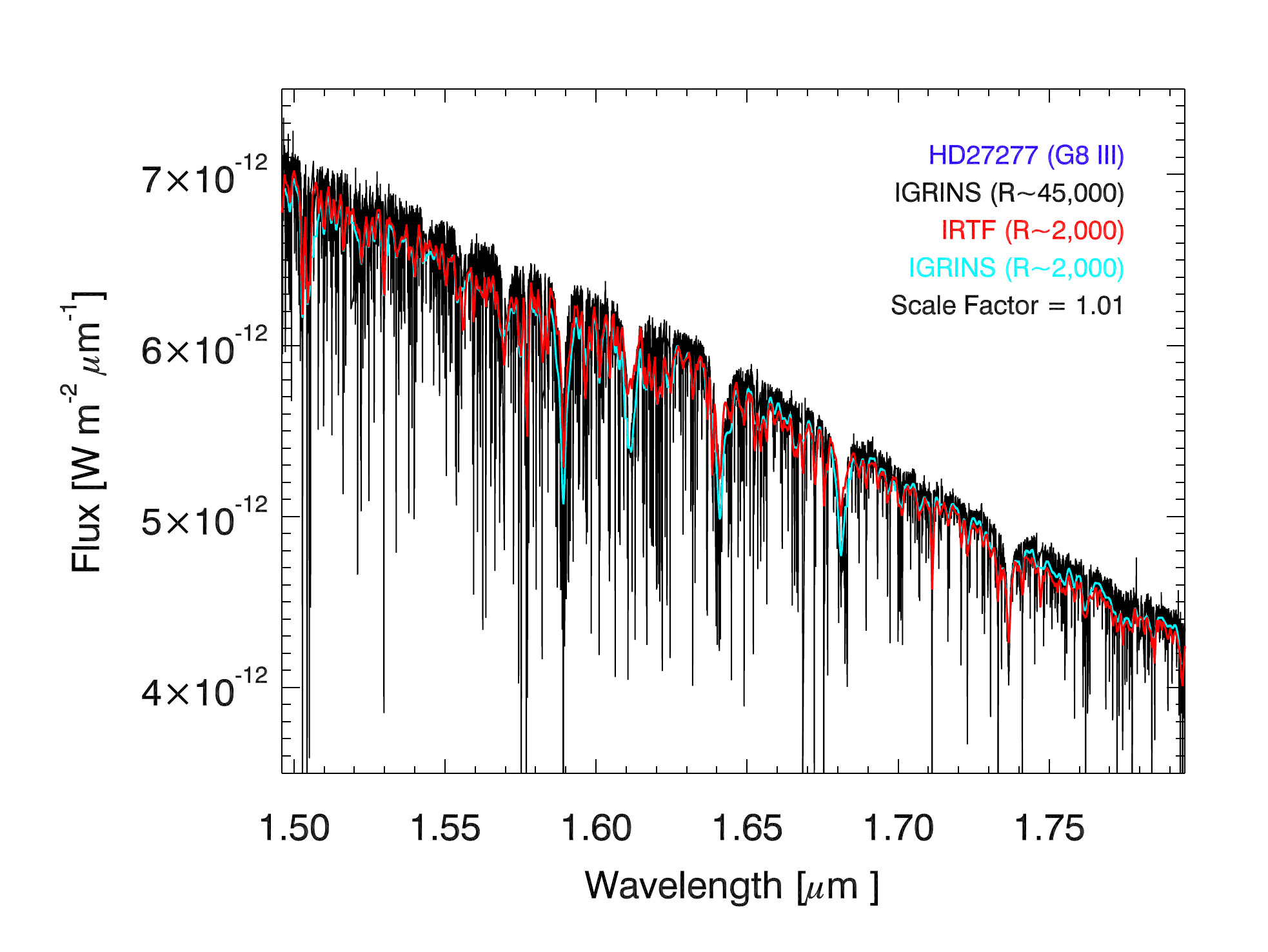}
 \includegraphics[width=9cm, height=7cm]{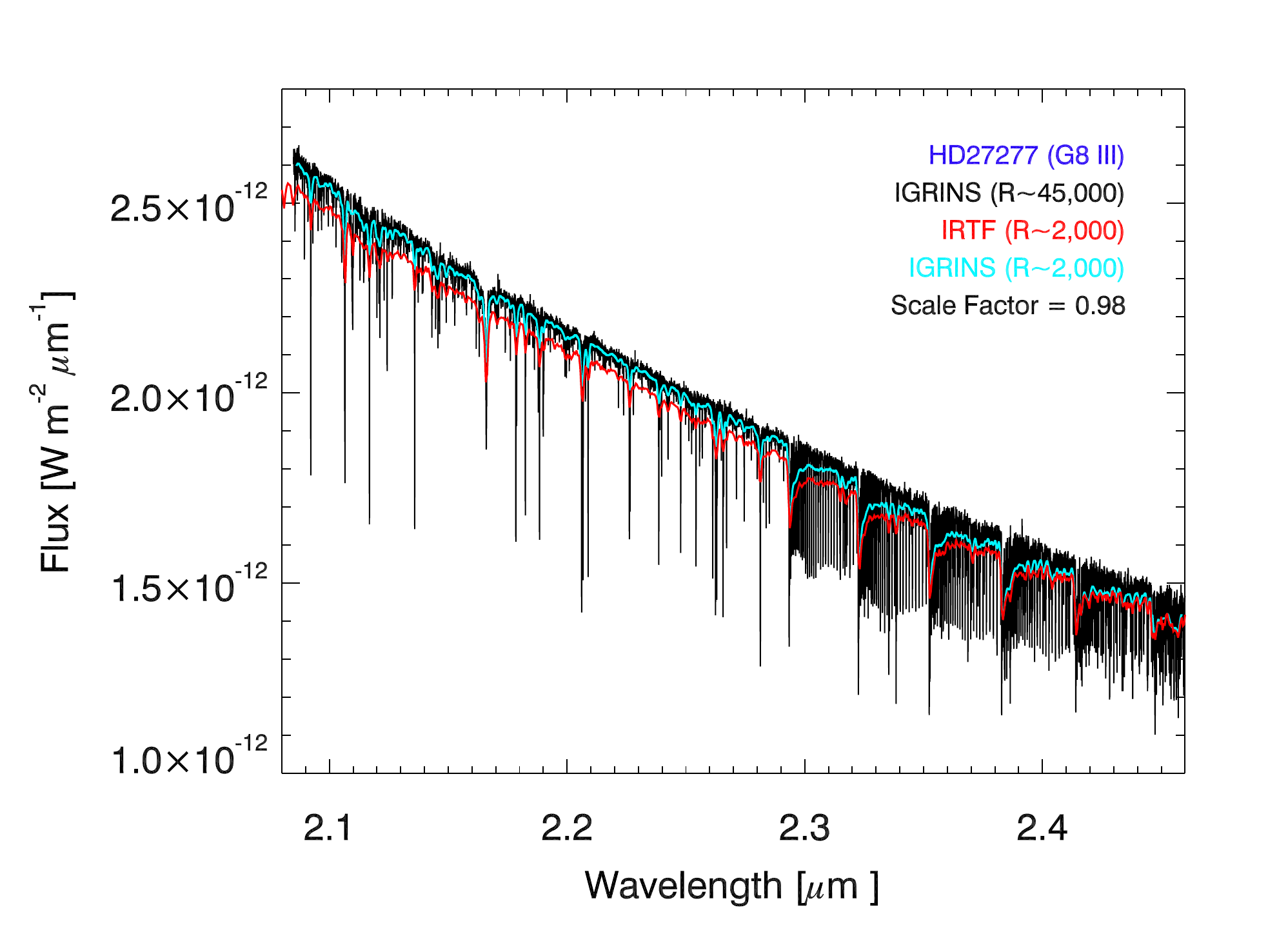}
 
 \includegraphics[width=9cm, height=7cm]{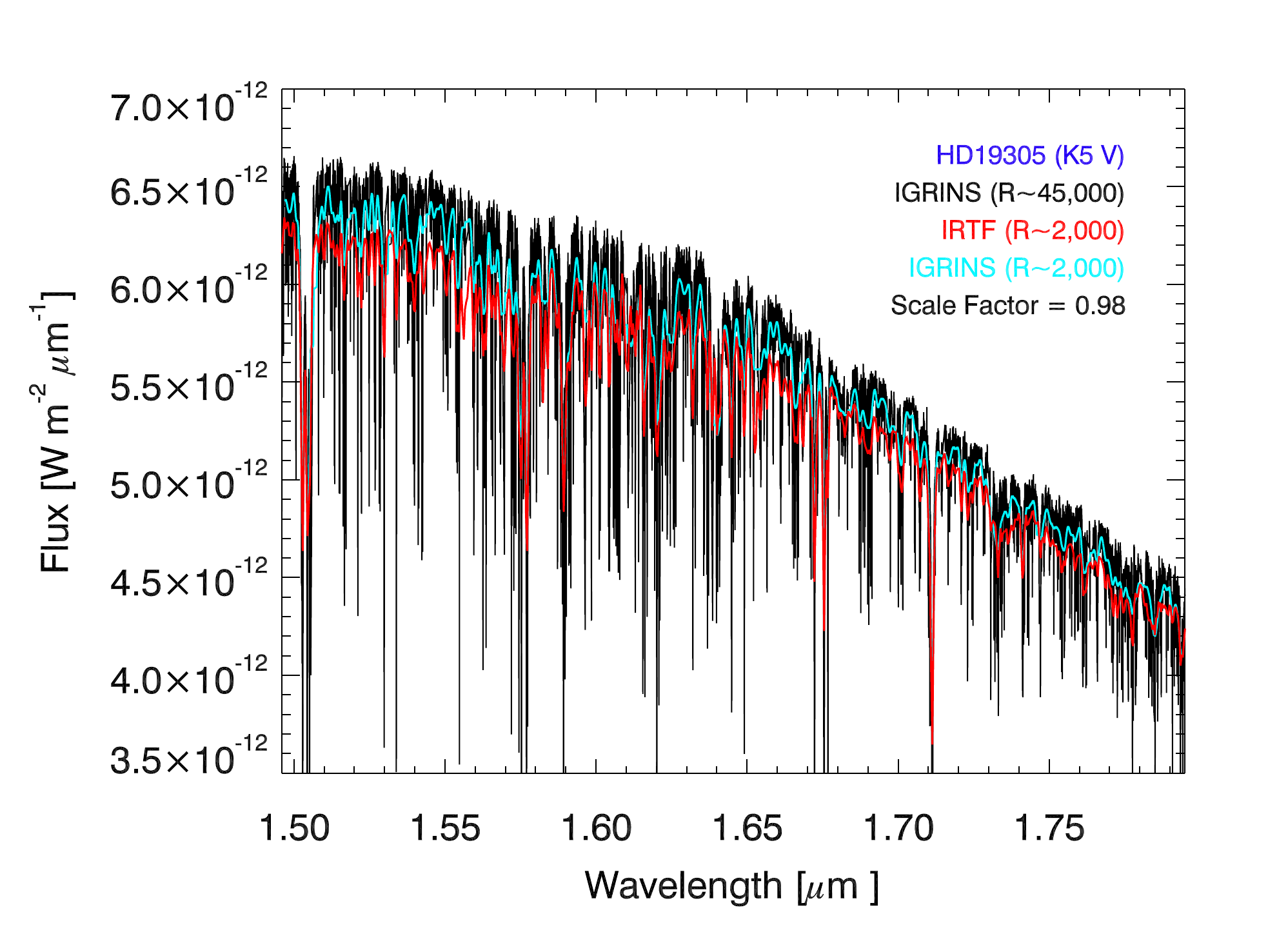}
 \includegraphics[width=9cm, height=7cm]{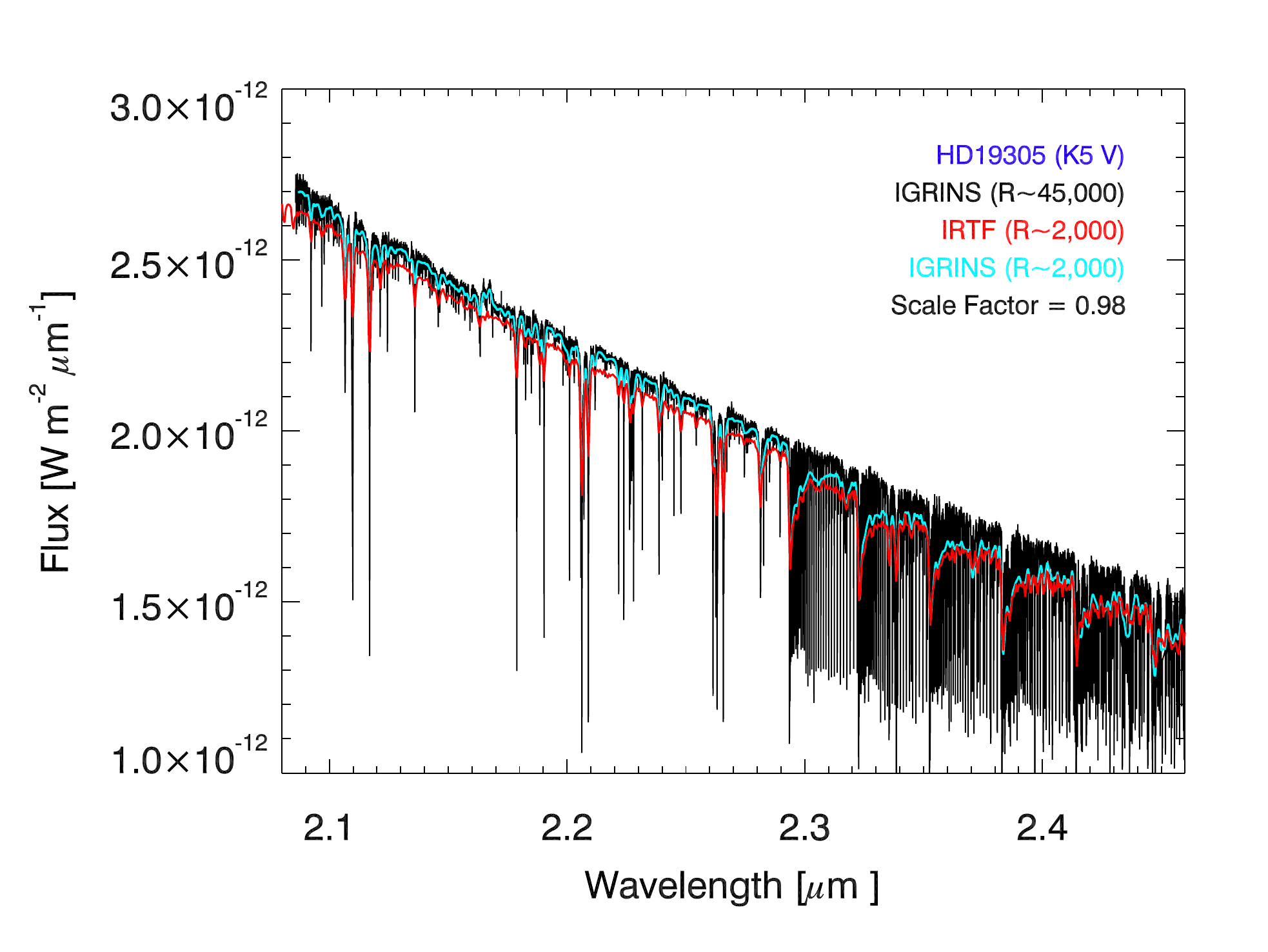}
 
\caption{
Results of flux calibration compared to the IRTF Spectral Library.
Top and bottom panels show the result of flux calibration of HD 27277 (G8 III) and HD 19305 (K5 V), while the left and right panels show the H- and K-band spectra, respectively.
Black lines indicate the observed spectra from IGRINS (R $\sim$ 45,000), while the red lines show the IRTF/SpeX spectra (R $\sim$ 2,000).
The cyan line represents the resolution-convolved IGRINS spectrum (R $\sim$ 2,000).
The flux ratio of the convolved IGRINS spectrum to the IRTF/SpeX spectrum is shown as the scale factor in the legend.
\label{flux_cal}
}
\end{figure}

\begin{figure}
\epsscale{1}
\plotone{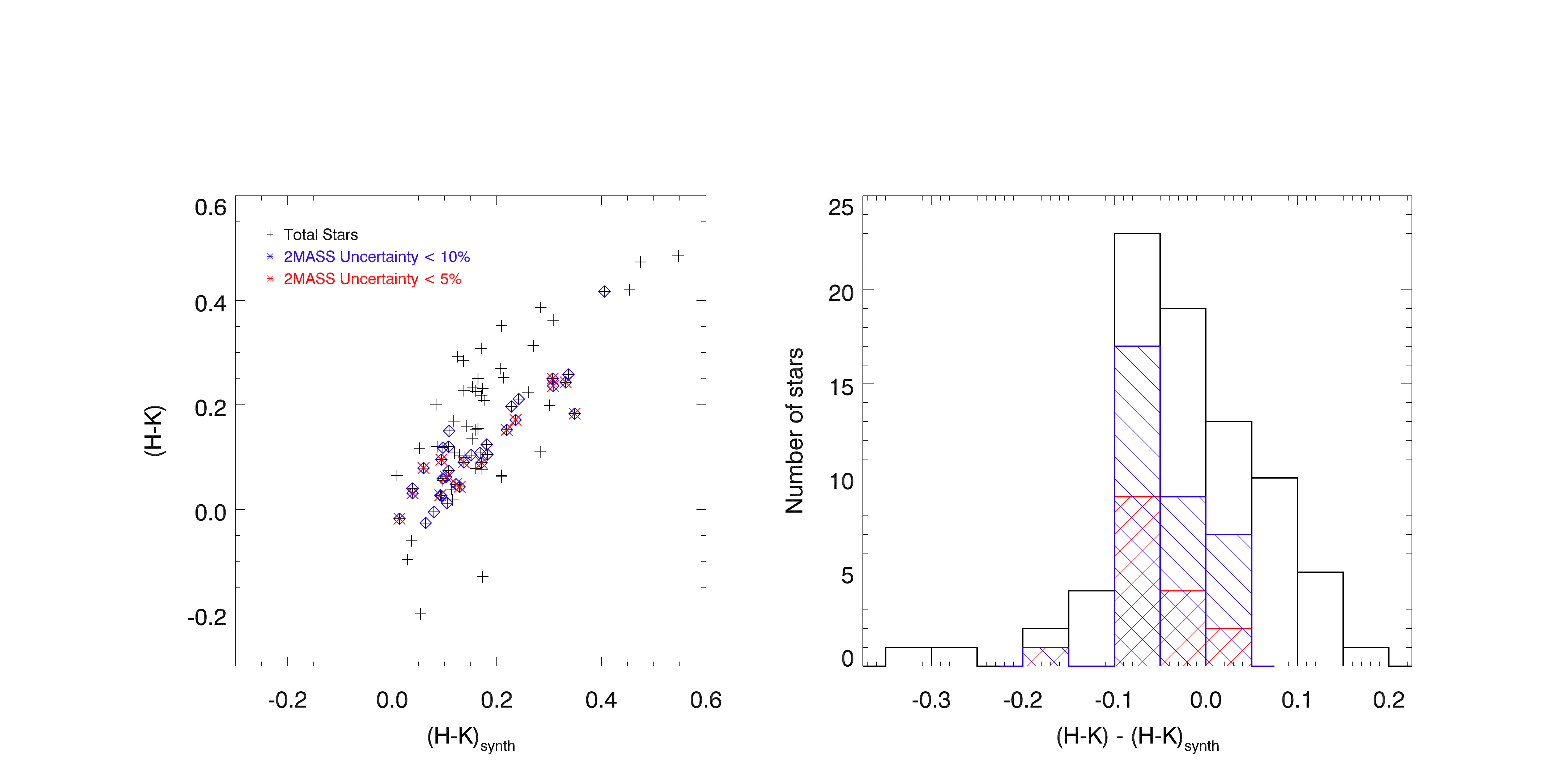}
\caption{
Computed synthetic colors (H-K$_{S}$) from flux calibrated IGRINS spectra.
Left: The comparison of synthetic colors between synthesized 2MASS values and observed values from the 2MASS Point Source Catalog \citep{cutri03}.
Right: The distribution of residuals between the points in the left panel. 
Black line represent the values of all the flux calibrated 80 stars, while blue and red symbols present 2MASS photometry data with $<$ 10\%\ and $<$ 5\%\ uncertainties. 
The average and rms of residuals are 0.02 and 0.06 for 80 stars which are in a good agreement with those of \citet{rayner09} ($< \Delta H-K_{S}>$~=~0.02~$\pm$~0.04).
The average and rms of residuals are ($< \Delta H-K_{S}>$~=~0.05~$\pm$~0.06) and ($< \Delta H-K_{S}>$~=~0.06~$\pm$~0.07) for the 2MASS photometry with 10\%\ and 5\%\ uncertainty, respectively.
\label{synth_color}
}
\end{figure}

Next, the spectra from all the individual IGRINS orders were combined to make a single continuous spectrum and flux-calibrated by scaling the spectra using 2MASS photometry.
For each target, we computed the scale factor as
\begin{equation} \label{eq_flux}
C = \frac{F_{2MASS} \times 10^{-0.4 \times m}\int{S_{\lambda}}d\lambda}{\int{F_{\lambda}S_{\lambda}}d\lambda},
\end{equation}
where $F_{2MASS}$ is the flux converted from the 2MASS zero magnitude, $m$ is the magnitude of the object, $F_{\lambda}$ is the observed flux, and $S_{\lambda}$ is the response function of the 2MASS filters given by \citet{cohen03}.  
The flux calibration was performed for the K-band first, and then extrapolated to the H-band.
For the 80 stars out of 84, we provide flux-calibrated spectra. 
For the remaining four stars, the blaze function of the target spectra is found to be significantly different from that of A0 V stars (likely caused by the target being located significantly off from the slit center), making flux calibration using the A0 V stars difficult. 
For these four stars, we instead only present their flattened spectra and they are marked with a star symbol ($\ast$) in Table~\ref{tbl_info}.
Figure~\ref{flux_cal} shows the result of flux calibration for HD~27277~(G8~III) and HD~19305~(K5~V), which are also included in the SpeX/IRTF Spectral Library \citep{rayner03,rayner09}.
As shown in Figure~\ref{flux_cal}, our flux calibration is consistent with that of the IRTF/SpeX spectra.
The flux difference between the IGRINS and SpeX spectra is about 3\%\ in the H- and K-bands, 
which is within the IRTF/SpeX flux calibration uncertainty of 10\%.
The uncertainty of our flux calibration includes error propagation.
The flux calibration was additionally checked by using the synthetic color calculated with Equation 5 of \citet{rayner09}. 
As shown in Figure~\ref{synth_color}, the calculated synthetic color is consistent with that of the IRTF Spectral Library, which demonstrates that our flux calibration is reliable and can be used in conjunction with this library. 

Finally, the spectra were shifted to heliocentric velocity using their radial velocity from the literature and heliocentric radial velocity calculated using the {\tt rvcorrect} package\footnote{\url{http://stsdas.stsci.edu/cgi-bin/gethelp.cgi?rvcorrect}} in the IRAF. 
The adopted velocities are listed in Table~\ref{tbl_info}.
For three stars without radial velocity information available in the literature (HD~228779, HD~216868, and BD+63~2073), we have measured the radial velocity either using the He~I~1.701~\um\ or Na~I~2.209~\um\ lines. 
They are marked with $\ddagger$ in Table~\ref{tbl_info}.

\section{Data and Analysis} \label{sec:data and analysis}
\subsection{Spectra}\label{subsec:spectra}
\begin{figure}
\epsscale{1}
\plotone{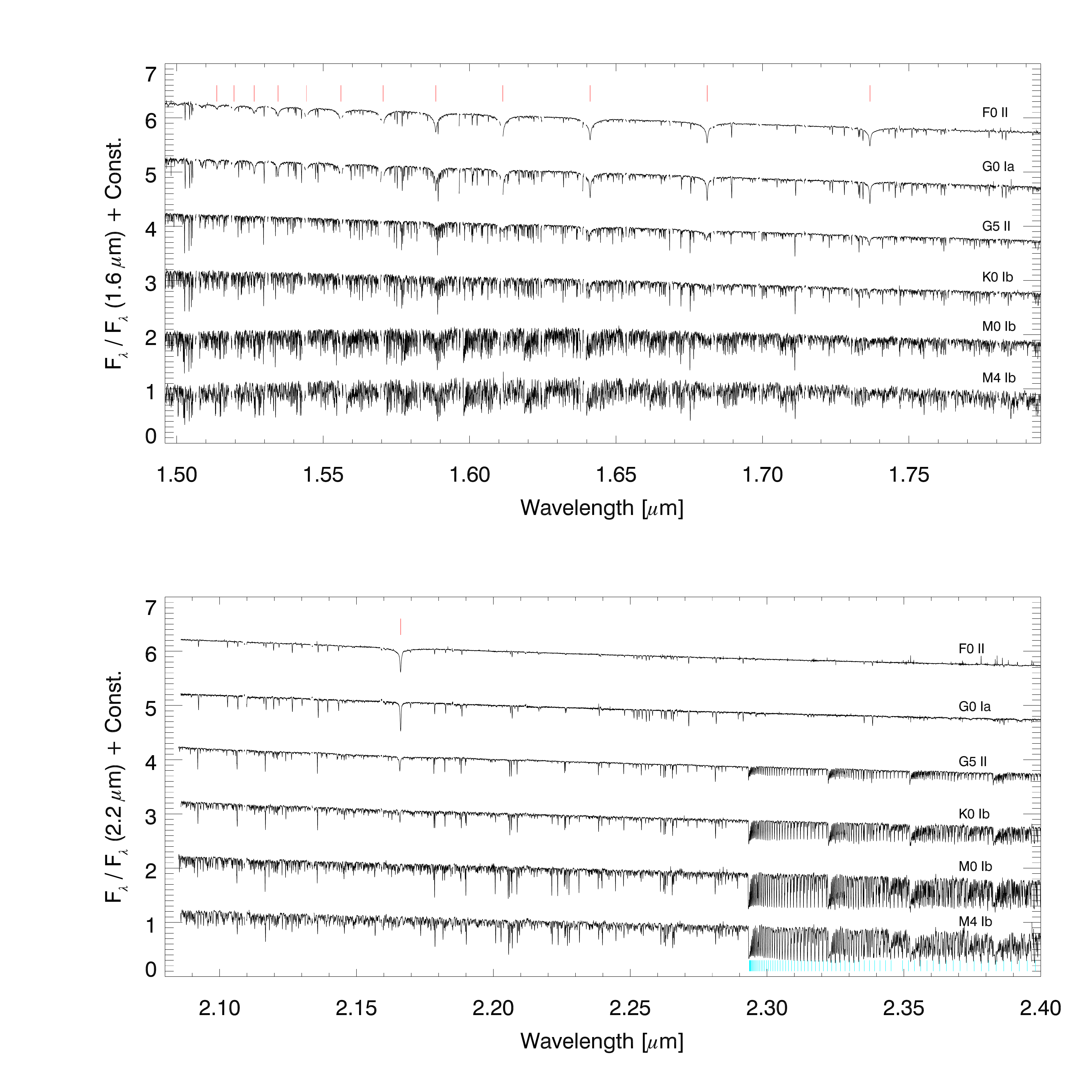}
\caption{
Supergiant stars observed with IGRINS. 
H-band spectra are shown in the top panel and K-band spectra are in the bottom panel.
The spectra are of HD 6130 (F0 II), HD 190323 (G0 Ia), HD 108477 (G5 II), HD 44391 (K0 Ib), BD+63 2073 (M0 Ib), and BD+56 512 (M4 Ib).
Red and cyan lines indicate the position of Brackett series and CO overtone rovibrational transitions.
\label{cont_supergiant}
}
\end{figure}

\begin{figure}
\epsscale{1}
\plotone{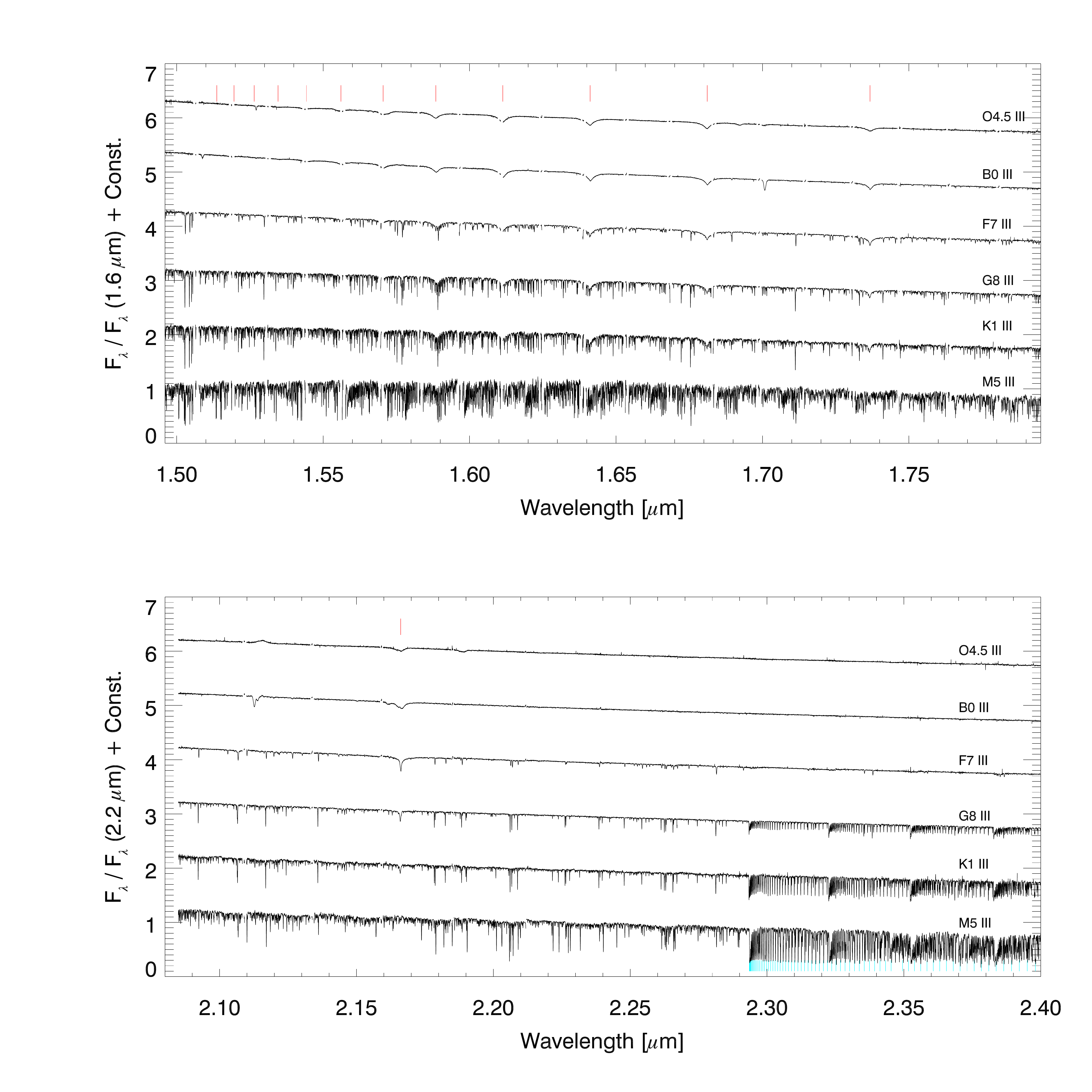}
\caption{
Giant stars observed with IGRINS. H-band spectra are shown in the top panel and K-band spectra are in the bottom panel.
The spectra are of HD 15558 (O4.5 III), HD 48434 (B0 III), HD 124850 (F7 III), HD 27277 (G8 III), HD 94600 (K1 III), and HD 132813 (M5 III).
Red and cyan lines indicate the position of Brackett series and CO overtone rovibrational transitions.
\label{cont_giant}
}
\end{figure}

\begin{figure}
\epsscale{1}
\plotone{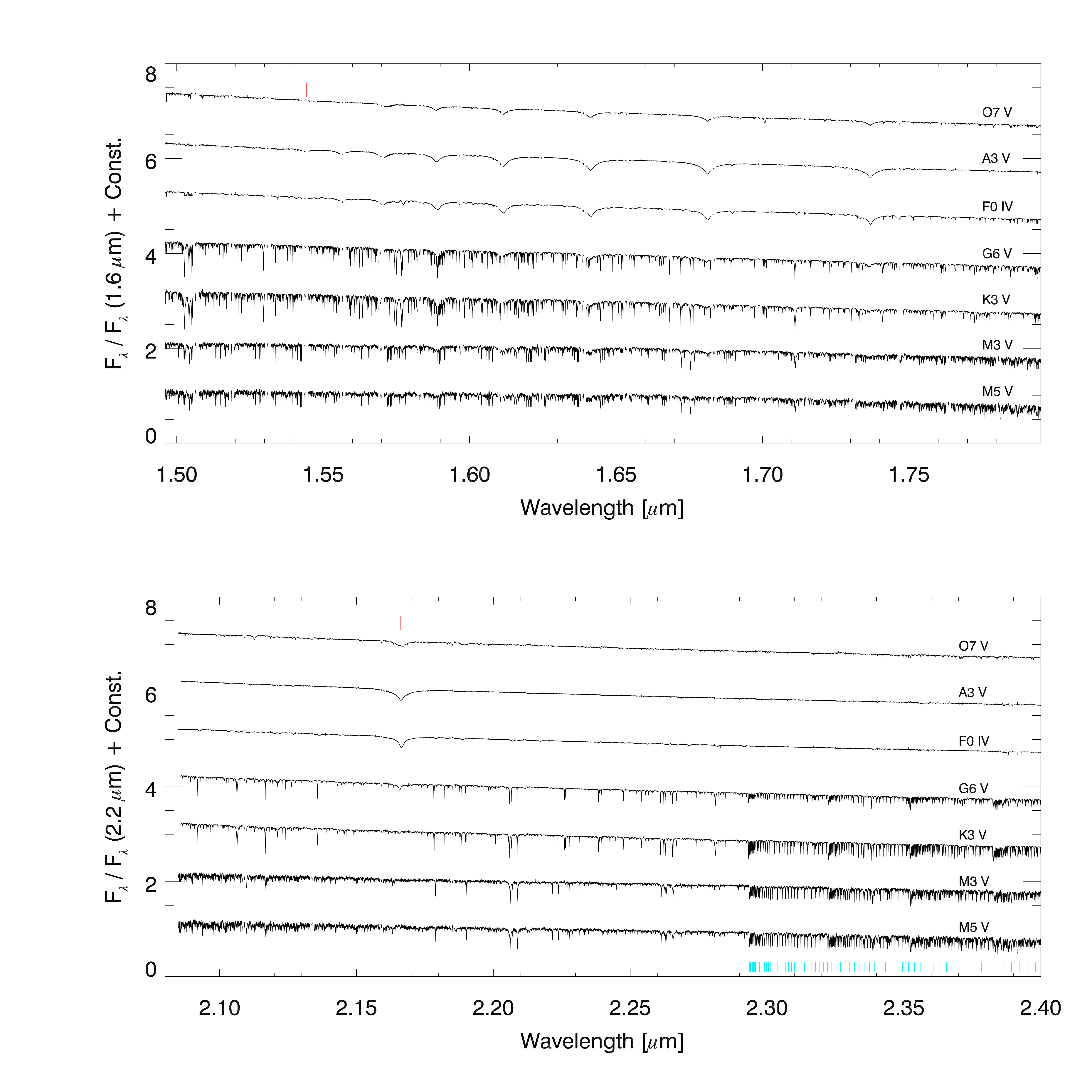}
\caption{
Dwarf stars observed with IGRINS. H-band spectra are shown in the top panel and K-band spectra are in the bottom panel.
The spectra are of HD 47839 (O7 V), HD 7804 (A3 V), HD 27397 (F0 IV), HD 117043 (G6 V), HD 122064 (K3 V), BD+44 3567 (M3 V), and BD-07 4003 (M5 V).
Red and cyan lines indicate the position of Brackett series and CO overtone rovibrational transitions.
\label{cont_dwarf}
}
\end{figure}

The digital version of the IGRINS Spectral Library is available at our website\footnote{\url{http://starformation.khu.ac.kr/IGRINS_spectral_library}}.
The pipeline-reduced data are available in FITS format and the telluric corrected data are available in ascii format.
The telluric corrected ascii file contains wavelength (in vacuum,~\um), flux (in cgs unit, W~m$^{-2}$~\um$^{-1}$), and total uncertainty (in cgs unit, W~m$^{-2}$~\um$^{-1}$).
The wavelength coverage of IGRINS Spectral Library is 1.496-1.780~\um\ and 2.080-2.460~\um\ for H and K bands, respectively.
Representative H- and K-bands spectra of supergiants, giants, and dwarfs are shown in Figures~\ref{cont_supergiant}-\ref{cont_dwarf}, which present the overall shape of the spectra with different spectral types and luminosity classes, as well as the general trend of spectral features with spectral type. 
The wide wavelength range of IGRINS spectra, and variations in the flux shape for different spectral types, make it difficult to present absolute flux spectra in a single plot.
Instead, we used the normalized flux at 1.6~\um\ and 2.2~\um\ for the H- and K-bands, respectively. 
The full spectral grasp and resolution of the IGRINS Spectral Library is shown with the representative spectrum of HD 44391 (K0 Ib) in the Appendix.

\subsection{Spectral Indices}\label{subsec:spectral indices}
The identification of atomic and molecular features in spectra allows for the determination of stellar atmospheric parameters such as effective temperature (\teff), surface gravity (\logg), and metallicity ([Fe/H]).
Early-type stars have relatively few spectral lines in their NIR spectra compared to those of late-type stars.
The representative lines in the early-type stars are neutral hydrogen and neutral and ionized helium lines and some of those lines are known to have relations with stellar atmospheric parameters \citep{lenorzer04, hanson05}.
In this section, we focus on the main spectral features of late-type stars and present spectral lines that can be used as indicators of particular stellar atmospheric parameters.

\begin{figure}
\epsscale{0.3}

\includegraphics[width=12cm, height=4cm]{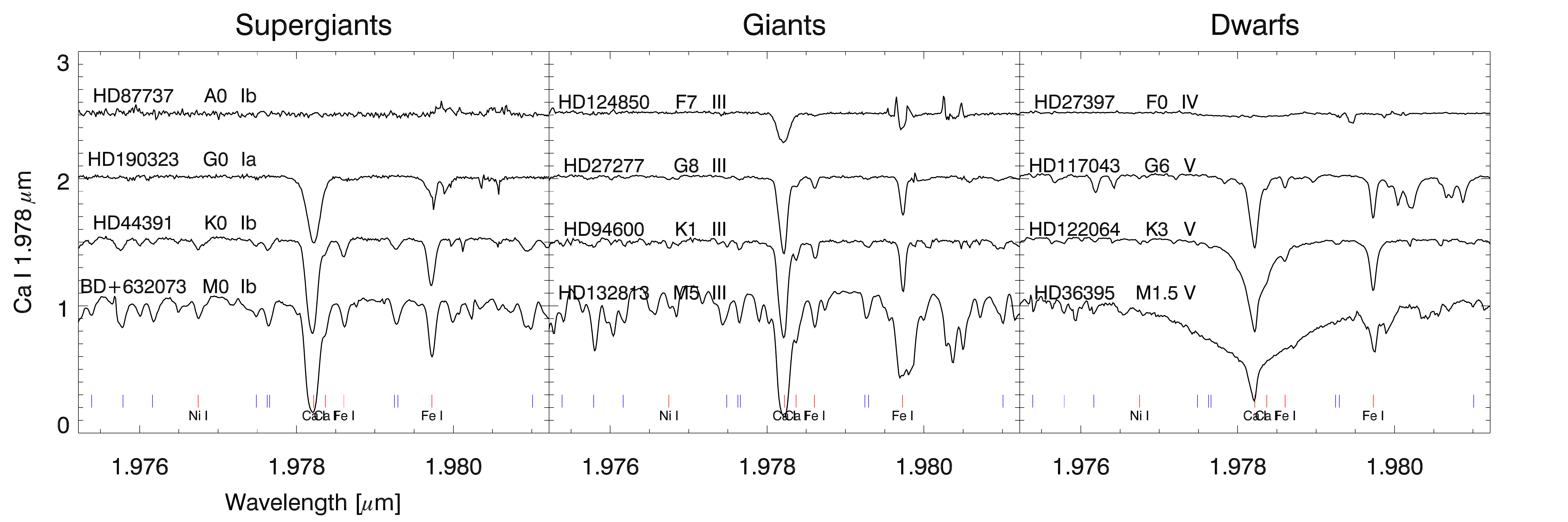}

\includegraphics[width=12cm, height=4cm]{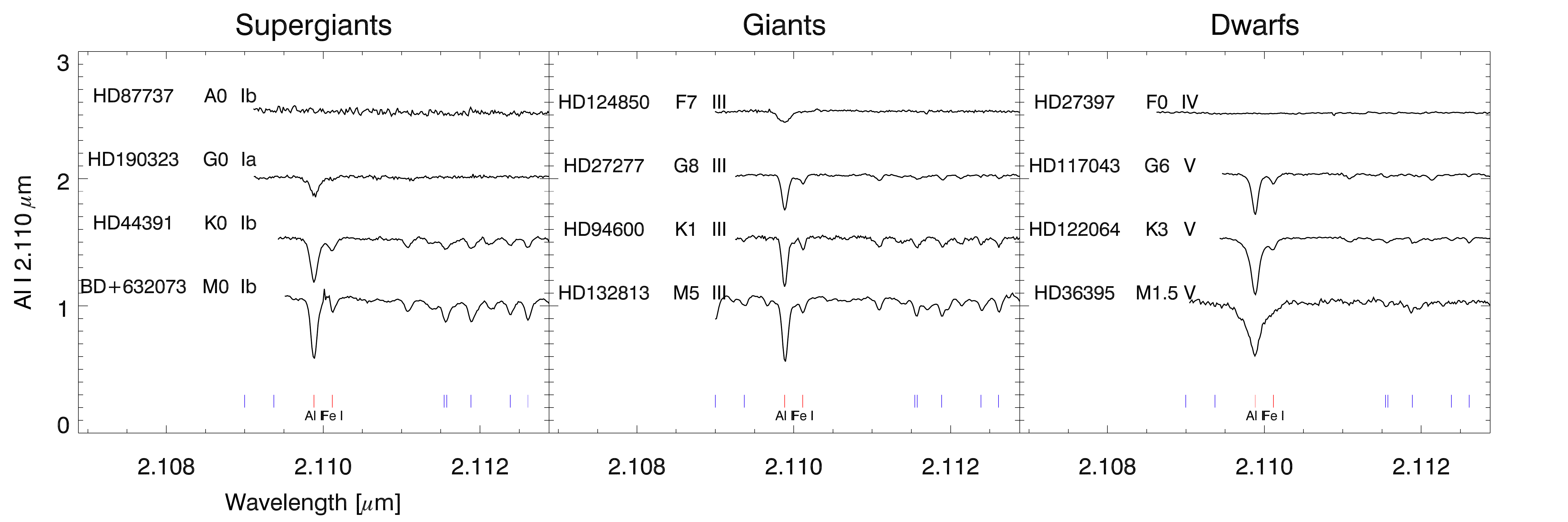}

\includegraphics[width=12cm, height=4cm]{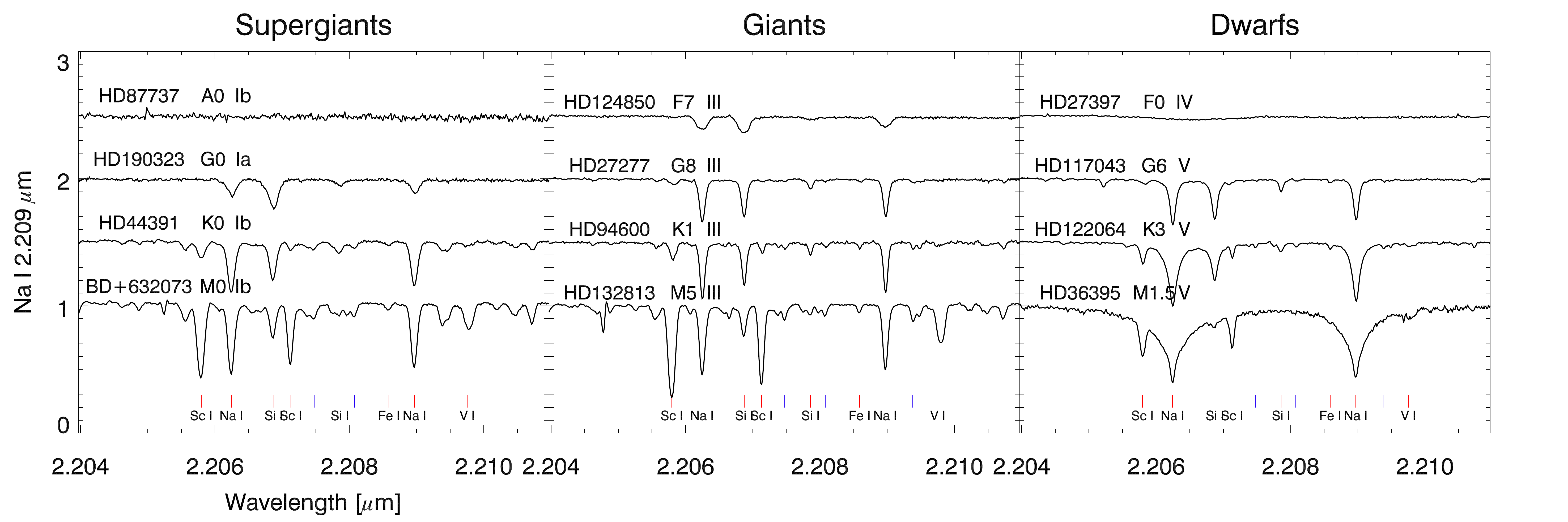}

\includegraphics[width=12cm, height=4cm]{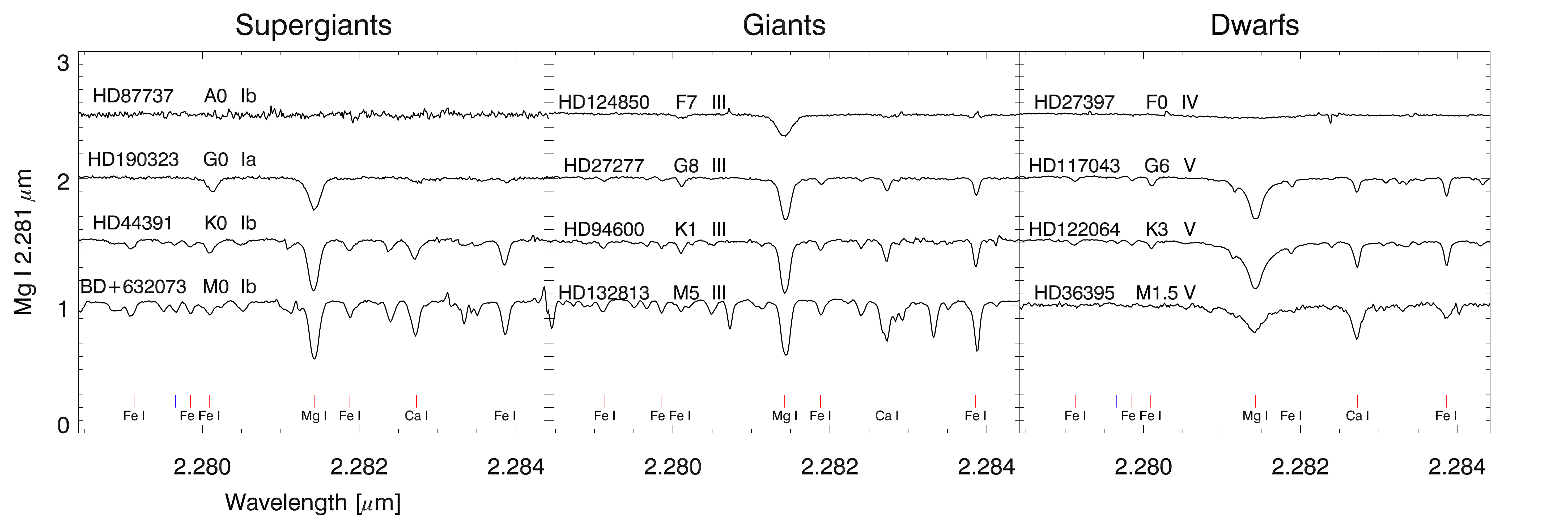}

\includegraphics[width=12cm, height=4cm]{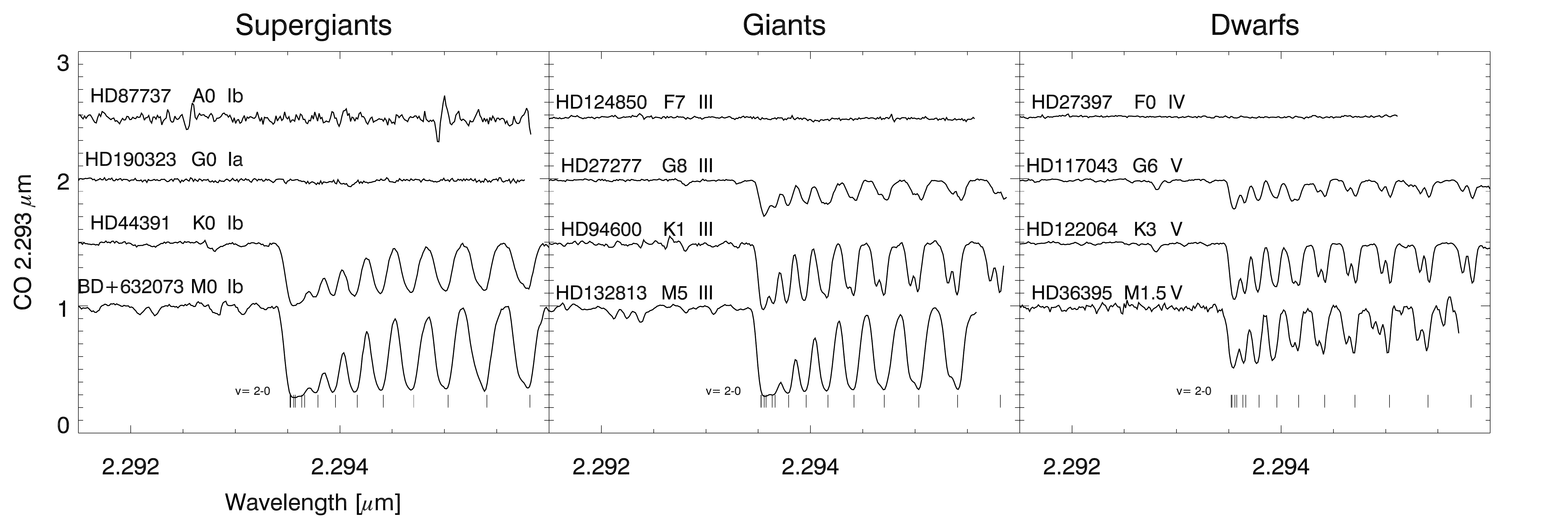}

\caption{
Comparing line strengths for different luminosity classes. 
From top to bottom, spectral features of Ca I 1.978, Al I 2.110, Na I 2.209, Mg I 2.281, and CO 2.293~\um\ are shown.
The strength of these lines increase with decreasing effective temperature. 
Line identification was made using the Arcturus line list from \citet{hinkle95}.
Red and blue sticks indicate the location of atomic lines and CN molecules, respectively.
The location of CO overtone rovibrational lines are marked with black sticks.
For the CO molecular features, HITRAN line lists \citep{HITRAN2016} were adopted.} \label{features}
\end{figure}

The most prominent lines of late-type stars in the K-band are Ca I at 1.978~\um, Al I at 2.110 ~\um, Na I doublet at 2.20~\um, Mg I at 2.281~\um, and the series of CO overtone bands starting at 2.293~\um\ (Figure~\ref{features}). 
Most of these lines are well-known to have relations between their EWs and stellar atmospheric parameters \citep{kleinmann86, ali95, ramirez97, ivanov00, ivanov04, cesetti13}. 
The IGRINS Spectral Library K-band spectra are of the highest quality between 2.08~\um\ to 2.46~\um\ because of dominant telluric absorption lines between 1.9-2.0~\um.
One exception is the Ca I 1.978~\um\ line, which is strong and resides in a relatively clear spectral region.
We used the EW of the Ca I 1.978~\um\ line for analysis, but we do not provide the spectra shortwards of 2.08~\um\ as part of the IGRINS Spectral Library.

\citet{cesetti13} defined spectral features as spectral indices if their EWs have correlations with stellar atmospheric parameters. 
Adopting those definitions, we measured the EWs of isolated lines to analyze the relation between the EWs and stellar atmospheric parameters.
We investigated some of the same features as \citet{cesetti13} but defined our own indices, with bandpasses suitable for the resolution of our spectra, and the bandpass of each spectral index is target-dependent.
We first used our spectra to test the usefulness of the spectral indices centered on Na I, Ca I, and CO first overtone bandhead, which show similar relations to what previous studies found (see Section 4.3).
We then searched for new spectral indices with the IGRINS spectra as listed in Table~\ref{tbl_indices}.

If the line corresponding to a spectral index of some star is blended with nearby spectral lines, we did not measure the EW of the spectral line for that star.
For example, the Fe I 2.226~\um\ line in HD 207089 is very strong, however, this line is contaminated by an adjacent spectral line. 
Because the spectral lines are intrinsically wide in this particular star (HD 207089), the contamination of this line is worse.
Therefore, we did not measure the EW of the Fe I 2.226~\um\ line for HD 207089.
The EW is defined as
\begin{equation} \label{eq_ew}
EW = \int{\Big(1-\frac{F_{line}}{F_{cont}}\Big)} d\lambda, 
\end{equation}
where $F_{line}$ is the observed line flux density and $F_{cont}$ is the continuum flux density at the central wavelength. 
Gaussian fitting was conducted to define the wavelength range (bandpass) of each line; the 3-sigma criterion was used for the bandpass, except for the CO band.
The EW was calculated by integrating the line flux over the bandpass defined by the 3-sigma criterion.
In the case of the CO~2.293~\um\ band, we used the integration range from 2.2933 to 2.2945~\um\ because this line is located at the edge of the order. 
The EWs were estimated with a Monte Carlo method in which we measured the EWs 100 times for each line, with random Gaussian errors multiplied by the observation errors in the spectrum.
The median value and the standard deviation derived from all 100 EW measurements were adopted as the EW of an individual line and the uncertainty of the EW, respectively.
For the analysis, we focused on 53 late-type stars (from F- to M-types), that have high S/N (mostly $\ge$ 200) and have atmospheric parameters available from the literature (43 stars) or have \teff\ calculated with the \teff-(B-V) relation (10 stars).
Table~\ref{tbl_ew_param} lists the computed EWs, errors, stellar parameters, and references of stellar parameters for late-type stars.

\subsection{Spectral Diagnostics}\label{subsec:spectral diagnostics}
\citet{kleinmann86} found several temperature and luminosity sensitive features of late-type stars in the K-band, such as Na I, Ca I, Br$\gamma$, CO and H$_{2}$O.
There are many other studies for spectral indices by \citet{lancon92}, \citet{origlia93}, \citet{ali95}, \citet{ramirez97}, \citet{meyer98}, \citet{ivanov04}, \citet{marmol08}, \citet{rayner09}, and \citet{cesetti13}.
Here we apply these known diagnostic tools and present some new ones at wavelengths covered by IGRINS.

\subsubsection{Early-Type Stars}\label{subsubsec:early-type stars}
\begin{figure}
\epsscale{1.0}
  \plotone{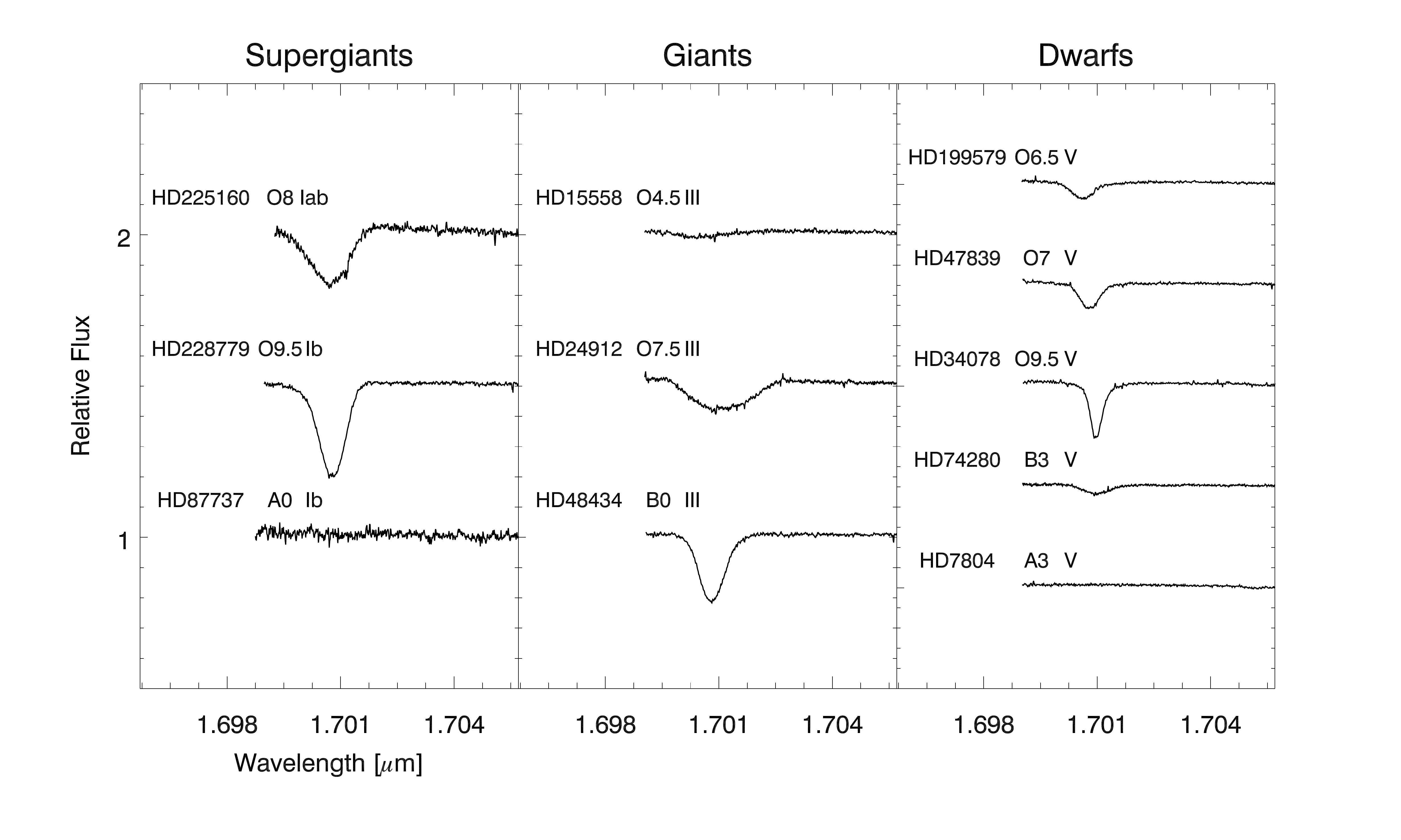} 
  \caption{
  The He I 1.701~\um\ line for different spectral types and luminosity classes of hot stars.
\label{heI1701}
}
\end{figure}

\begin{figure}
\epsscale{1.0}
  \plotone{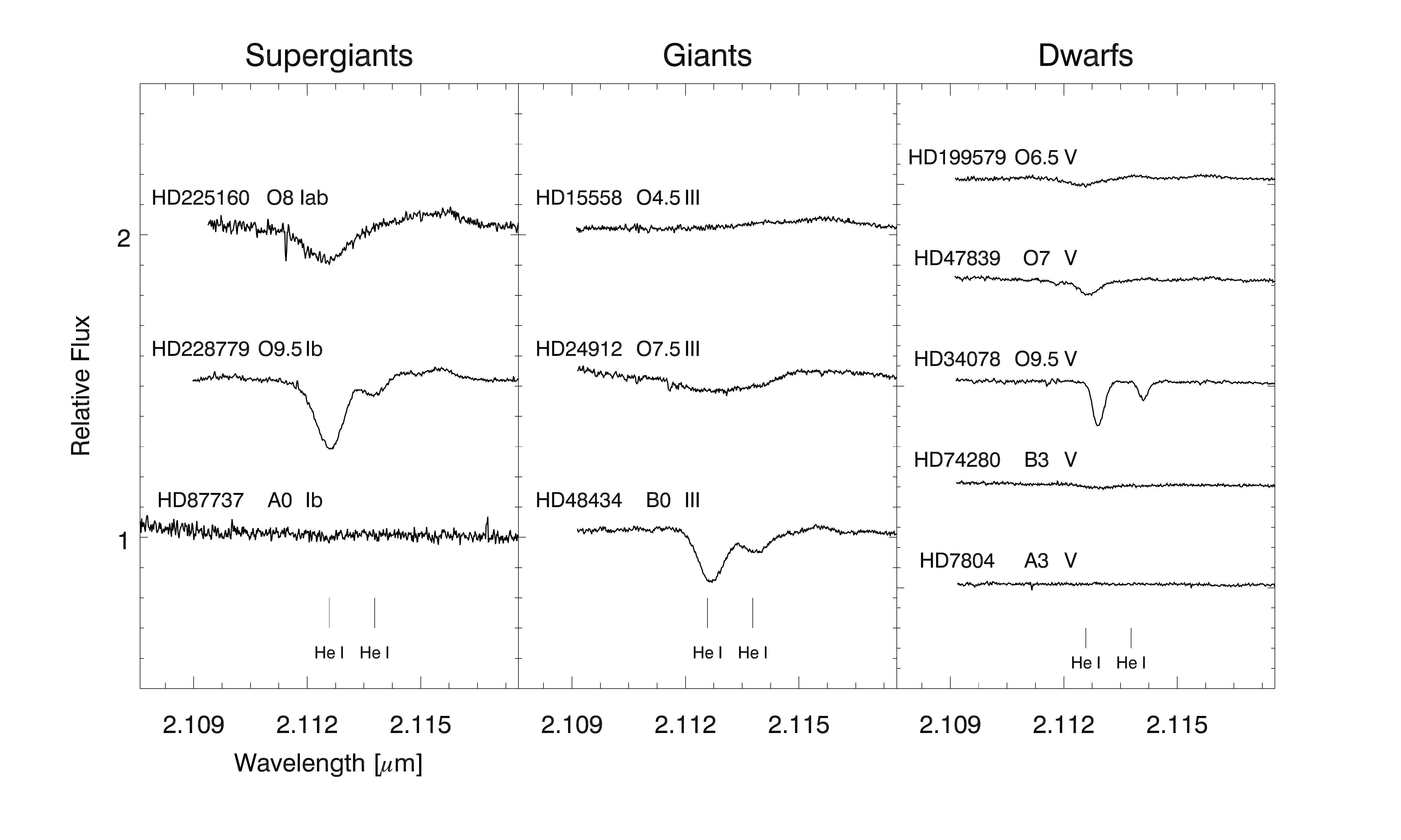} 
  \caption{
  The He I 2.113~\um\ and 2.114~\um\ lines for different spectral types and luminosity classes of hot stars.
\label{heI2113}
}
\end{figure}

\begin{figure}
\epsscale{1.0}
  \plotone{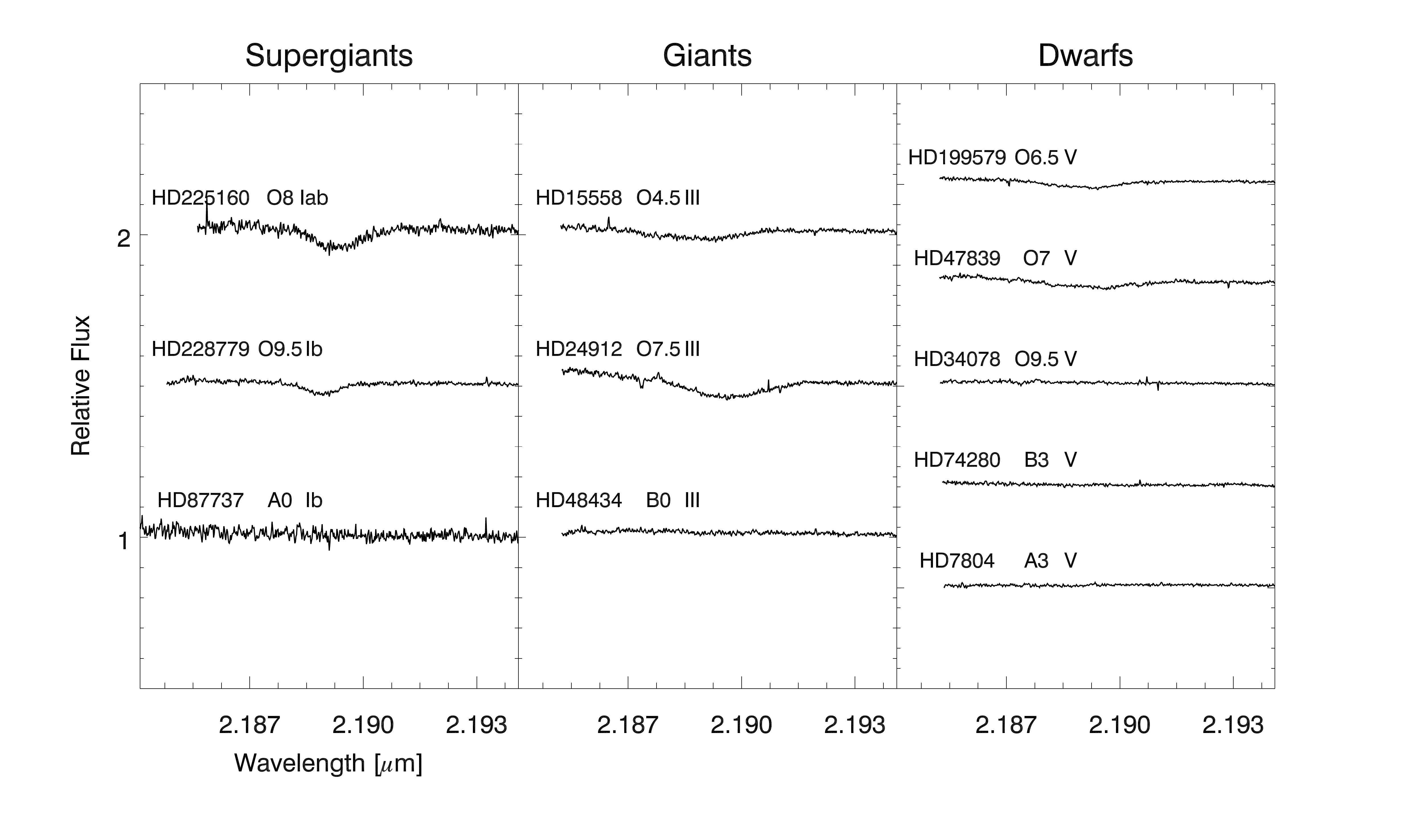} 
  \caption{
 The He II 2.189~\um\ line for different spectral types and luminosity classes of hot stars.
\label{heII2189}
}
\end{figure}

Although our main focus is late-type stars, our sample covers about 20 early-type stars. 
Here we provide a summary of dominant spectral features in early-type stars, covered by IGRINS, without detailed analysis.
\citet{lancon92} presented a NIR spectral library for early-type stars, and since then numerous libraries of early-type stars have been added to the literature \citep{dallier96, hanson96, repolust05}.
More recently, \citet{hanson05} presented an intermediate-resolution NIR library of O- and early-B type stars and identified a number of spectral indices in the H- and K-bands.
For example, in the O-type stars, He I and He II lines are reliable temperature indicators. 
The strongest He I lines observed with IGRINS are at 1.701~\um\ and 2.113~\um.
\citet{lenorzer04} and \citet{hanson05} found that the strengths of these lines increase with decreasing \teff\ in the temperature range of 25,000-50,000 K. 
Moreover, the spectral features of He I at 1.701~\um\ and 2.113~\um\ change with luminosity class \citep{lenorzer04}.
The IGRINS spectra of early-type stars show the same trends as previous studies.
Figures~\ref{heI1701} and \ref{heI2113} show the IGRINS spectra of the two He I lines at 1.701~\um\ and 2.113~\um\ in different luminosity classes of O-, B-, and A-type stars.
In addition, one of the strongest ionized helium lines in the K-band, He~II~2.189~\um, is present in all O-type stars in absorption \citep{conti93, hanson05} whereas this line does not appear in the B- and A-type stars (Figure~\ref{heII2189}).

\begin{figure}
\epsscale{0.7}
\plotone{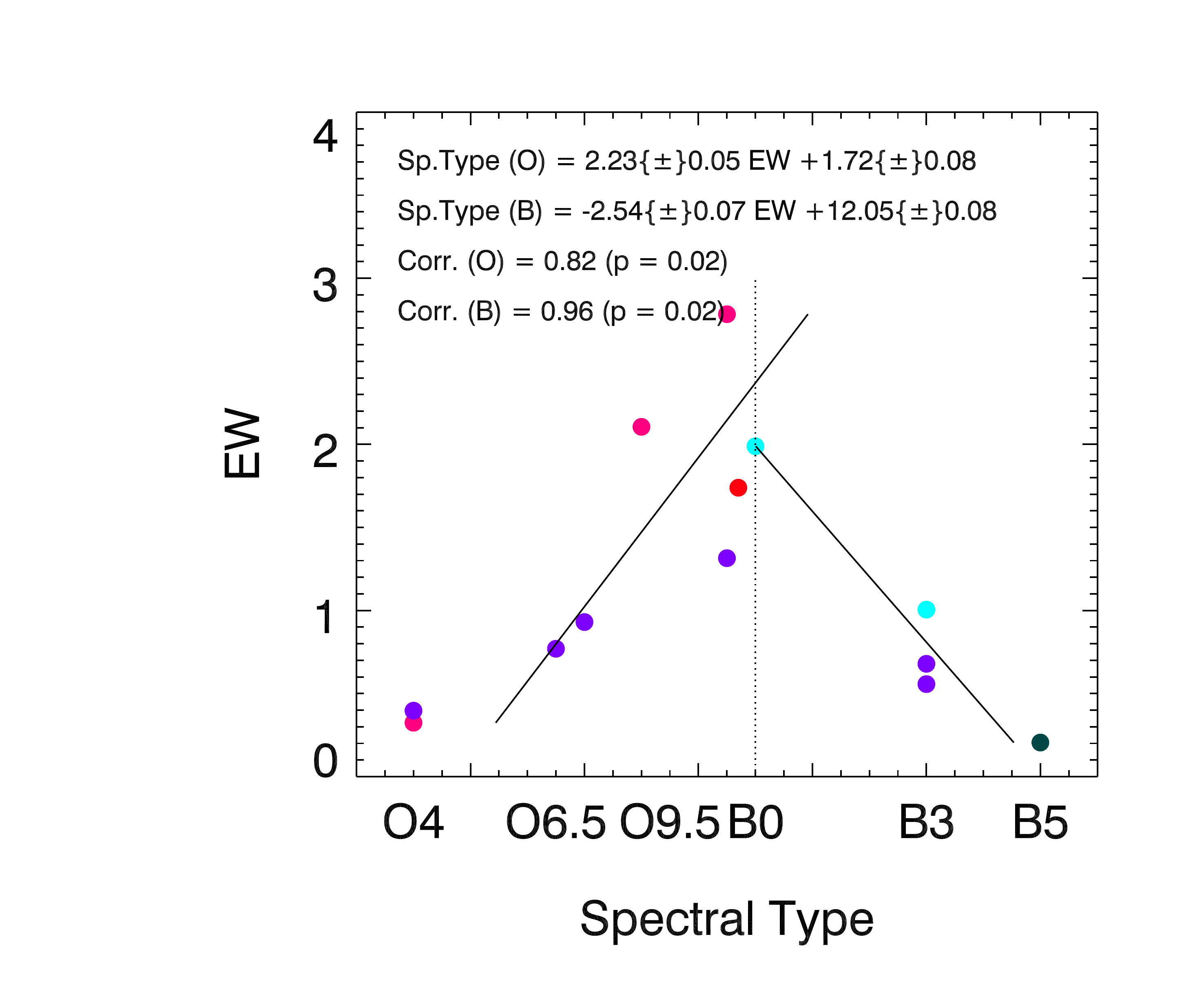}
\caption{
Equivalent widths of He I 1.701~\um\ as a function of spectral type.
The values in the parentheses indicate p-value. 
The different colors correspond to the supergiant (magenta), bright giant (red), giant (cyan), subgiant (green), and dwarf stars (purple).
The dotted line indicates the border between O- and B-type stars. 
\label{heI_ew}
}
\end{figure}

The relation between the EW of He~I~1.701~\um\ and spectral type is shown in Figure~\ref{heI_ew}.
Thirteen stars (O4 - B5) show the He I 1.701~\um\ line in absorption and only five stars have known stellar parameters. 
Therefore, we used spectral type instead of \teff.
The EW of He~I~1.701~\um\ increases towards later spectral type in O-type stars, however, the relation shows a turnover at B-type stars.
Both of these lines have strong correlations with \teff\ (Pearson's correlation coefficient of r=0.82 (O-type) and r=0.96 (B-type)).
Furthermore, we calculated p-values because of the small number of data points, obtaining 0.02 for both of O- and B-type stars.
The p-value is used for testing null hypothesis to quantify the evidence of statistical significance. 
The smaller p-value indicates that the significance is higher \citep{moore08}.
These values provide sufficient evidence to conclude that the EW of He~I~1.701~\um\ line is well correlated with \teff.
The measured EWs and errors are listed in Table~\ref{tbl_hei}.

\subsubsection{Late-Type Stars}\label{subsubsec:late-type stars}
\begin{figure}
\epsscale{.9}
\plotone{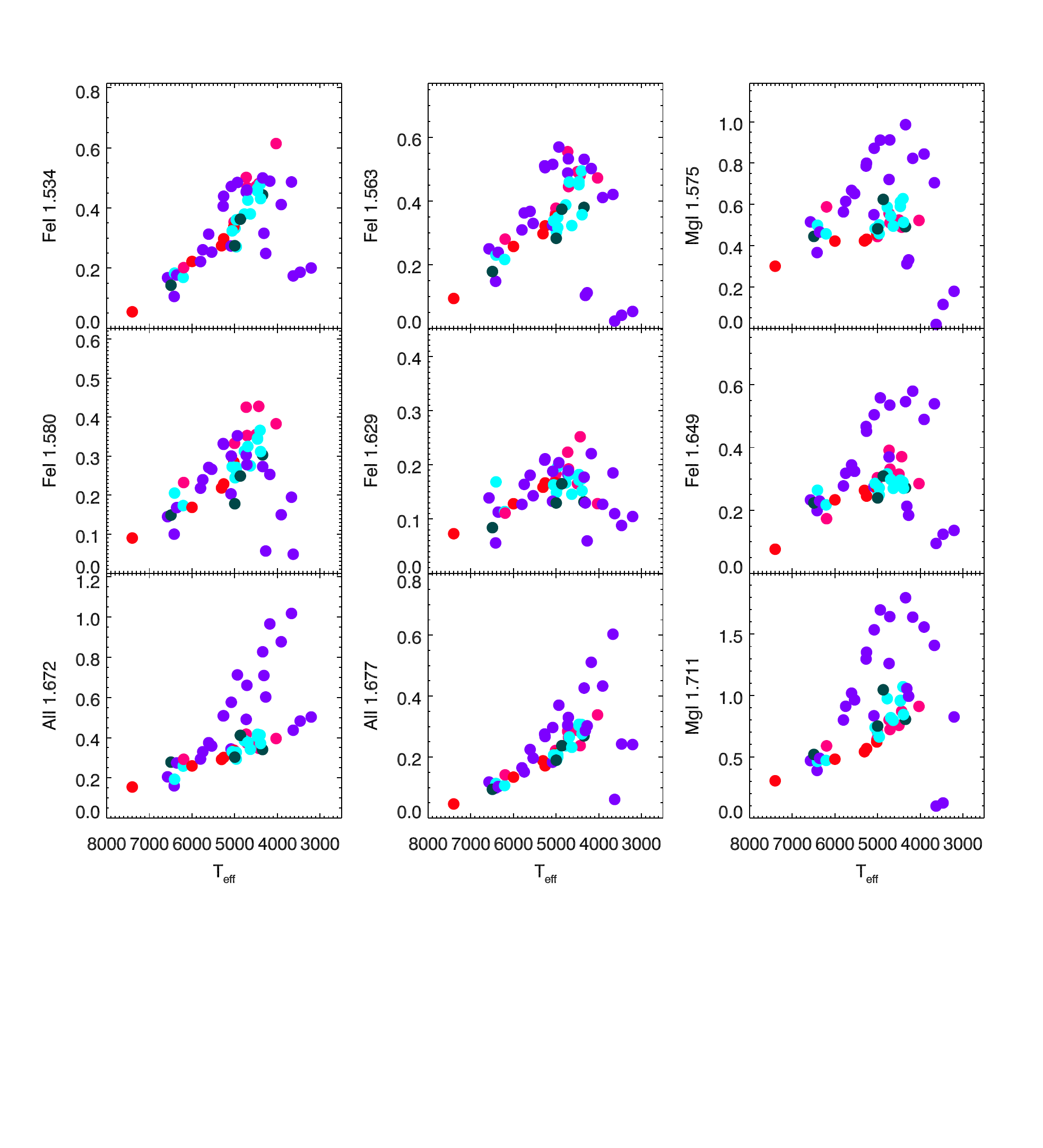}			
\caption{
Equivalent widths of the lines corresponding to H-band spectral indices as a function of effective temperature.
The different colors correspond to the supergiant (magenta), bright giant (red), giant (cyan), subgiant (green), and dwarf stars (purple).
The EWs of Fe I, Mg I, and Al I lines increase as \teff\ decreases although some cool dwarfs (\teff\ $\le$ 5000~K) are off the relation.
\label{ew_H_Teff}
}
\end{figure}

The EWs of lines corresponding to spectral indices for late-type stars in the H-band are plotted as a function of \teff\ and \logg\ in Figure~\ref{ew_H_Teff} and Figure~\ref{ew_H_logg}, respectively.
Previous studies of H-band spectra are more limited than those in the K-band.
\citet{origlia93} presented the H-band spectral indices corresponding to Si~I~1.59~\um\ and the CO~v=6-3~bandhead at 1.62~\um.
\citet{meyer98} identified numerous atomic and molecular spectral indices (see their Table 4) corresponding to Mg~I~1.575~\um, Al~I~1.672~\um,~1.677~\um, and Mg~I~1.711~\um.
\citet{ivanov04} also studied spectral indices corresponding to Fe~I~1.580~\um, Mg~I~1.711~\um\ and other lines (see their Table 4).

As shown in Figure~\ref{ew_H_Teff}, the EWs of Fe~I, Mg~I, and Al~I~lines increase as \teff\ decreases although some cool dwarfs (\teff\ $\le$ 5000~K) are off the relation. 
Figure~\ref{ew_H_logg} shows a plot of the EWs as a function of \logg.
Especially, the EWs of Fe~I~1.534~\um, Al~I~1.672~\um\ and Al~I~1.677~\um\ lines show tight correlations, except dwarfs $\leq$ 4000~K, with \teff\ while they show marginal correlation with \logg\ (see Section 4.4 for details). 
Therefore, Fe~I~1.534~\um, Al~I~1.672~\um\ and Al~I~1.677~\um\ lines can be used as \teff\ indicators of late-type stars from F- to M-types.
Most of the spectral indices show no notable trend with \logg.

\begin{figure}
\plotone{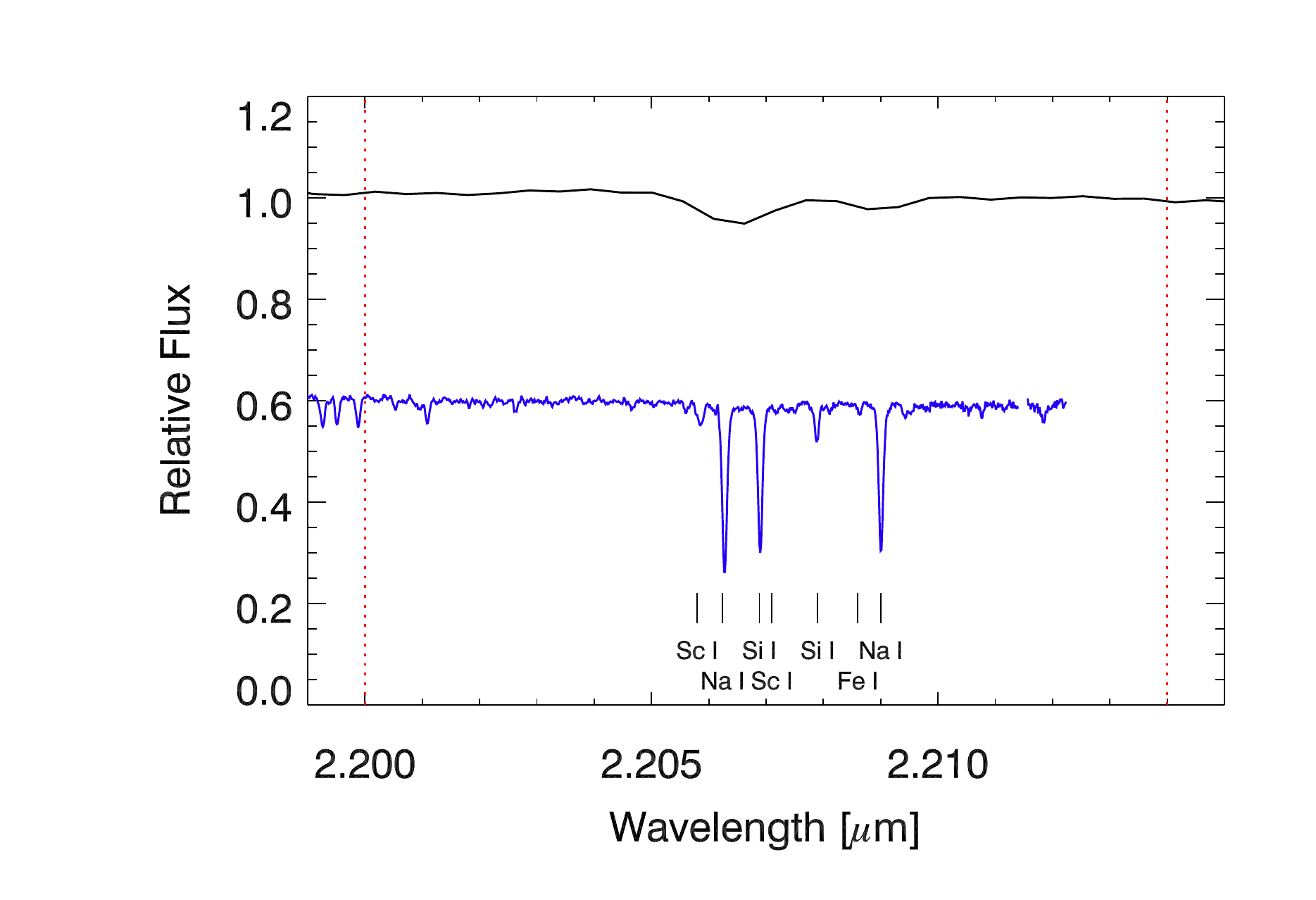}
\caption{
Comparison of IRTF/SpeX (upper) and IGRINS (lower) spectra of HD 27277 (G8 III).
The moderate-resolution IRTF/SpeX spectrum does not resolve the Na I doublet and nearby spectral lines such as Sc I at 2.2058~\um\ and Si I at 2.2068~\um, while the high-resolution IGRINS spectrum shows each spectral feature clearly.
The red dashed line indicates the EW bandpass employed to study the IRTF/SpeX spectrum \citep{cesetti13}.
\label{comp_spec}
}
\end{figure}

\begin{figure}
\epsscale{1.1}
\plottwo{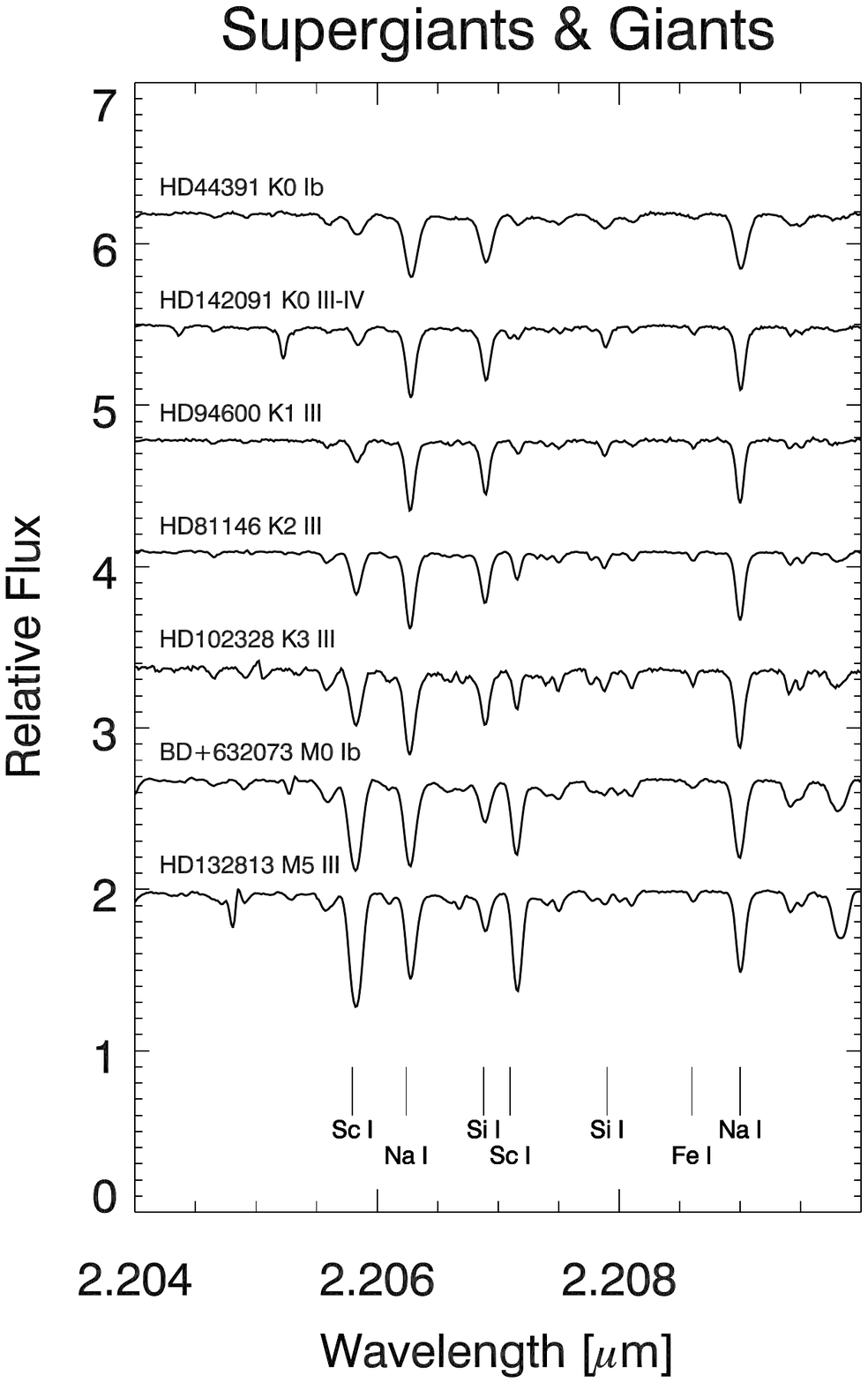}{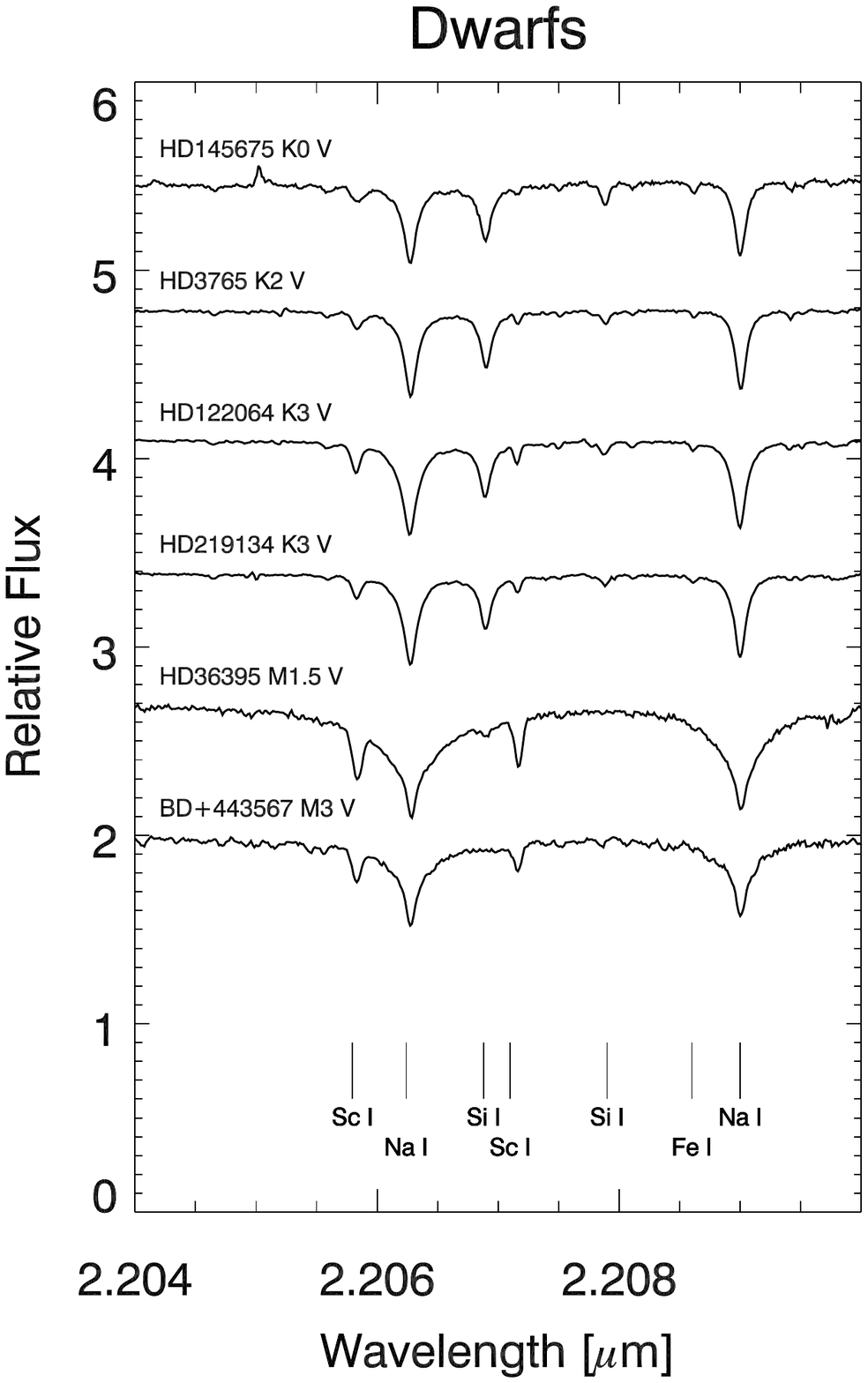}
\caption{
Si and Sc lines in the K-band IGRINS spectra for different spectral types.
The strength of the Si~I~2.2069~\um\ line relative to that of the Sc~I~2.2071~\um\ line decreases with decreasing effective temperature \citep{ramirez97, doppmann03}.
 \label{sisc}
 }
 \end{figure}

The EWs of lines corresponding to the spectral indices in the K-band are plotted as a function of \teff\ and \logg\ in Figures~\ref{ew_K_Teff}-\ref{ew_K_logg}.
As shown in Figure~\ref{ew_K_Teff}, the EWs of Ca I, Al I, Na I, Ti I, Mg I, and CO overtone features increase as \teff\ decreases from F- to M-type.
Based on the trend, these spectral indices can be used as approximate indicators of \teff\ (spectral type).
Especially, relatively weak Ti I lines (2.222, 2.224, 2.232, and 2.245~\um) can be used as \teff\ indicators for stars with \teff\ $\le$ 6000~K.
On the other hand, strong Ca I 1.978~\um, Al I 2.110~\um\ and Na I 2.209~\um\ lines can be applied for the wide range of \teff\ from 3000 to 7000~K.
In the case of Ca I 2.266~\um, the spectral line is contaminated at spectral types later than K4.5. 
Therefore, we only provide the EW of the Ca I 2.266~\um\ line of stars whose spectral types are earlier than K4.5.
The Na I doublet at 2.20~\um\ becomes stronger with spectral type from F- to M-type \citep{kleinmann86, ali95, ramirez97, ivanov00, ivanov04, rayner09, rojas12, cesetti13}.
In the previous studies with intermediate-resolutions, the Na~I~doublet features were not resolved, therefore, they provided a blended Na~I~2.20~\um\ feature as an indicator of \teff.
In addition, \citet{cesetti13} measured the EW of Na~I~2.20~\um\, as shown in Figure~\ref{comp_spec}, however, there are several metal features around the Na I doublet that must be considered \citep[Figure~\ref{comp_spec}; see their Table 7 in][]{ramirez97}. 
With high-resolution IGRINS spectra, the Na I doublet is resolved and we measure the EW of each Na~I~line~(2.206~\um\ and 2.209~\um). 
Both of the Na~I~2.206~\um\ and~2.209~\um\ lines show similar trends with respect to \teff, but we present only Na~I~2.209~\um\ here.
In the late K- or M-type dwarfs, Na I 2.206~\um\ is contaminated by nearby Si/Sc lines, whose ratio decreases as \teff\ decreases in dwarfs \citep{doppmann03}. 
The IGRINS spectra in Figure~\ref{sisc} show the same line proximities for all luminosity classes.
Therefore, by resolving adjacent spectral lines we can provide more accurate spectral indicators of stellar atmospheric parameters.

Figure~\ref{ew_K_logg} shows the plot of EWs as a function of \logg.
Except for dwarfs with \logg\ $>$ 4, most of the spectral indices tend to decrease with \logg. 
The CO~overtone~band~2.293~\um\ shows the strongest \logg\ dependency.

\subsection{New Spectral Indicators}\label{subsec:new spectral indicators}
\begin{figure}
\epsscale{.9}
\plotone{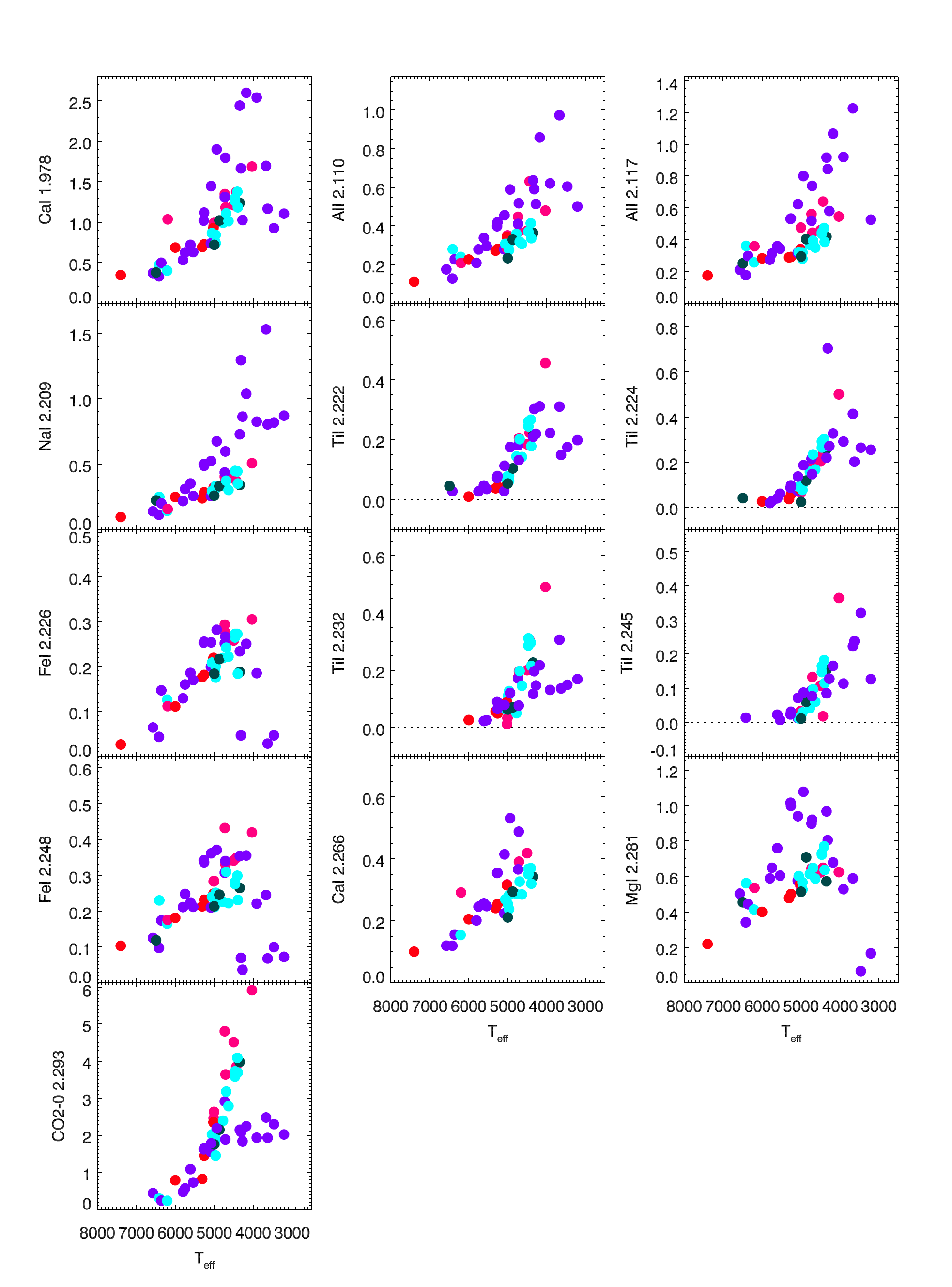}	
\caption{Equivalent widths of the lines corresponding to K-band spectral indices as a function of effective temperature.
The different colors correspond to the supergiant (magenta), bright giant (red), giant (cyan), subgiant (green), and dwarf stars (purple).
Zero level is denoted by a dashed line.
The EWs of Ca I, Al I, Na I, Ti I, Mg I, and CO overtone features increase as \teff\ decreases.
\label{ew_K_Teff}
}
\end{figure}

\begin{figure}
\epsscale{1.0}
  \plotone{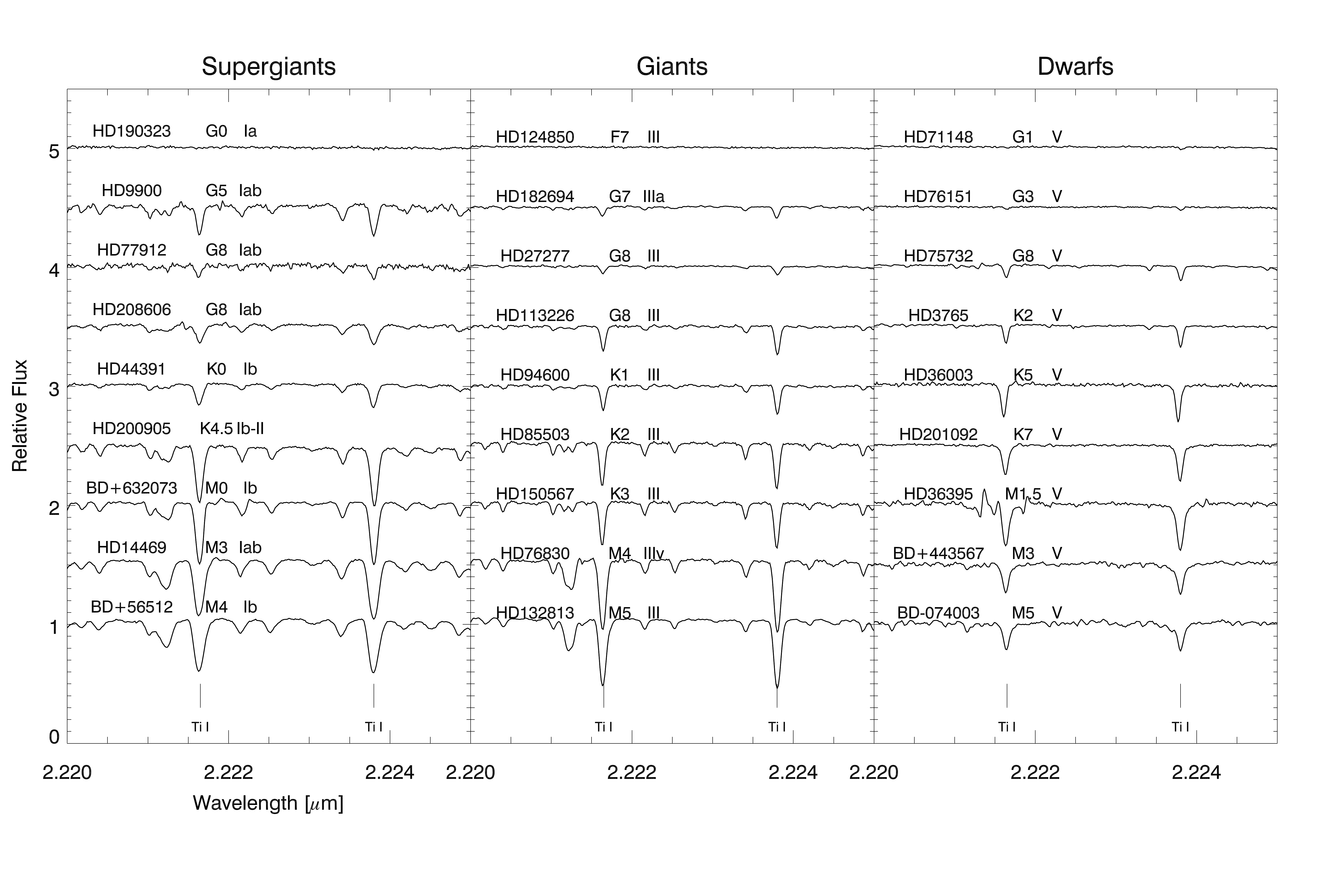}
  \caption{
  Neutral titanium features at 2.222 and 2.224~\um\ for different spectral types and luminosity classes.
  As shown in Figure~\ref{ew_K_Teff}, the strength of Ti I lines increases with decreasing effective temperature.
\label{ti_spectra}
}
\end{figure}

By combining the broad spectral grasp of IGRINS with our investigation of spectral indices in the literature, we found new H- and K-band spectral indicators for \teff\ and \logg.
We note that while the strengths of metal lines are sensitive to the \teff\ \citep{gray05, takeda02}, the EWs of strong lines can be overestimated by variations in microturbulence \citep{gray05}.
For this reason, the EW measurements of weak lines are important for providing a more accurate \teff\ indicator.
To detect weak lines, we need to use high-resolution spectra, like those from IGRINS, to avoid line blending.
We have detected several Ti~I~lines (2.190, 2.201, 2.222, 2.224, 2.228, 2.232, and 2.245~\um) in the K-band that are relatively weaker and narrower than other neutral metal lines.
In addition, each of the Ti~I~lines shows a strong correlation with \teff. 
Ti~I~2.222~\um\ and Ti~I~2.224~\um\ show the tightest correlations (Figure~\ref{ew_K_Teff}).
Figure~\ref{ti_spectra} shows the Ti~I~2.222 and 2.224~\um\ features for different spectral types and luminosity classes.

\begin{figure}
\epsscale{1.0}
  \plotone{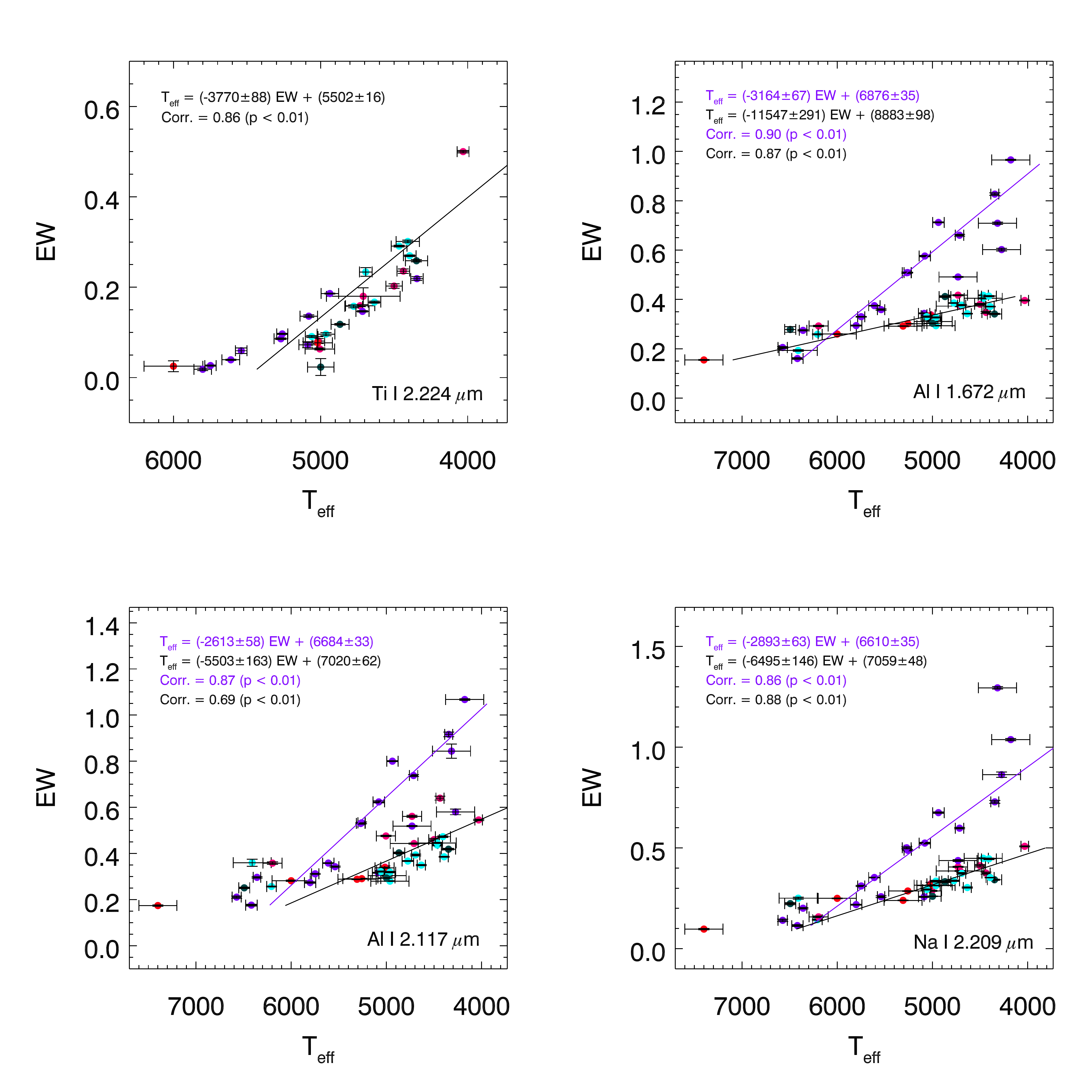}
    \caption{ 
Equivalent widths of the lines corresponding to four spectral indices (Ti I~2.224, Al I~1.672, Al I~2.117, and Na I~2.209~\um) as a function of effective temperature.
The different colors correspond to the supergiant (magenta), bright giant (red), giant (cyan), subgiant (green), and dwarf stars (purple).
These four spectral indices have tight correlations with \teff\ (see Table~\ref{tbl_corr}).
\label{teff_indicators}  
}
\end{figure}

Based on 53 stars with physical parameters we derive empirical relations between the EWs.
We adopted 31 stars which have measurable and unblended Ti I lines and \teff\ \textgreater~4000 K.
As depicted in Figure~\ref{ti_spectra}, Ti I lines appear in cool stars, whose spectral types are later than G2 (\teff\ $\le$ 6000~K). 
Here, we present only the representative result of Ti I 2.224~\um, the most isolated Ti I line.
Figure~\ref{teff_indicators} (upper left) shows the relation between the EW of Ti I 2.224 \um\ and \teff.
The sudden increase in the EW of Ti~I~2.224~\um\ with decreasing \teff\ implies that this line can be used as a \teff\ indicator.
We obtained the relation between the EW of Ti~I~2.224~\um\ and the \teff\ using the linear least square fitting,
\begin{equation}\label{eq_tiI}
{ T_\mathrm{eff} = (- 3770 \pm 88) EW(Ti I 2.224) + (5502 \pm 16). }
\end{equation}
The EW of Ti~I~2.224~\um\ and \teff\ show a strong correlation with the Pearson's correlation coefficient (r) of 0.86 (p-value \textless~0.01) while the correlation between the EW of Ti~I~2.224~\um\ and \logg\ is moderate (r=0.48, p-value=0.01).
Moreover, according to the regression fitting analysis among the EW of the Ti~I~2.224~\um, \teff, and \logg ($ EW(Ti) = -2.32 \log T_\mathrm{eff} -0.003 \log g_\mathrm{Fit} + 8.69$), \teff\ affects the line strength of Ti~I~2.224~\um\ much more significantly than \logg\ by a factor of 800.
Therefore, by measuring the EW of Ti~I~2.224~\um\ we can empirically estimate the \teff\ of a star.

We have applied the same analysis to several other spectral features such as Al~I~1.672, Al~I~2.117, Na~I~2.209 (Figure~\ref{teff_indicators}), and CO~2.293~\um\ (Figure~\ref{co_ew}).
We find that the atomic metal lines, such as Fe~I, Al~I, Ca~I and Na~I show a turnover around 4000~K, especially for dwarfs (Figure~\ref{ew_H_Teff} and Figure~\ref{ew_K_Teff}).
The strength of these lines increases with decreasing \teff\ until around 4000~K, at which point they decrease with decreasing \teff.
As a result, we adopted 48 stars which have \teff~\textgreater~4000~K for the analysis.
Furthermore, the relation between \teff\ and EWs varies depending on the luminosity class.
However, the overall trend in supergiants, bright giants, giants and subgiants is similar.
Therefore, we derived two separate empirical equations for dwarfs and the other luminosity classes (from supergiants to subgiants).
The derived empirical equations for \teff\ are listed in Table~\ref{tbl_diagnostics}. 

\begin{figure}
\epsscale{1.0}
  \plotone{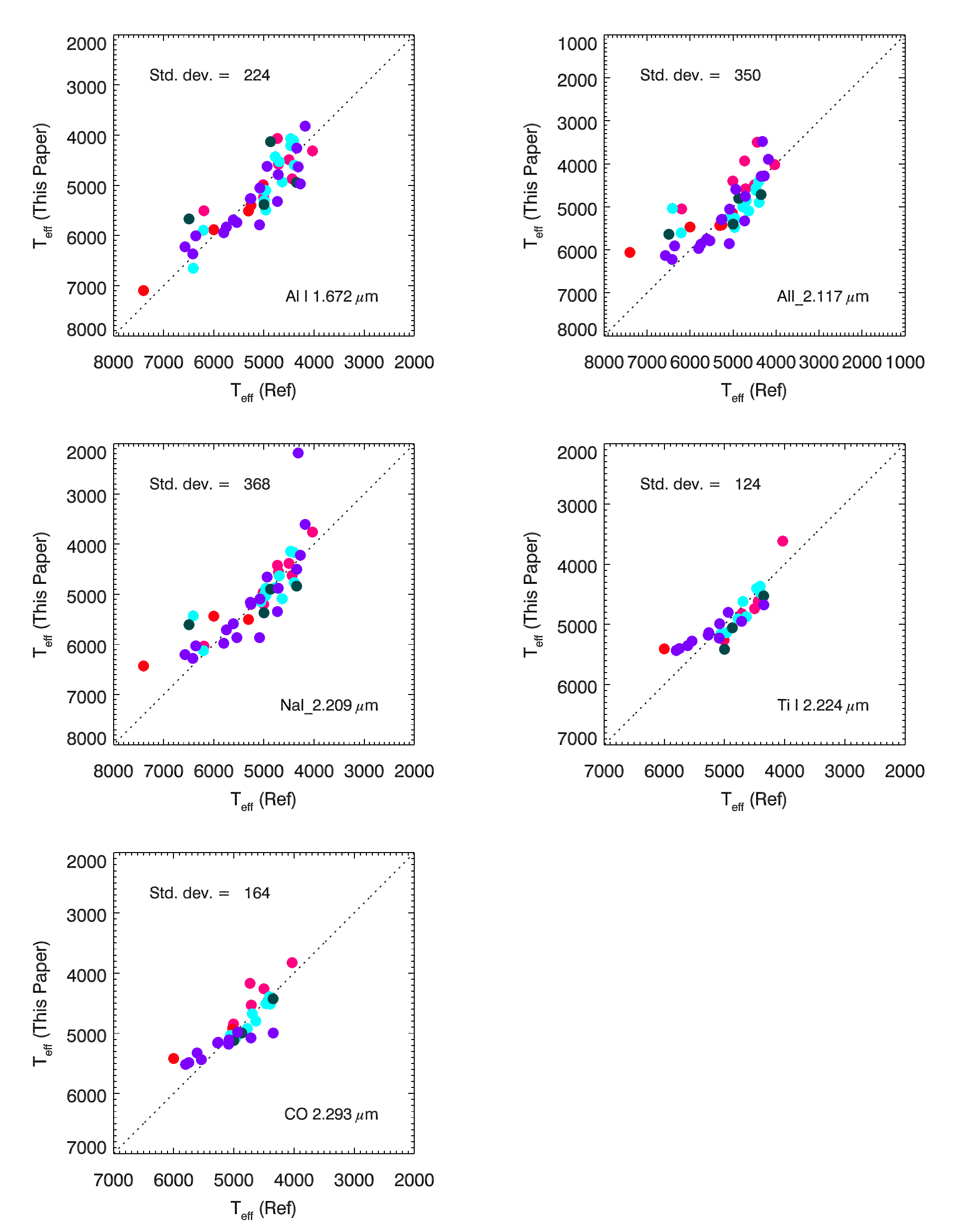}
  \caption{Comparisons between the effective temperatures calculated by equations in Table~\ref{tbl_diagnostics} ($T_{eff}$ (This Paper)) and those adopted from references ($T_{eff}$ (Ref)).
  Ti I 2.224~\um\ shows the smallest difference between two values.
  Therefore, among these spectral indices, Ti I 2.224~\um\ is the most sensitive line to the \teff.
  The dotted line in each panel indicates equivalent $T_{eff}$ (This Paper) and $T_{eff}$ (Ref) values.
\label{comp_teff}  
  }
\end{figure}

Figure~\ref{teff_indicators} shows the relations between the EWs of several lines and \teff.
In cases of the Al~I~1.672~\um, Al~I~2.117~\um, and Na~I~2.209~\um\ lines, we used 48 stars for the relation between the EW and \teff\ because more stars show clear detections of those lines even in hotter stars (up to 7000~K).
However, in case of the CO~2.293~\um\ feature (Figure~\ref{co_ew}), we used the same 31 stars as in the Ti~I~2.224~\um\ line analysis because this bandhead appears in cooler stars. 
As shown in Figures~\ref{teff_indicators} and~\ref{co_ew}, these spectral indices have tight correlations with \teff. 
The calculated correlation coefficients and p-values are listed in Table~\ref{tbl_corr}.
The correlation coefficients and p-values show that these spectral indices are strongly correlated with \teff.
As a result, these four lines (Al~I~1.672~\um, Al~I~2.117~\um, Na~I~2.209~\um, and CO~2.293~\um) can be used as \teff\ indicators of late-type stars.
To confirm the derived \teff\ from equations in Table~\ref{tbl_diagnostics} we compared the calculated \teff\ with that adopted from references (Figure~\ref{comp_teff}).
According to Figure~\ref{comp_teff}, the average standard deviation is a few hundred~K.
Among them, \teff\ derived by the EW of the Ti~I~2.224~\um\ shows the smallest difference (standard deviation of 124~K) from the reference \teff.
Thus, Ti~I~2.224~\um\ is the most sensitive line to the \teff.

Furthermore, the CO~overtone band (v=2-0) is a useful indicator of luminosity class \citep{baldwin73, kleinmann86, lancon92, cesetti13}.
Figure~\ref{co_ew} shows a clear correlation between the EW of CO~2.293~\um\ and \logg\ as well as \teff, unlike other metal lines.
Therefore, we derived an empirical equation below for \logg\ using the measured EW of the CO~2.293~\um\ and \teff\ from the literatures:
\begin{equation}\label{eq_regress}
	\log g_\mathrm{Fit} = (- 2.130 \pm 0.003) EW(CO) + (- 41.79 \pm 0.05) \log T_\mathrm{eff} + (162.54 \pm 0.18).
\end{equation}

\begin{figure}
\epsscale{0.8}
\plotone{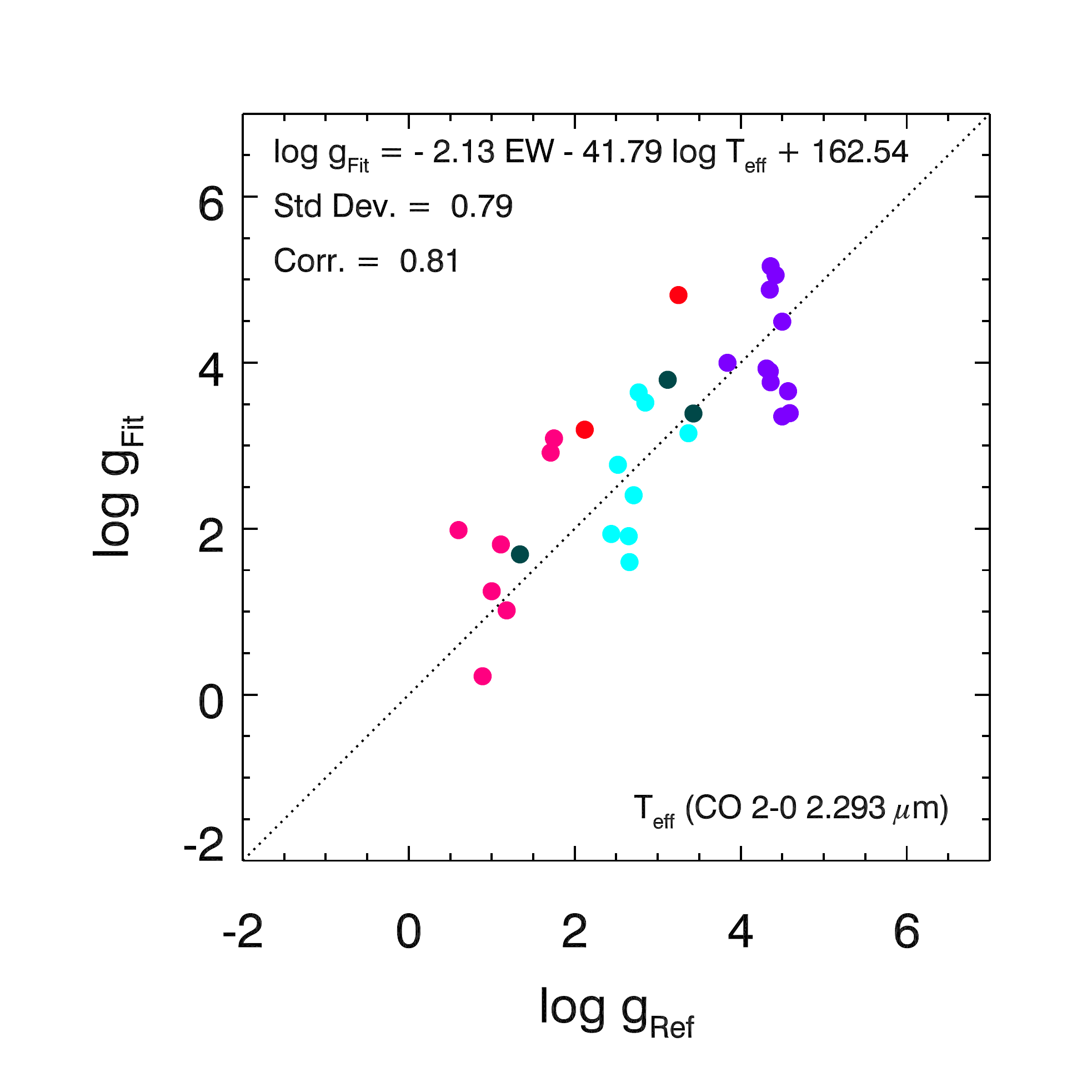}
  \caption{Plot of $\log g_\mathrm{Ref}$ vs. $\log g_\mathrm{Fit}$, where
  $\log g_\mathrm{Ref}$ was adopted from the literature and $\log g_\mathrm{Fit}$ was calculated using Equation~\ref{eq_regress}. 
 The \teff\ used to calculate $\log g_\mathrm{Fit}$ is derived from the EW of CO 2.293~\um.
Once we measure the EWs of Ti I 2.224~\um\ or other \teff\ indicators in addition to CO 2.293~\um, we can empirically estimate both of \teff\ and \logg\ of a star without a stellar atmospheric model using the equations in Table~\ref{tbl_diagnostics} and Equation~\ref{eq_regress}, respectively.
  Magenta, red, cyan, green, and purple colors represent supergiants, bright giants, giants, subgiants, and dwarfs, respectively.
  The dotted line indicates equivalent $\log g_\mathrm{Fit}$ and $\log g_\mathrm{Ref}$ values.
\label{co_regress}
}
\end{figure}

Figure~\ref{co_regress} shows the relation between the reference \logg\ ($\log g_\mathrm{Ref}$) and the \logg\ ($\log g_\mathrm{Fit}$) estimated with Equation~\ref{eq_regress}.
In Figure~\ref{co_regress} we have adopted the \teff\ derived from the equations in Table~\ref{tbl_diagnostics}, along with the EW of CO~2.293~\um, to calculate the $\log g_\mathrm{Fit}$.
The standard deviation of the differences between $\log g_\mathrm{Ref}$ and $\log g_\mathrm{Fit}$ is 0.79.
Alternatively, a more accurate \teff\ can be derived by using Ti~I~2.224~\um\ (Equation~\ref{eq_tiI}) to calculate $\log g_\mathrm{Fit}$. 
In this case, the standard deviation of the differences between $\log g_\mathrm{Ref}$  and $\log g_\mathrm{Fit}$ is 0.83.
Therefore, once we measure the EWs of Ti~I~2.224~\um\ or other \teff\ indicators in addition to CO 2.293~\um, we can empirically estimate both of \teff\ and \logg\ of a star without a stellar atmospheric model using the equations in Table~\ref{tbl_diagnostics} and Equation~\ref{eq_regress}, respectively. 

Unlike the CO feature, other lines do not show a strong correlation with \logg\ for all luminosity classes. 
However, the Al~I~1.672 \um\ and Na~I~2.209 \um\ line correlate weakly with \logg\ for dwarfs. 
The correlation coefficients between the EWs and \logg\ are listed separately for dwarfs and the other luminosity classes in Table~\ref{tbl_corr}. 
We summarized the derived empirical equations for \teff\ and \logg, and the applicable \teff\ ranges in Table~\ref{tbl_diagnostics}.
The derived equations are appropriate for the stars with near-solar abundance because our targets have typical solar metallicities.

\section{Conclusions}\label{sec:conclusions}
We present the IGRINS Spectral Library of NIR spectra for 84 stars with high-resolution (R $\sim$ 45,000) and high signal-to-noise ratio (S/N $\geq$~200), as part of the IGRINS Legacy Project.
The library covers O- to M-type stars with luminosity classes between I and V.
The observed spectra were reduced using the IGRINS pipeline version 2, and hydrogen lines were removed by fitting the Kurucz Vega model.
Then, we performed a correction for the telluric absorption lines and absolute flux calibration using the 2MASS photometry and bandpass profiles.
Finally, the spectra were shifted to heliocentric velocity.

\begin{figure}
\epsscale{.9}
\plotone{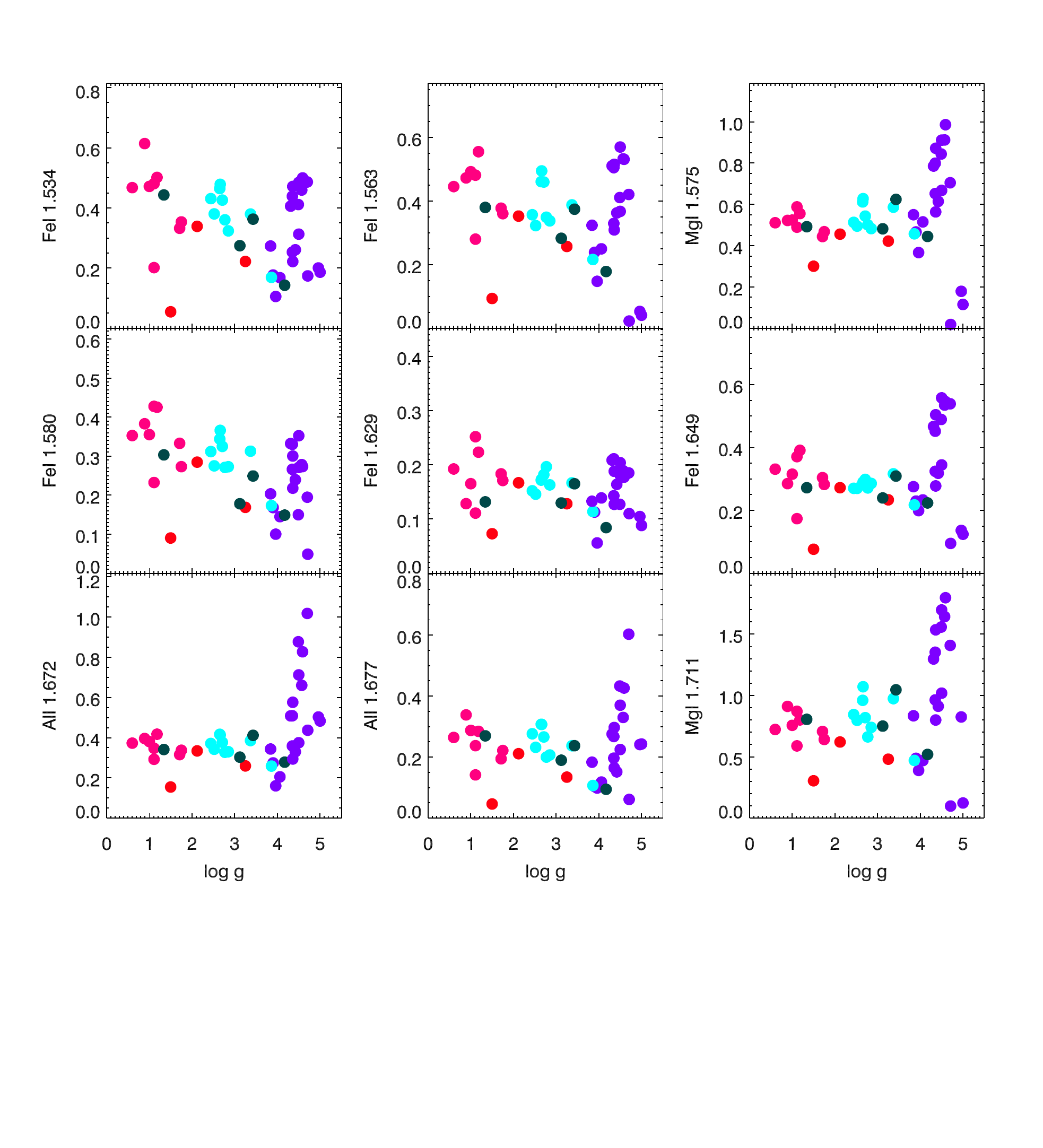}
\caption{Equivalent widths of the lines corresponding to H-band spectral indices as a function of surface gravity.
The different colors correspond to the supergiant (magenta), bright giant (red), giant (cyan), subgiant (green), and dwarf stars (purple). 
Most of the spectral indices show no notable trend with \logg. 
\label{ew_H_logg}
}
\end{figure}

\begin{figure}
\epsscale{.9}
\plotone{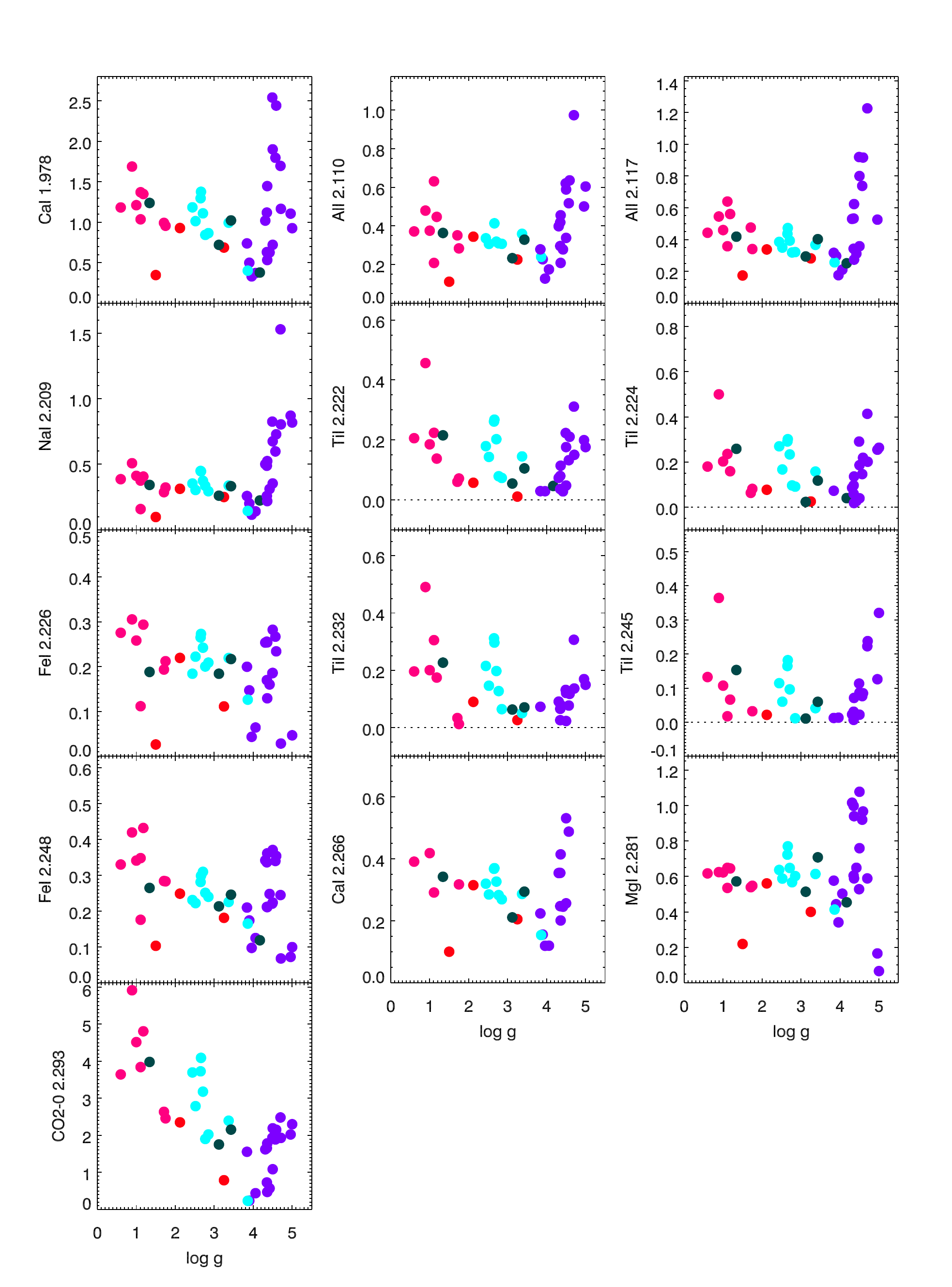}
\caption{Equivalent widths of the lines corresponding to K-band spectral indices as a function of surface gravity.
The different colors correspond to the supergiant (magenta), bright giant (red), giant (cyan), subgiant (green), and dwarf stars (purple).
Zero level is denoted by a dashed line.
Most of the spectral indices tend to decrease with \logg, except for dwarfs with \logg\ $>$ 4.
The CO overtone band 2.293~\um\ shows the strongest \logg\ dependency.
\label{ew_K_logg}
}
\end{figure}

\begin{figure}
\plotone{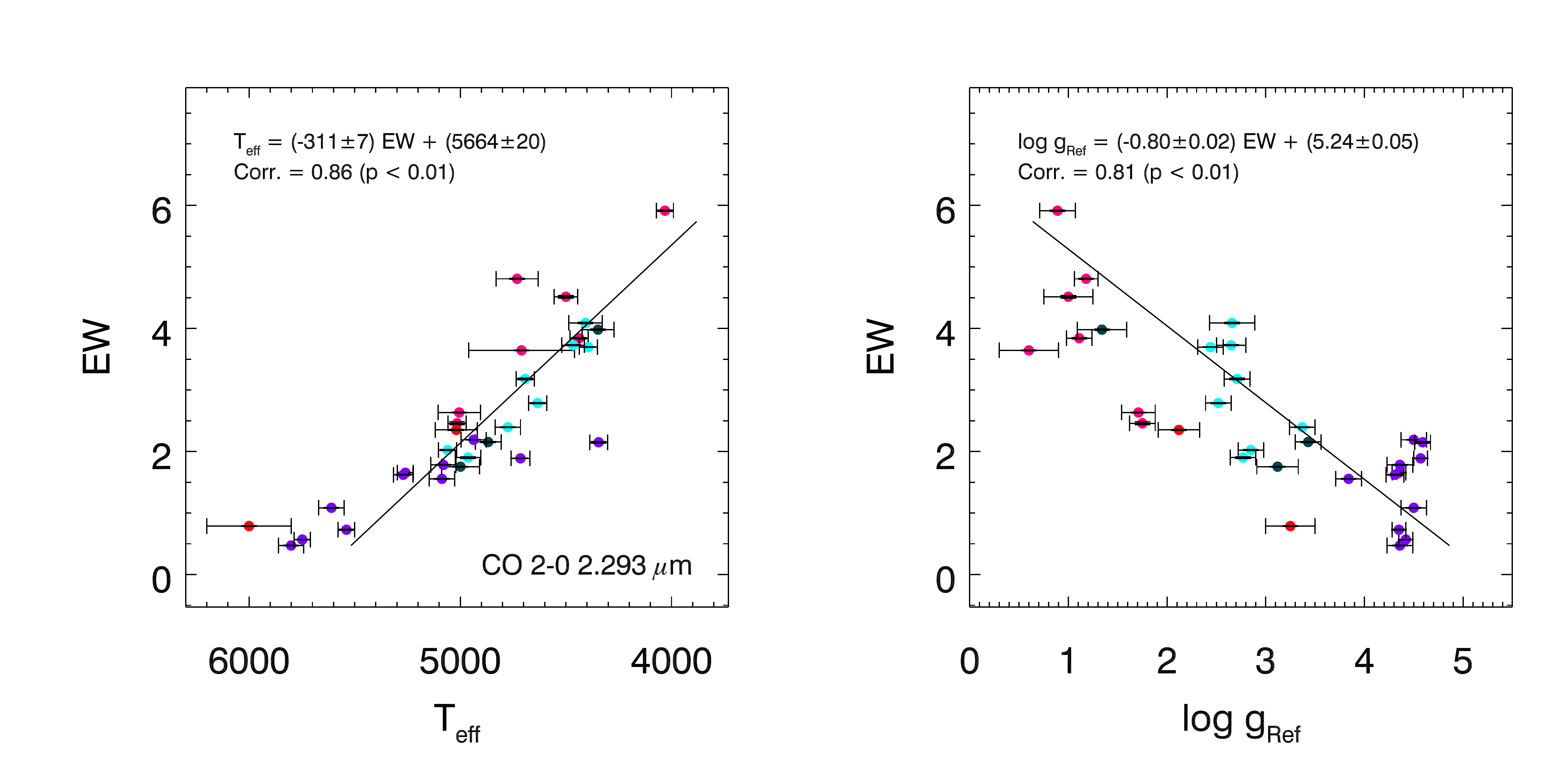}
  \caption{
Left and right panels show the equivalent widths of CO~2.293~\um\ feature as a function of effective temperature and surface gravity, respectively.
Different colors indicate different type of stars: supergiant (magenta), bright giant (red), giant (cyan), subgiant (green), and dwarf (purple). 
The EWs of CO~2.293~\um\ feature have tight correlations with not only \teff\ but also \logg. 
\label{co_ew} 
}
\end{figure}

In this paper we selected isolated spectral lines and measured their EWs to find spectral indicators of stellar physical parameters (\teff\ and \logg).
However, our targets do not cover a sufficiently wide range of [Fe/H] to investigate the relation between EWs and metallicity.
Therefore, the relations between EWs and stellar parameters (\teff\ and \logg) derived from their indicators are applicable exclusively for the stars with the near-solar abundances.
We have found several spectral indicators of \teff\ such as Al I 1.672, Al I 2.117, Na I 2.209, Ti I 2.224, and CO 2.293~\um.
Among the five spectral features, Al I 2.117 and Ti I 2.224~\um\ lines are newly found spectral indicators of \teff.
Most of these spectral lines do not have significant correlations with \logg\ (Figure~\ref{ew_H_logg} and Figure~\ref{ew_K_logg}), but CO 2.293~\um\ EWs correlate with \teff\ and also \logg\ (Figure~\ref{co_ew}).
Therefore, we found a relation among the EW of the CO feature, \teff, and \logg\ using a regression analysis (Equation~\ref{eq_regress}). 
Surface gravity can be calculated by measuring the EW of the CO overtone band 2.293~\um\ and the \teff\ derived from other lines (Table~\ref{tbl_diagnostics}) including the CO feature itself.
The proposed method to estimate the stellar \teff\ and \logg\ relies on the features in the K-band that are less affected by reddening than shorter wavelength NIR and optical spectroscopy. 
In addition, the spectral indicators are close in wavelength, so they can be observed quickly and efficiently with spectrographs that have high-resolution, but shorter wavelength coverage than IGRINS, making this diagnostic method easily accessible.
In conclusion, the derived relations (Table~\ref{tbl_diagnostics}) provide empirical ways to estimate the \teff\ and the \logg\ of late-type stars without direct comparison to stellar atmospheric models for all the sources, which is a valuable tool for population studies.

\clearpage

\acknowledgments

This work used the Immersion Grating Infrared Spectrograph (IGRINS) that was developed under a collaboration between the University of Texas at Austin and the Korea Astronomy and Space Science Institute (KASI) with the financial support of the US National Science Foundation under grant AST-1229522, of the University of Texas at Austin, and of the Korean GMT Project of KASI. 
This work was supported by the BK21 plus program through the National Research Foundation (NRF) funded by the Ministry of Education of Korea. 
This research was also supported by the Basic Science Research Program through the National Research Foundation of Korea (grant No. NRF-2018R1A2B6003423) and the Korea Astronomy and Space Science Institute under the R\&D program supervised by the Ministry of Science, ICT and Future Planning.

\bibliographystyle{aasjournal}
\bibliography{ms}

\clearpage

 \clearpage

\clearpage

\clearpage

\clearpage

\clearpage


\begin{deluxetable}{ccccccc}
\tablecaption{Spectral Composition of Library\label{tbl_comp}}
\tablewidth{0pt}
\tablehead{
\colhead{Spectral Type} & \multicolumn{6}{c}{Luminosity Class} \\
\cline{2-7}
\colhead{} & \colhead{I} & \colhead{II} & \colhead{III} & \colhead{IV} & \colhead{V} & \colhead{Unknown$^{a}$}
}

\startdata
 O & 4 & 1 & 2 & 0 & 4 & 0 \\
 B & 0 & 0 & 3 & 1 & 3 & 0 \\
 A & 1 & 0 & 0 & 0 & 2 & 0 \\
 F & 0 & 1 & 3 & 2 & 3 & 0 \\
 G & 5 & 4 & 4 & 1 & 6 & 1 \\
 K & 3 & 0 & 7 & 2 & 8 & 0 \\
 M & 3 & 0 & 2 & 0 & 8 & 0 \\
\enddata
\tablenotetext{a}{ Targets with no information of luminosity class. Among the 84 stars, HD 216868 (G5) is the only source with no information on its luminosity class.}
\end{deluxetable}
\clearpage

\begin{longrotatetable}
\begin{deluxetable}{lllllrrlccrc}
\tabletypesize{\scriptsize} 
\rotate
\tablecaption{Target Information and Observation Log\label{tbl_info}}
\tablewidth{0pt}
\tablehead{
\colhead{Object} & \colhead{Spectral Type} & \colhead{H} & \colhead{K} & \colhead{Obs. Date} & \colhead{Exp. Time$^{a}$} & \colhead{S/N$^{b}$} & \colhead{A0 V$^{c}$} & \colhead{V$_{helio}$} & \colhead{V$_{radial}$} & \colhead{Ref.$^{d}$} & \colhead{Ref.$^{e}$} \\
\colhead{} & \colhead{} & \colhead{mag} & \colhead{mag} & \colhead{UT Date} & \colhead{sec} & \colhead{2.2~\um} & \colhead{} & \colhead{\kms} & \colhead{\kms} & \colhead{RV} & \colhead{Target}
}
\startdata
           HD190429A$^\dagger$  &        O4 If  &  6.162 $\pm$ 0.031  &  6.150 $\pm$ 0.021  &    2014 Nov 19  & 300  $\times$   4  =    1200  & 373  &       HIP99719  & -17.33  & -16.00  &  1  &                1 \\
             HD46223$^\dagger$  &    O4 V((f))  &  6.703 $\pm$ 0.017  &  6.676 $\pm$ 0.021  &    2016 Feb 28  & 360  $\times$   4  =    1440  & 374  &       HIP33297  & -24.90  &  43.40  &  1  &             1, 2 \\
             HD15558$^\dagger$  &     O4.5 III  &  6.609 $\pm$ 0.018  &  6.519 $\pm$ 0.021  &    2015 Nov 24  & 300  $\times$   4  =    1200  & 261  &        HIP9690  &  -1.37  & -45.10  &  2  &                1 \\
            HD210839$^\dagger$  &        O6 If  &  4.618 $\pm$ 0.036  &  4.500 $\pm$ 0.02   &    2015 Nov 22  &  40  $\times$   8  =     320  & 334  &      HIP106393  &  -9.95  & -75.10  &  3  &             2, 3 \\
                      HD199579  &       O6.5 V  &  5.831 $\pm$ 0.051  &  5.857 $\pm$ 0.026  &    2015 Dec 07  &  70  $\times$   6  =     420  & 359  &      HIP103159  & -16.00  &  -6.40  &  2  &          1, 2, 3 \\
             HD47839$^\dagger$  &         O7 V  &  5.322 $\pm$ 0.021  &  5.340 $\pm$ 0.021  &    2015 Nov 22  & 180  $\times$   4  =     720  & 331  &       HIP33297  &  18.62  &  22.00  &  2  &             2, 4 \\
             HD24912$^\dagger$  &     O7.5 III  &  4.070 $\pm$ 0.224  &  3.953 $\pm$ 0.036  &    2015 Nov 24  &  30  $\times$   8  =     240  & 297  &       HIP15925  &   1.16  &  65.40  &  3  &                1 \\
            HD225160$^\dagger$  &       O8 Iab  &  7.431 $\pm$ 0.031  &  7.336 $\pm$ 0.023  &    2014 Nov 19  & 420  $\times$   4  =    1680  & 121  &         HIP582  &  -6.03  & -45.40  &  3  &                1 \\
            HD228779$^\dagger$  &      O9.5 Ib  &  5.834 $\pm$ 0.015  &  5.663 $\pm$ 0.021  &    2015 Nov 22  & 180  $\times$   4  =     720  & 303  &       HIP99719  & -18.34  &   1.39 $\pm$ 0.10$^\ddagger$  &  $\ldots$  &                3 \\
             HD34078$^\dagger$  &       O9.5 V  &  5.355 $\pm$ 0.018  &  6.641   &    2015 Nov 24  & 180  $\times$   4  =     720  & 334  &       HIP15925  &   9.22  &  56.70  &  4  &             1, 2 \\
             HD10125$^\dagger$  &      O9.7 II  &  7.570 $\pm$ 0.029  &  7.491 $\pm$ 0.018  &    2014 Nov 18  & 600  $\times$   4  =    2400  & 126  &       HIP10316  &  -1.26  & -38.00  &  3  &                1 \\
             HD48434$^\dagger$  &       B0 III  &  6.011 $\pm$ 0.024  &  5.980 $\pm$ 0.018  &    2015 Nov 22  & 180  $\times$   4  =     720  & 315  &       HIP33297  &  18.60  &  34.50  &  1  &                3 \\
             HD21483$^\dagger$  &       B3 III  &  6.190 $\pm$ 0.026  &  6.128 $\pm$ 0.018  &    2015 Nov 22  & 300  $\times$   4  =    1200  & 381  &       HIP15925  &  -1.37  &  -3.60  &  3  &             2, 3 \\
                       HD74280  &         B3 V  &  4.840 $\pm$ 0.033  &  4.800 $\pm$ 0.023  &    2015 May 06  &  70  $\times$   7  =     490  & 355  &       HIP41307  & -28.74  &  16.10  &  3  &             2, 3 \\
             HD32630$^\dagger$  &         B3 V  &  3.761 $\pm$ 0.238  &  3.857 $\pm$ 0.292  &    2015 Nov 24  &  30  $\times$   8  =     240  & 346  &       HIP21823  &   8.36  &   7.30  &  3  &          2, 3, 4 \\
            HD184930$^\dagger$  &       B5 III  &  4.422 $\pm$ 0.244  &  4.482 $\pm$ 0.018  &    2015 May 07  &  30  $\times$   7  =     210  & 243  &       HIP95793  &  26.14  & -21.40  &  3  &             2, 3 \\
                      HD147394  &        B5 IV  &  4.085 $\pm$ 0.236  &  4.285 $\pm$ 0.017  &    2016 Feb 28  &  70  $\times$   4  =     280  & 397  &       HIP78017  &  11.38  & -15.50  &  3  &             2, 3 \\
             HD70011$^\dagger$  &       B9.5 V  &  5.910 $\pm$ 0.033  &  5.915 $\pm$ 0.021  &    2015 Jan 21  & 180  $\times$   4  =     720  & 328  &       HIP38722  &   0.45  &  23.00  &  3  &                $\ldots$ \\
             HD87737$^\dagger$  &        A0 Ib  &  3.499 $\pm$ 0.212  &  3.299 $\pm$ 0.256  &    2015 Jan 24  &  30  $\times$   6  =     180  & 104  &       HIP50459  &  12.10  &   1.40  &  3  &          2, 3, 4 \\
                      HD111397  &         A1 V  &  5.625 $\pm$ 0.026  &  5.566 $\pm$ 0.029  &    2015 Dec 23  & 200  $\times$   4  =     800  & 350  &           bVir  &  28.93  &  -5.90  &  3  &                $\ldots$ \\
                        HD7804  &         A3 V  &  5.041 $\pm$ 0.096  &  4.921 $\pm$ 0.027  &    2014 Nov 19  &  70  $\times$   4  =     280  & 338  &        HIP5164  & -18.63  &   2.00  &  3  &                3 \\
                        HD6130  &        F0 II  &  4.690 $\pm$ 0.254  &  4.406 $\pm$ 0.02   &    2014 Nov 18  & 120  $\times$   4  =     480  & 354  &        HIP6002  &  -3.89  &  -0.90  &  1  &          2, 3, 5 \\
                       HD27397  &        F0 IV  &  4.927 $\pm$ 0.047  &  4.853 $\pm$ 0.015  &    2016 Feb 28  &  70  $\times$   4  =     280  & 415  &       HIP17453  & -29.98  &  42.00  &  3  &             2, 5 \\
             HD91752$^\dagger$  &         F3 V  &  5.228 $\pm$ 0.031  &  5.202 $\pm$ 0.026  &    2015 Jan 21  & 180  $\times$   8  =    1440  & 189  &       HIP50303  &  11.80  & -26.50  &  3  &                6 \\
                       HD87822  &         F4 V  &  5.251 $\pm$ 0.033  &  5.131 $\pm$ 0.021  &    2015 Jan 24  & 400  $\times$   4  =    1600  & 169  &       HIP50459  &   9.19  &  -8.00  &  5  &                5 \\
                       HD75555  &    F5 III-IV  &  7.035 $\pm$ 0.016  &  6.992 $\pm$ 0.018  &    2015 Nov 18  & 300  $\times$   6  =    1800  & 309  &       HIP41798  &  24.76  &  22.00  &  3  &                5 \\
                       HD55052  &    F5 III-IV  &  4.983 $\pm$ 0.036  &  4.833 $\pm$ 0.018  &    2015 Jan 21  &  60  $\times$   4  =     240  & 285  &       HIP35345  &  -7.53  &  13.00  &  1  &             2, 3 \\
               HD218804$^\ast$  &        F5 IV  &  4.738 $\pm$ 0.031  &  4.674 $\pm$ 0.018  &    2015 Nov 19  &  70  $\times$   6  =     420  & 231  &      HIP109452  & -15.78  & -32.40  &  3  &                5 \\
                       HD87141  &         F5 V  &  4.730 $\pm$ 0.23   &  4.503 $\pm$ 0.015  &    2016 Feb 28  &   5  $\times$   6  =      30  & 333  &       HIP52478  & -10.66  & -19.10  &  3  &                6 \\
                      HD124850  &       F7 III  &  2.909 $\pm$ 0.236  &  2.801 $\pm$ 0.266  &    2016 May 01  &  11  $\times$   8  =      88  & 320  &       HIP77516  &  -3.53  &  12.51  &  6  &          5, 6, 7 \\
                      HD190323  &        G0 Ia  &  4.992 $\pm$ 0.038  &  4.888 $\pm$ 0.018  &    2015 May 07  &  40  $\times$   7  =     280  & 232  &       HIP95793  &  24.09  &  25.00  &  3  &                5 \\
                      HD119605  &        G0 II  &  3.710 $\pm$ 0.23   &  3.606 $\pm$ 0.206  &    2015 Dec 24  &  30  $\times$   4  =     120  & 377  &          85Vir  &  26.87  &   0.60  &  3  &             2, 6 \\
             HD71148$^\dagger$  &         G1 V  &  4.878 $\pm$ 0.023  &  4.830 $\pm$ 0.02   &    2015 Jan 21  &  70  $\times$   4  =     280  & 291  &       HIP41798  &  -1.52  & -32.30  &  7  &                6 \\
             HD76151$^\dagger$  &         G3 V  &  4.625 $\pm$ 0.276  &  4.456 $\pm$ 0.023  &    2015 Nov 19  &  70  $\times$   8  =     560  & 373  &       HIP41307  &  27.46  &  32.08  &  7  &          2, 5, 6 \\
                      HD216868  &         G5   &  6.771 $\pm$ 0.045  &  6.647 $\pm$ 0.02   &    2015 Dec 05  & 420  $\times$   4  =    1680  & 320  &      HIP114745  &  -4.16  &  21.93 $\pm$ 0.04$^\ddagger$  &  $\ldots$  &                $\ldots$ \\
              HD9900$^\dagger$  &       G5 Iab  &  2.683 $\pm$ 0.174  &  2.391 $\pm$ 0.234  &    2014 Nov 18  &  30  $\times$   4  =     120  & 115  &       HIP10316  &  -3.24  & -10.20  &  3  &                6 \\
                      HD108477  &        G5 II  &  4.497 $\pm$ 0.232  &  4.271 $\pm$ 0.036  &    2015 Dec 24  &  90  $\times$   4  =     360  & 410  &         HR4911  &  29.25  &  -8.30  &  1  &             2, 5 \\
                      HD219477  &    G5 II-III  &  4.771 $\pm$ 0.204  &  4.537 $\pm$ 8.888  &    2015 Nov 19  &  70  $\times$   6  =     420  & 396  &      HIP114714  & -21.36  &   1.20  &  3  &                5 \\
                      HD202314  &     G6 Ib-II  &  4.098 $\pm$ 0.25   &  4.020 $\pm$ 0.27   &    2015 Nov 22  &  70  $\times$   8  =     560  & 351  &       HIP99719  & -22.20  &  -7.50  &  3  &                5 \\
                      HD117043  &         G6 V  &  4.895 $\pm$ 0.053  &  4.795 $\pm$ 0.018  &    2015 Dec 25  &  90  $\times$   4  =     360  & 374  &       HIP66198  &  12.34  & -30.93  &  7  &                6 \\
            HD182694$^\dagger$  &      G7 IIIa  &  3.994 $\pm$ 0.206  &  3.835 $\pm$ 0.286  &    2015 Nov 21  &  30  $\times$   8  =     240  & 316  &       HIP99719  & -12.35  &  -0.70  &  3  &             2, 5 \\
                      HD115617  &         G7 V  &  2.974 $\pm$ 0.176  &  2.956 $\pm$ 0.236  &    2015 Dec 24  &   5  $\times$   8  =      40  & 235  &          85Vir  &  27.71  &  -8.13  &  8  &          5, 7, 8 \\
             HD67767$^\dagger$  &         G7 V  &  3.711 $\pm$ 0.182  &  3.840 $\pm$ 0.036  &    2015 Jan 21  &  30  $\times$   3  =      90  & 283  &       HIP38722  &  -0.87  & -44.48  &  9  &                6 \\
                       HD77912  &       G8 Iab  &  2.462 $\pm$ 0.258  &  2.400 $\pm$ 0.282  &    2014 Dec 29  &  11  $\times$   4  =      44  &  99  &       HIP41798  &  13.92  &  16.60  &  3  &             2, 6 \\
                      HD208606  &       G8 Iab  &  3.124 $\pm$ 0.188  &  2.811 $\pm$ 0.28   &    2015 Nov 22  &  10  $\times$   8  =      80  & 309  &      HIP106393  &  -8.86  & -27.20  &  3  &          2, 3, 5 \\
             HD25877$^\dagger$  &       G8 IIa  &  3.973 $\pm$ 0.172  &  3.819 $\pm$ 0.22   &    2015 Nov 24  &  30  $\times$   8  =     240  & 204  &       HIP15086  &   4.34  & -12.50  &  3  &             2, 5 \\
                       HD27277  &       G8 III  &  5.785 $\pm$ 0.029  &  5.680 $\pm$ 0.018  &    2015 Nov 24  & 180  $\times$   4  =     720  & 344  &       HIP21823  &   4.53  & -13.46  & 10  &                5 \\
            HD108225$^\dagger$  &       G8 III  &  2.991 $\pm$ 0.222  &  2.856 $\pm$ 0.304  &    2015 May 06  &   5  $\times$   6  =      30  & 215  &       HIP61471  & -19.51  &  -4.30  &  3  &             2, 6 \\
    HD113226$^{\ast, \dagger}$  &       G8 III  &  0.733 $\pm$ 0.244  &  0.664 $\pm$ 0.278  &    2016 Mar 01  &   1  $\times$   6  =       9  & 158  &       HD101060  &  14.16  & -14.29  &  9  &          2, 4, 6 \\
                       HD75732  &         G8 V  &  4.265 $\pm$ 0.234  &  4.015 $\pm$ 0.036  &    2015 Nov 18  &  70  $\times$   6  =     420  & 455  &       HIP41798  &  28.00  &  27.58  &  8  &                5 \\
                      HD114946  &        G9 IV  &  3.169 $\pm$ 0.192  &  3.114 $\pm$ 0.256  &    2015 Dec 24  &  30  $\times$   4  =     120  & 416  &          85Vir  &  27.67  & -48.12  &  8  &                5 \\
                      HD207089  &       K0 Iab  &  2.620 $\pm$ 0.202  &  2.396 $\pm$ 0.238  &    2014 Nov 22  &   4  $\times$   4  =      16  & 202  &      HIP108060  & -24.99  & -11.43  & 11  &                3 \\
             HD44391$^\dagger$  &        K0 Ib  &  4.759 $\pm$ 0.036  &  4.548 $\pm$ 0.017  &    2014 Nov 18  &  70  $\times$   4  =     280  & 325  &       HIP31434  &  18.51  & -12.83  & 11  &             5, 6 \\
                      HD142091  &    K0 III-IV  &  2.575 $\pm$ 0.18   &  2.423 $\pm$ 0.242  &    2016 Mar 01  &  11  $\times$   4  =      44  & 316  &       HIP78017  &  15.75  & -25.16  &  9  &          2, 5, 7 \\
            HD165438$^\dagger$  &        K0 IV  &  3.788 $\pm$ 0.254  &  3.749 $\pm$ 0.228  &    2016 Apr 30  &  10  $\times$   8  =      80  & 238  &       HIP90967  &  21.95  & -26.75  &  8  &             2, 5 \\
                      HD136442  &         K0 V  &  4.215 $\pm$ 0.188  &  3.946 $\pm$ 0.016  &    2015 May 03  &  30  $\times$   4  =     120  & 262  &       HIP77516  &   3.36  & -46.81  &  7  &                6 \\
            HD145675$^\dagger$  &         K0 V  &  4.803 $\pm$ 0.016  &  4.714 $\pm$ 0.016  &    2015 May 07  &  30  $\times$   7  =     210  & 248  &       HIP79332  &  -0.52  & -13.73  &  7  &                5 \\
                       HD94600  &       K1 III  &  2.551 $\pm$ 0.252  &  2.334 $\pm$ 0.316  &    2016 Mar 01  &   5  $\times$   6  =      30  & 249  &       HIP50459  &  -4.15  & -23.47  & 12  &                6 \\
               HD108381$^\ast$  &       K1 III  &  2.048 $\pm$ 0.196  &  1.810 $\pm$ 0.25   &    2016 Apr 29  &   1  $\times$   7  =      11  & 171  &       HIP60327  & -18.42  &   3.38  & 11  &             2, 6 \\
            HD218356$^\dagger$  &        K1 IV  &  1.830 $\pm$ 0.196  &  1.765 $\pm$ 0.22   &    2014 Nov 18  &  30  $\times$   4  =     120  & 233  &      HIP115250  & -22.54  & -27.55  & 11  &             6, 7 \\
             HD81146$^\dagger$  &       K2 III  &  1.765 $\pm$ 0.174  &  1.688 $\pm$ 0.186  &    2014 Nov 21  &   4  $\times$   4  =      16  & 401  &       HIP44512  &  28.72  &  27.94  & 11  &             2, 6 \\
     HD85503$^{\ast, \dagger}$  &       K2 III  &  1.327 $\pm$ 0.174  &  1.364 $\pm$ 0.224  &    2015 Jan 21  &   5  $\times$   6  =      30  & 236  &       HIP50303  &  10.36  &  13.63  & 11  &             2, 6 \\
              HD3765$^\dagger$  &         K2 V  &  5.272 $\pm$ 0.051  &  5.164 $\pm$ 0.016  &    2015 Jan 21  & 150  $\times$   4  =     600  & 337  &         HD1561  & -25.64  & -63.11  &  7  &                5 \\
                      HD150567  &       K3 III  &  5.136 $\pm$ 0.024  &  4.984 $\pm$ 0.018  &    2015 May 07  &  45  $\times$   7  =     315  & 257  &       HIP79332  &   5.52  & -52.26  & 11  &                6 \\
                      HD102328  &       K3 III  &  2.884 $\pm$ 0.214  &  2.632 $\pm$ 0.282  &    2015 Dec 23  &   5  $\times$   8  =      40  & 195  &       HIP56147  &  16.88  &   0.55  & 11  &             2, 6 \\
                      HD219134  &         K3 V  &  3.469 $\pm$ 0.226  &  3.261 $\pm$ 0.304  &    2015 Nov 19  &  30  $\times$  12  =     360  & 429  &      HIP114714  &  -9.65  & -18.83  &  8  &    2, 3, 5, 7, 8 \\
                      HD122064  &         K3 V  &  4.402 $\pm$ 0.206  &  4.094 $\pm$ 8.888  &    2015 Dec 25  &  90  $\times$   4  =     360  & 442  &       HIP66198  &  12.38  & -26.54  &  7  &                6 \\
            HD200905$^\dagger$  &   K4.5 Ib-II  &  0.027 $\pm$ 0.178  & -0.038 $\pm$ 0.202  &    2015 Nov 22  &   1  $\times$   8  =      13  & 252  &      HIP106393  & -16.51  & -19.10  & 11  &             2, 3 \\
             HD19305$^\dagger$  &         K5 V  &  5.843 $\pm$ 0.034  &  5.646 $\pm$ 0.021  &    2015 Jan 21  & 180  $\times$   4  =     720  & 418  &       HIP16322  & -28.37  & -30.78  & 13  &                5 \\
             HD36003$^\dagger$  &         K5 V  &  5.111 $\pm$ 0.071  &  4.880 $\pm$ 0.024  &    2015 Nov 21  &  60  $\times$   4  =     240  & 155  &       HIP26812  &   9.91  & -55.53  &  8  &                5 \\
            HD201092$^\dagger$  &         K7 V  &  2.895 $\pm$ 0.218  &  2.544 $\pm$ 0.328  &    2015 Nov 22  &  10  $\times$   8  =      80  & 305  &      HIP106393  & -18.69  & -64.07  &  8  &    2, 3, 5, 7, 8 \\
                     BD+632073  &        M0 Ib  &  4.345 $\pm$ 0.023  &  3.925 $\pm$ 0.358  &    2015 Nov 21  &  30  $\times$   6  =     180  & 316  &      HIP117450  &  -5.94  & -56.83 $\pm$ 0.02$^\ddagger$  &  $\ldots$  &                6 \\
             HD36395$^\dagger$  &       M1.5 V  &  4.149 $\pm$ 0.212  &  4.039 $\pm$ 0.26   &    2015 Nov 21  &  20  $\times$   8  =     160  & 156  &       HIP26812  &  10.19  &   8.67  &  8  &          2, 5, 8 \\
             HD95735$^\dagger$  &         M2 V  &  3.640 $\pm$ 0.202  &  3.254 $\pm$ 0.306  &    2015 Dec 06  &  30  $\times$   4  =     120  & 243  &       HIP41798  &  26.35  & -84.69  &  8  &    2, 3, 5, 8, 9 \\
                       HD14469  &       M3 Iab  &  1.928 $\pm$ 0.184  &  1.455 $\pm$ 0.222  &    2015 Nov 21  &  10  $\times$   8  =      80  & 366  &        HIP9690  &  -1.94  & -44.00  & 14  &          2, 3, 5 \\
                     BD+443567  &         M3 V  &  6.769 $\pm$ 0.023  &  6.533 $\pm$ 0.016  &    2015 Nov 21  & 360  $\times$   4  =    1440  & 176  &      HIP103159  & -15.68  & -24.70  &  8  &                5 \\
                     BD+051668  &       M3.5 V  &  5.219 $\pm$ 0.063  &  4.857 $\pm$ 0.023  &    2015 Nov 22  &  70  $\times$   6  =     420  & 321  &       HIP37478  &  22.68  &  18.22  &  8  &                5 \\
                      BD+56512  &        M4 Ib  &  2.679 $\pm$ 0.204  &  2.194 $\pm$ 0.23   &    2015 Nov 21  &  10  $\times$   8  =      80  & 335  &        HIP9690  &  -1.99  & -32.24  & 10  &                6 \\
                       HD76830  &      M4 IIIv  &  1.487 $\pm$ 0.164  &  1.288 $\pm$ 0.214  &    2015 Nov 22  &   2  $\times$  10  =      20  & 270  &       HIP37478  &  28.50  &  22.99  & 11  &             2, 6 \\
               GJ388$^\dagger$  &      M4.5 Ve  &  4.843 $\pm$ 0.02   &  4.593 $\pm$ 0.017  &    2015 May 06  &  70  $\times$   7  =     490  & 327  &       HIP50459  & -28.39  &  12.50  & 15  &             5, 9 \\
                      HD132813  &       M5 III  & -0.774 $\pm$ 0.182  & -0.957 $\pm$  0.2   &    2016 Mar 01  &   1  $\times$   3  =       4  & 378  &       HIP78017  &   0.03  &  13.60  & 16  &             2, 3 \\
           BD-074003$^\dagger$  &         M5 V  &  6.095 $\pm$ 0.033  &  5.837 $\pm$ 0.023  &    2016 May 01  & 150  $\times$   4  =     600  & 312  &       HIP77516  &   4.51  &  -9.21  &  8  &                5 \\
                         GJ866  &         M5 V  &  5.954 $\pm$ 0.031  &  5.537 $\pm$ 0.02   &    2015 Dec 06  & 200  $\times$   4  =     800  & 405  &      HIP114371  & -29.84  & -60.00  &  1  &                5 \\
                       HD1326B  &         M6 V  &  6.191 $\pm$ 0.016  &  5.948 $\pm$ 0.024  &    2014 Dec 05  &  60  $\times$   4  =     240  & 149  &       HD219290  & -17.94  &  10.96  & 17  &                3 \\
 \enddata
\tablenotetext{\ast}{ Targets with a reduction problem; order shapes are not flattened well. The spectra of these targets will be provided as normalized spectra.}
\tablenotetext{\dagger}{ Targets with a reduction problem; Brackett lines are not removed completely.} 
\tablenotetext{\ddagger}{ Estimated radial velocity in this work.} 
\tablenotetext{a}{ Total integration time of each target (exposure time $\times$ the number of exposures = total integration time).}
\tablenotetext{b}{ The S/N for each star is the medians of the S/N values per resolution element for $\lambda$=2.21~$\sim$~2.24~\um.}
\tablenotetext{c}{ Telluric standard star of each target.}
\tablenotetext{d}{ Reference of the radial velocity.}
\tablenotetext{e}{ Reference of the target selection.}
\tablerefs{
(1) \citet{wilson53}; (2) \citet{pourbaix04}; (3) \citet{gontcharov06}; 
(4) \citet{kharchenko07}; (5) \citet{holmberg07}; (6) \citet{maldonado10}; 
(7) \citet{soubiran13}; (8) \citet{nidever02}; (9) \citet{jofre15};
(10) \citet{mermilloid08}; (11) \citet{famaey05}; (12) \citet{massarotti08}; 
(13) \citet{tokovinin92}; (14) \citet{gahm71}; (15)\citet{shkolnik12}; 
(16) \citet{campbell28}; (17) \citet{terrien15}
}

\tablerefs{
(1) \citet{walborn73}; (2) \citet{garcia89}; (3) \citet{munari99}; 
(4) \citet{garrison94}; (5) \citet{rayner09}; (6) \citet{marrese03}; 
(7) \citet{montes97}; (8) \citet{montes98}; (9) \citet{cushing05}
}

\end{deluxetable}
\end{longrotatetable}
\clearpage


\begin{deluxetable}{ccccc}
\tabletypesize{\scriptsize}%
 \tablecaption{Comparisions of uncertainties \label{tbl_uncertainty}}
 \tablehead{\colhead{Library} & \colhead{Spectral Range} & \colhead{\teff} & \colhead{\logg} & \colhead{\feoh} \\
	    \colhead{} & \colhead{} & \colhead{K} & \colhead{dex} & \colhead{dex}
 }
 \startdata
ELODIE$^{1}$ & OBA & $>$ 250 & 0.15 & 0.12 \\
ELODIE & FGK & $<$ 100 & 0.15 & 0.12 \\
ELODIE & M & $<$ 100 & 0.15 & 0.12 \\
MILES$^{2}$ & OBA & 3.5\% & 0.17 & 0.13 \\
MILES & FGK & 60 & 0.13 & 0.05 \\
MILES & M & 38 & 0.26 & 0.12 \\
CFLIB$^{3}$ & OBA & 5.1\% & 0.19 & 0.16 \\
CFLIB & FGK & 43 & 0.13 & 0.05 \\
CFLIB & M & 82 & 0.22 & 0.28 \\
 \enddata

 \tablerefs{
(1) \citet{prugniel07}; (2) \citet{prugniel11}; (3) \citet{wu11}
}
\end{deluxetable}


\begin{deluxetable}{cccc}
\tabletypesize{\scriptsize}%
 \tablecaption{Spectral Indices\label{tbl_indices}}
 \tablewidth{0pt}
 \tablehead{\colhead{Element} & \colhead{Central line} & \colhead{Band} & \colhead{Reference}
 }
 \startdata
 Fe I & 1.5339 & H & 1\\
 Fe I & 1.5625 & H & 1\\
 Mg I & 1.5753 & H & 1, 2\\
 Fe I & 1.5803 & H & 1, 3\\
 Fe I & 1.6289 & H & 1\\
 Fe I & 1.6491 & H & 1\\
 Al I & 1.6723 & H & 1, 2\\
 Al I & 1.6768 & H & 1, 2\\
 He I & 1.7007 & H & 1, 4, 5 \\
 Mg I & 1.7113 & H & 1, 2, 3\\
 Ca I & 1.9782 & K & 1, 6, 7\\
 Al I & 2.1098 & K & 1\\
 He I & 2.1126 & K & 1, 4, 5\\
 Al I & 2.1169 & K & 1 \\
 He II & 2.1885 & K & 1, 4, 5 \\
 Na I & 2.2089 & K & 1, 3, 6, 7, 8, 9, 10\\
 Ti I & 2.2217 & K & 1\\
 Ti I & 2.2238 & K & 1\\
 Fe I & 2.2263 & K & 1\\
 Ti I & 2.2318 & K & 1\\
 Ti I & 2.2450 & K & 1\\
 Fe I & 2.2479 & K & 1\\
 Ca I & 2.2657 & K & 1, 8, 9\\
 Mg I & 2.2814 & K & 1, 3, 6, 7, 9\\
 CO 2-0 & 2.2935 & K & 1, 3, 8, 9, 11, 12 \\
 \enddata

 \tablerefs{
(1) This work (new indices); (2) \citet{meyer98}; (3) \citet{ivanov04};
(4) \citet{lenorzer04}; (5) \citet{hanson05};
(6) \citet{rayner09}; (7) \citet{cesetti13};
(8) \citet{kleinmann86}; (9) \citet{ali95}; (10) \citet{rojas12}; (11) \citet{lancon92}; (12) \citet{origlia93}
}
 
\end{deluxetable}

\begin{longrotatetable}
\begin{deluxetable}{cccccccccccc}
\tabletypesize{\tiny}
\rotate
\tablecaption{Stellar atmospheric parameters and EWs of spectral lines  of late-type stars \label{tbl_ew_param}}
\tablewidth{0pt}
\tablehead{
\colhead{Object} & \colhead{Spectral Type} & \colhead{\teff} & \colhead{\logg} & \colhead{\feoh} & \colhead{Fe I} & \colhead{Fe I} & \colhead{Mg I} & \colhead{Fe I} & \colhead{Fe I} & \colhead{Fe I} & \colhead{Reference$^{a}$} \\
\colhead{} & \colhead{} & \colhead{} & \colhead{} & \colhead{} & \colhead{1.534~\um} & \colhead{1.563~\um} & \colhead{1.575~\um} & \colhead{1.580~\um} & \colhead{1.629~\um} & \colhead{1.649~\um} & \colhead{}
}

\startdata
    HD6130 &   F0   II  &  7400$\pm$200 & 1.50$\pm$0.25  &  0.02  & 0.055$\pm$0.002 & 0.094$\pm$0.002 & 0.301$\pm$0.001 & 0.090$\pm$0.003 & 0.073$\pm$0.003  & 0.076$\pm$0.001  &      1 \\ 
   HD91752 &    F3   V  &  6418$\pm$ 60 & 3.96$\pm$0.13  & -0.23  & 0.106$\pm$0.003 & 0.148$\pm$0.002 & 0.367$\pm$0.003 & 0.100$\pm$0.003 & 0.056$\pm$0.003  & 0.199$\pm$0.002  &      2 \\ 
   HD87822 &    F4   V  &  6573$\pm$ 48 & 4.06$\pm$0.06 &  0.10$\pm$0.03 & 0.168$\pm$0.004 & 0.250$\pm$0.013 & 0.515$\pm$0.025 & 0.145$\pm$0.004 & 0.139$\pm$0.009  & 0.233$\pm$0.003  &      3 \\
   HD75555 & F5   III-IV &  6409$^{b}$$\pm$ 200 & $\ldots$ &  $\ldots$ & 0.183$\pm$0.003 & 0.230$\pm$0.002 & 0.499$\pm$0.002 & 0.205$\pm$0.003 & 0.168$\pm$0.004  & 0.264$\pm$0.002  & $\ldots$      \\
  HD218804 &   F5   IV  &  6493$\pm$ 58 & 4.17$\pm$0.07 & -0.13$\pm$0.04 & 0.143$\pm$0.002 & 0.178$\pm$0.002 & 0.445$\pm$0.002 & 0.149$\pm$0.003 & 0.084$\pm$0.002  & 0.224$\pm$0.003  &      3 \\
   HD87141 &    F5   V  &  6359$\pm$ 40 & 3.90$\pm$0.06 &  0.09$\pm$0.03 & 0.177$\pm$0.002 & 0.239$\pm$0.002 & 0.467$\pm$0.002 & 0.168$\pm$0.001 & 0.113$\pm$0.002  & 0.229$\pm$0.002  &      3 \\
  HD124850 &  F7   III  &  6207$\pm$ 50 & 3.86$\pm$0.08 & -0.06$\pm$0.04 & 0.169$\pm$0.002 & 0.216$\pm$0.002 & 0.458$\pm$0.002 & 0.173$\pm$0.002 & 0.114$\pm$0.002  & 0.217$\pm$0.001  &      3 \\
  HD190323 &   G0   Ia  &  6196$\pm$100 & 1.11$\pm$0.11  &  0.12  & 0.202$\pm$0.003 & 0.280$\pm$0.002 & 0.588$\pm$0.003 & 0.232$\pm$0.003 & 0.111$\pm$0.002  & 0.173$\pm$0.002  &      4 \\
  HD119605 &   G0   II  &  6000$\pm$200 & 3.25$\pm$0.25  &  0.11  & 0.222$\pm$0.001 & 0.257$\pm$0.002 & 0.422$\pm$0.002 & 0.169$\pm$0.002 & 0.128$\pm$0.001  & 0.233$\pm$0.001  &      5 \\
   HD71148 &    G1   V  &  5801$\pm$ 60 & 4.36$\pm$0.13 &  -0.04 &  0.222$\pm$0.002 & 0.309$\pm$0.002 & 0.564$\pm$0.002 & 0.218$\pm$0.013 & 0.127$\pm$0.001  & 0.278$\pm$0.001  &      2 \\
\enddata
\tablenotetext{a}{ References of stellar atmospheric parameters.}
\tablenotetext{b}{ \teff\ was calculated by $T_\mathrm{eff}-(B-V)$ relation \citep{flower96, torres10}. 
The standard deviation ($\sim$~200~K) of the differences between $T_\mathrm{eff}~(Ref.)$ and $T_\mathrm{eff}~(B-V)$ for common stars is adopted as error of \teff\ derived by $T_\mathrm{eff}-(B-V)$ relation.}
\tablecomments{Table~\ref{tbl_ew_param} is published in its entirety in the machine-readable format. A portion of it is shown here for guidance regarding its form and content.}
\tablerefs{
(1) \citet{venn95}
(2) \citet{prugniel11}
(3) \citet{prugniel07}
(4) \citet{luck14} 
(5) \citet{giridhar97}
(6) \citet{luck95}
(7) \citet{wu11}
(8) \citet{luck80}
(9) \citet{bubar10}
(10) \citet{gaidos14}
(11) \citet{lepine13}
(12) \citet{santos13}
(13) \citet{koleva12}
}
\end{deluxetable}
\end{longrotatetable}

\clearpage

\begin{deluxetable}{lll}
\tabletypesize{\scriptsize}%
 \tablecaption{EW of He I 1.701~\um\ of early-type stars \label{tbl_hei}}
 \tablehead{\colhead{Object} & \colhead{Spectral Type} & \colhead{EW} 
 }
 \startdata
 HD190429A &      O4 If & 0.324 $\pm$ 0.153 \\
   HD46223 &  O4 V((f)) & 0.397 $\pm$ 0.007 \\
  HD199579 &     O6.5 V & 0.770 $\pm$ 0.005 \\
   HD47839 &       O7 V & 0.930 $\pm$ 0.006 \\
  HD225160 &     O8 Iab & 2.106 $\pm$ 0.834 \\
  HD228779 &    O9.5 Ib & 2.783 $\pm$ 0.007 \\
   HD34078 &     O9.5 V & 1.315 $\pm$ 0.004 \\
   HD10125 &    O9.7 II & 1.739 $\pm$ 0.016 \\
   HD48434 &     B0 III & 1.989 $\pm$ 0.006 \\
   HD21483 &     B3 III & 1.005 $\pm$ 0.004 \\
   HD74280 &       B3 V & 0.557 $\pm$ 0.006 \\
   HD32630 &       B3 V & 0.679 $\pm$ 0.007 \\
  HD147394 &      B5 IV & 0.205 $\pm$ 0.005 \\
\enddata

\end{deluxetable}


\begin{longrotatetable}
\begin{deluxetable}{cccc}
\tabletypesize{\scriptsize}
 \tablecaption{Spectral Diagnostics \label{tbl_diagnostics}}
 \tablehead{\colhead{Spectral Index} & \colhead{\teff\ [K]$^{a}$} & \colhead{Dwarfs} & \colhead{Other Luminosity Classes$^{b}$}}
 \startdata
 AI 1.672 & 4000-7000 & $T_\mathrm{eff} = (- 3164 \pm 67) EW + (6876 \pm 35) $ & $T_\mathrm{eff} = (- 11547 \pm 291) EW + (8883 \pm 98) $ \\
 Al I 2.117 & 4000-7000 & $T_\mathrm{eff} = (- 2613 \pm 58) EW + (6684 \pm 33) $ & $T_\mathrm{eff} = (- 5503 \pm 163) EW + (7020 \pm 62) $ \\
 Na I 2.209 & 4000-7000 & $T_\mathrm{eff} = (- 2893 \pm 63) EW + (6610 \pm 35) $ & $T_\mathrm{eff} = (- 6495 \pm 146) EW + (7059 \pm 48) $ \\
 Ti I 2.224$^{c}$ & 3000-6000 & \multicolumn{2}{c}{$T_\mathrm{eff} = (- 3770 \pm 88) EW + (5502 \pm 16) $} \\
 CO 2.293$^{c}$ & 3000-6000 & \multicolumn{2}{c}{$T_\mathrm{eff} = (- 311 \pm 7) EW + (5664 \pm 20) $} \\
\cline{0-3}
Al I 1.672$^{d}$ & 4000-7000 & $\log g_\mathrm{Fit} = (0.97 \pm 0.12) EW + (3.87 \pm 0.06) $ &  $\ldots$ \\
Na I 2.209$^{d}$ & 4000-7000 & $\log g_\mathrm{Fit} = (0.98 \pm 0.12) EW + (3.92 \pm 0.05) $ &  $\ldots$ \\
CO 2.293$^{c}$ & 3000-6000 &  \multicolumn{2}{c}{$\log g_\mathrm{Fit} = (- 0.80 \pm 0.02) EW + (5.24 \pm 0.05) $}  \\
 \cline{0-3}
 CO 2.293$^{c}$ &  3000-6000 & \multicolumn{2}{c}{$\log g_\mathrm{Fit} = (- 2.130 \pm 0.003) EW + (- 41.79 \pm 0.05) \log T_\mathrm{eff} + (162.54 \pm 0.18) $}  \\
\enddata
\tablenotetext{a}{ Applicable \teff\ ranges.}
\tablenotetext{b}{ Luminosity classes of supergiant, bright giant, giants, and subgiant.}
\tablenotetext{c}{ Whole luminosity classes are applicable for this equation.}
\tablenotetext{d}{ Only dwarfs can be applied for this equation.}

\end{deluxetable}
\end{longrotatetable}

\clearpage


\begin{deluxetable}{ccccc}
\tabletypesize{\scriptsize}%
 \tablecaption{Correlation coefficients and p-values of spectral indicators \label{tbl_corr}}
 \tablehead{\colhead{Spectral Index} & \multicolumn{2}{c}{\teff} & \multicolumn{2}{c}{\logg} \\
 \cline{2-5}
	    \colhead{} & \colhead{Dwarfs} & \colhead{Others$^{a}$} & \colhead{Dwarfs} & \colhead{Others$^{a}$}
 }
 \startdata
Al I 1.672 & 0.90 (p \textless~0.01) & 0.87 (p \textless~0.01) & 0.72 (p \textless~0.01) & 0.15 (p = 0.96) \\
Al I 2.117 & 0.87 (p \textless~0.01) & 0.69 (p \textless~0.01) & 0.69 (p = 0.01) & 0.57 (p \textless~0.01) \\
Na I 2.209 & 0.86 (p \textless~0.01) & 0.88 (p \textless~0.01) & 0.74 (p \textless~0.01) & 0.31 (p = 0.27) \\
\cline{1-5}
Ti I 2.224$^{b}$ & \multicolumn{2}{c}{0.86 (p \textless~0.01)} &  \multicolumn{2}{c}{0.48 (p = 0.01)} \\
CO 2-0 2.293$^{b}$ &  \multicolumn{2}{c}{0.86 (p \textless~0.01)} & \multicolumn{2}{c}{0.81 (p \textless~0.01)} \\
\enddata
\tablenotetext{a}{ Luminosity classes of supergiant, bright giant, giants, and subgiant.}
\tablenotetext{b}{ For the Ti~I~2.224~\um\ and CO~2-0~2.293~\um\ features, total stars (whole luminosity classes) are used for the fitting.}
\tablecomments{The values in the parentheses are p-values.}
\end{deluxetable}


\appendix

\section{Spectra}
In Figures~\ref{hband_appendix}-\ref{kband_appendix} we provide the full H- and K-band spectra of HD 44391 (K0 Ib).

\section{Sky Spectrum}
\begin{figure}
 \includegraphics[width=18cm, height=5cm]{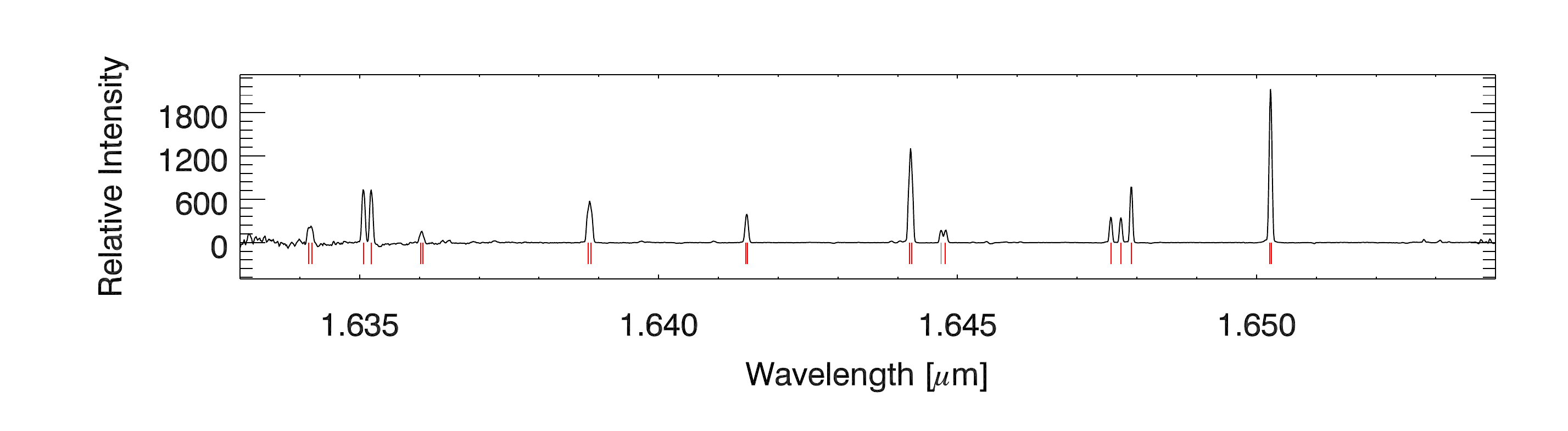}
 \includegraphics[width=18cm, height=5cm]{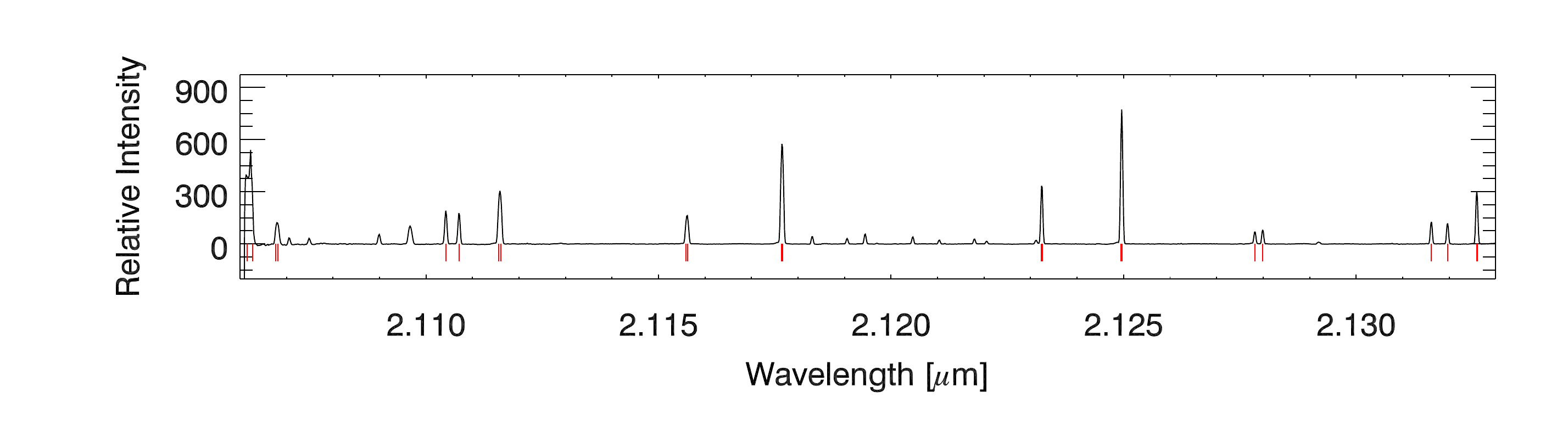}
\caption{ 
Average wavelength calibrated sky spectrum.
The top and bottom panels show example sky spectrum of H- and K-bands, respectively.
Red marks indicate OH sky emission lines.
\label{sky_spec}}
\end{figure}

The median sky spectrum was generated by median-combining 297 sky spectra that were observed between September 2015 and July 2016. 
The spectra suffer from contamination by other thermal continuum background from the instrument and the telescope. 
To show only the line emission from the sky, we present highpass filtered spectra (the original spectra were subtracted by median-filtered ones with a kernel size of 128 pixels). 
We note that this procedure also removes the thermal continuum background from the sky itself which becomes significant in the longer wavelength regime.
Thus, the presented spectrum is not suitable for any quantitative analysis, but good enough for qualitative investigation.
Figure~\ref{sky_spec} shows an example sky spectrum of H- and K-bands.
The full spectrum is available at our web page (\url{http://starformation.khu.ac.kr/IGRINS_spectral_library.htm}).


\begin{figure*}
\begin{tabbing}
\epsscale{0.95}
\plotone{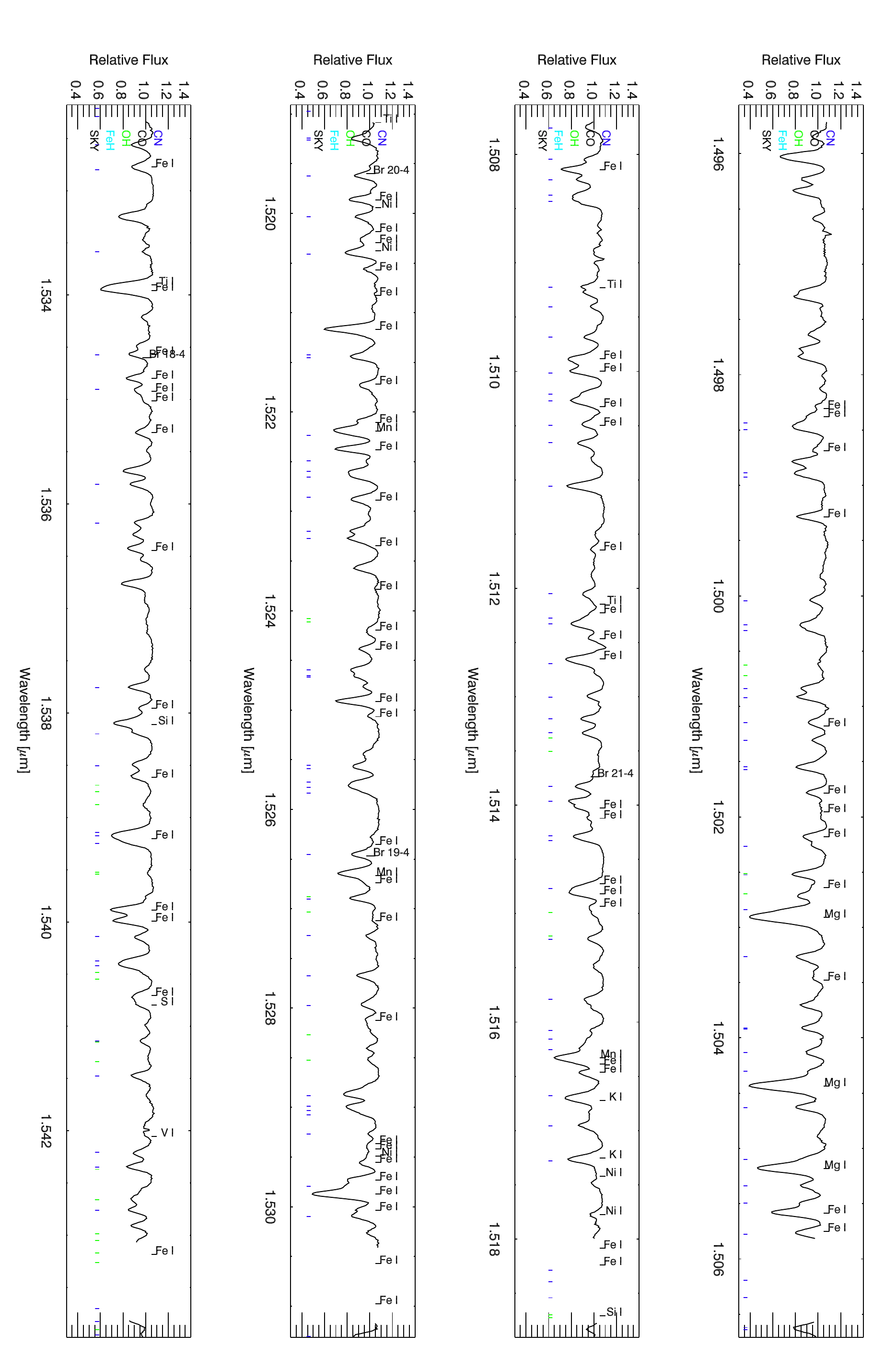}
\end{tabbing}
\caption{
The IGRINS H-band spectrum of HD 44391 (K0 Ib). 
The some narrow lines are the result of bad telluric correction.
Line identification was made using the Arcturus line list from \citet{hinkle95} and HITRAN database \citep{HITRAN2016}.
\label{hband_appendix}
}
\end{figure*}

\begin{figure*}
\begin{tabbing}
\epsscale{0.95}
\plotone{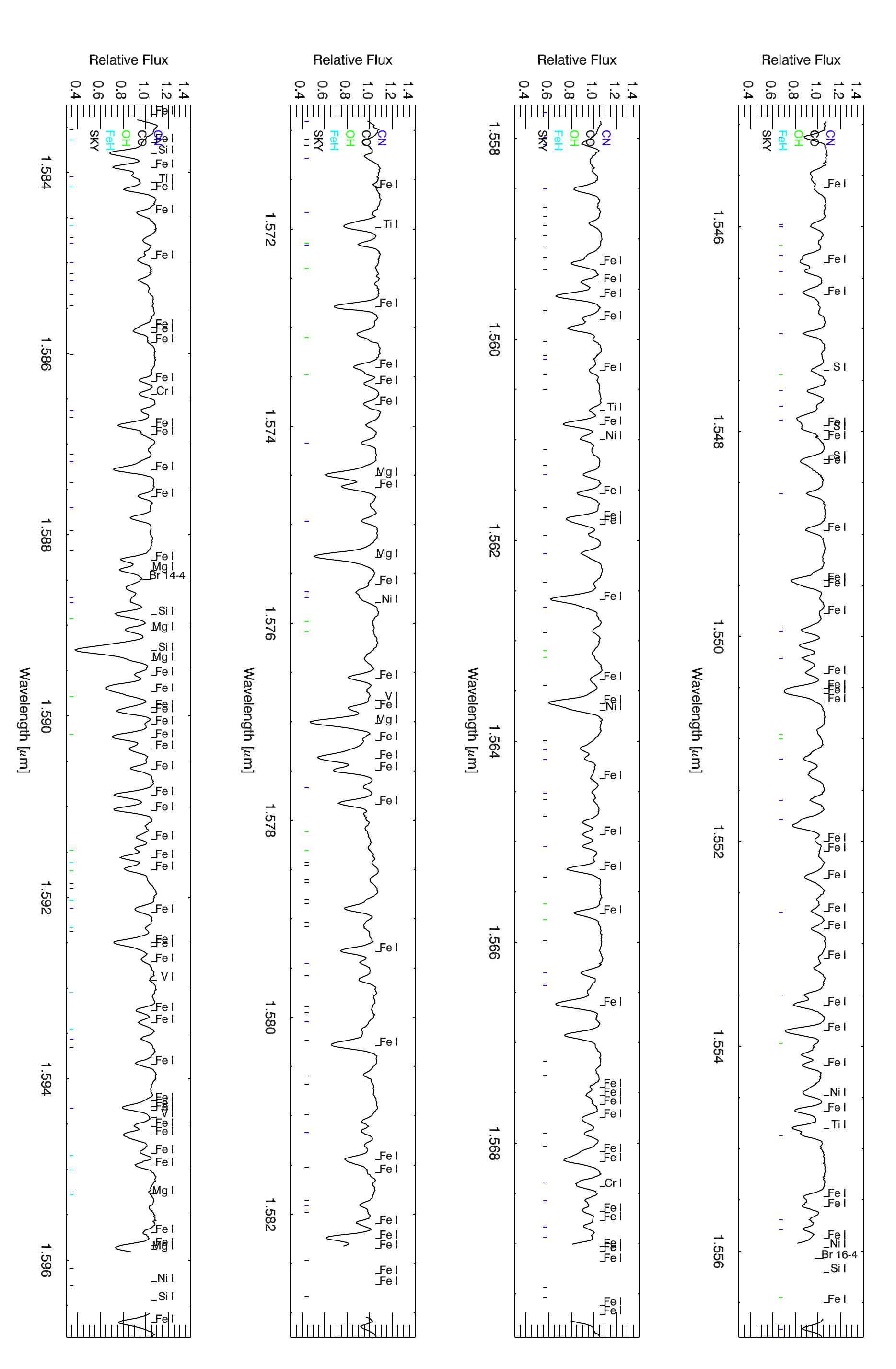}
\end{tabbing}
\center{\textbf{Figure~\ref{hband_appendix}.} Continued.}
\label{hband_appendix}
\end{figure*}

\begin{figure*}
\begin{tabbing}
\epsscale{0.95}
\plotone{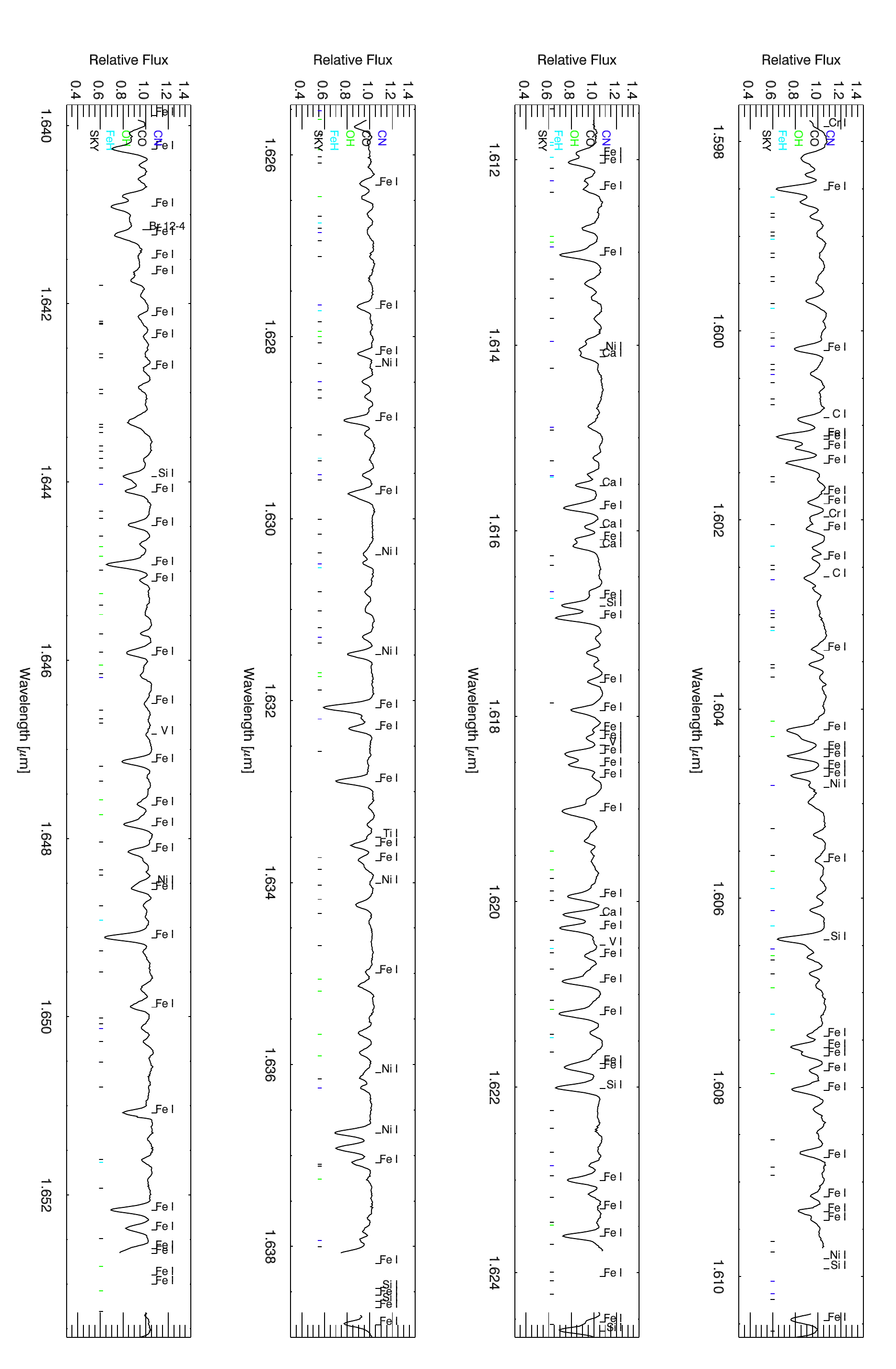}
\end{tabbing}
\center{\textbf{Figure~\ref{hband_appendix}.} Continued.}
\label{hband_appendix}
\end{figure*}

\begin{figure*}
\begin{tabbing}
\epsscale{0.95}
\plotone{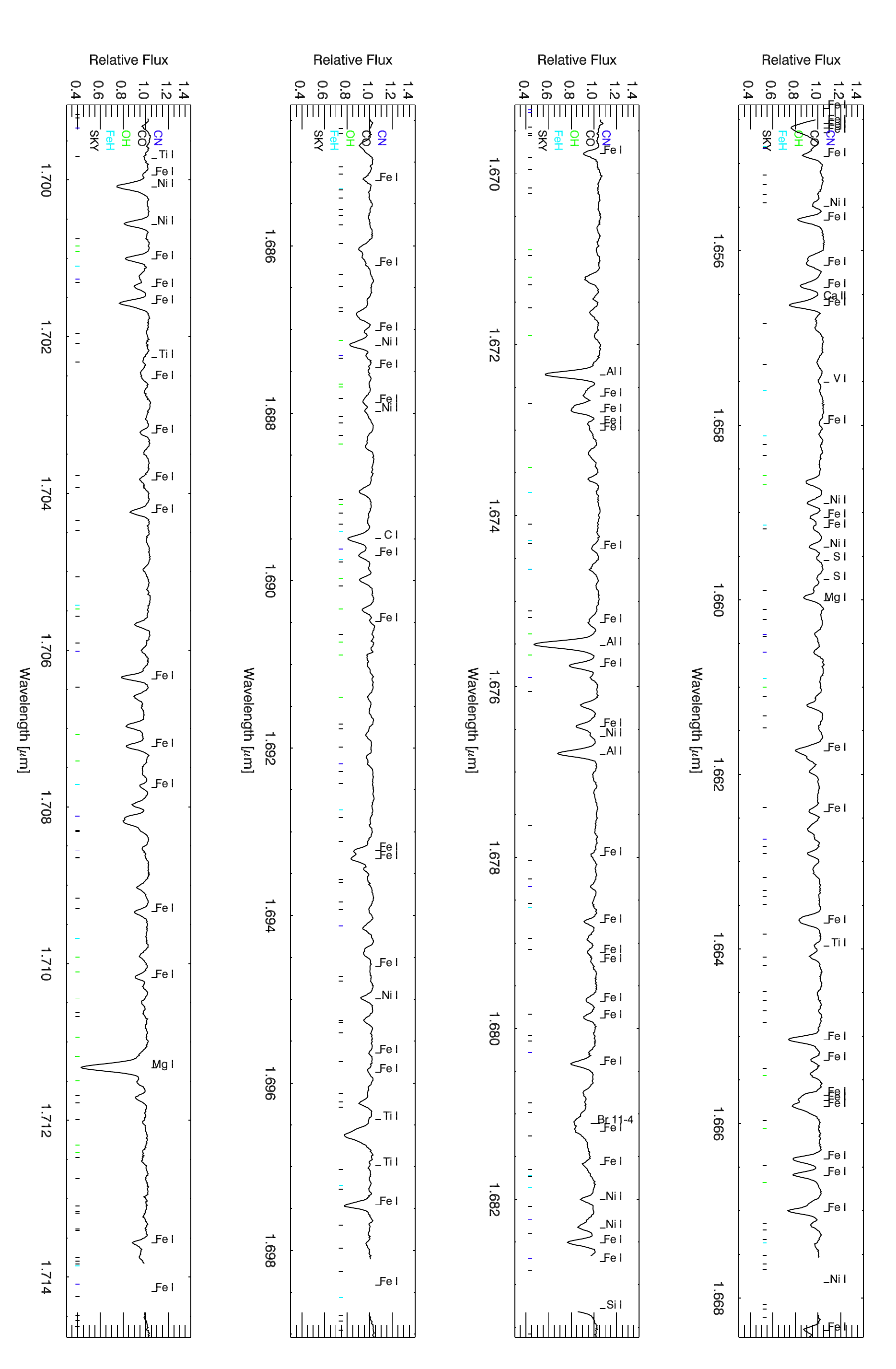}
\end{tabbing}
\center{\textbf{Figure~\ref{hband_appendix}.} Continued.}
\label{hband_appendix}
\end{figure*}

\begin{figure*}
\begin{tabbing}
\epsscale{0.95}
\plotone{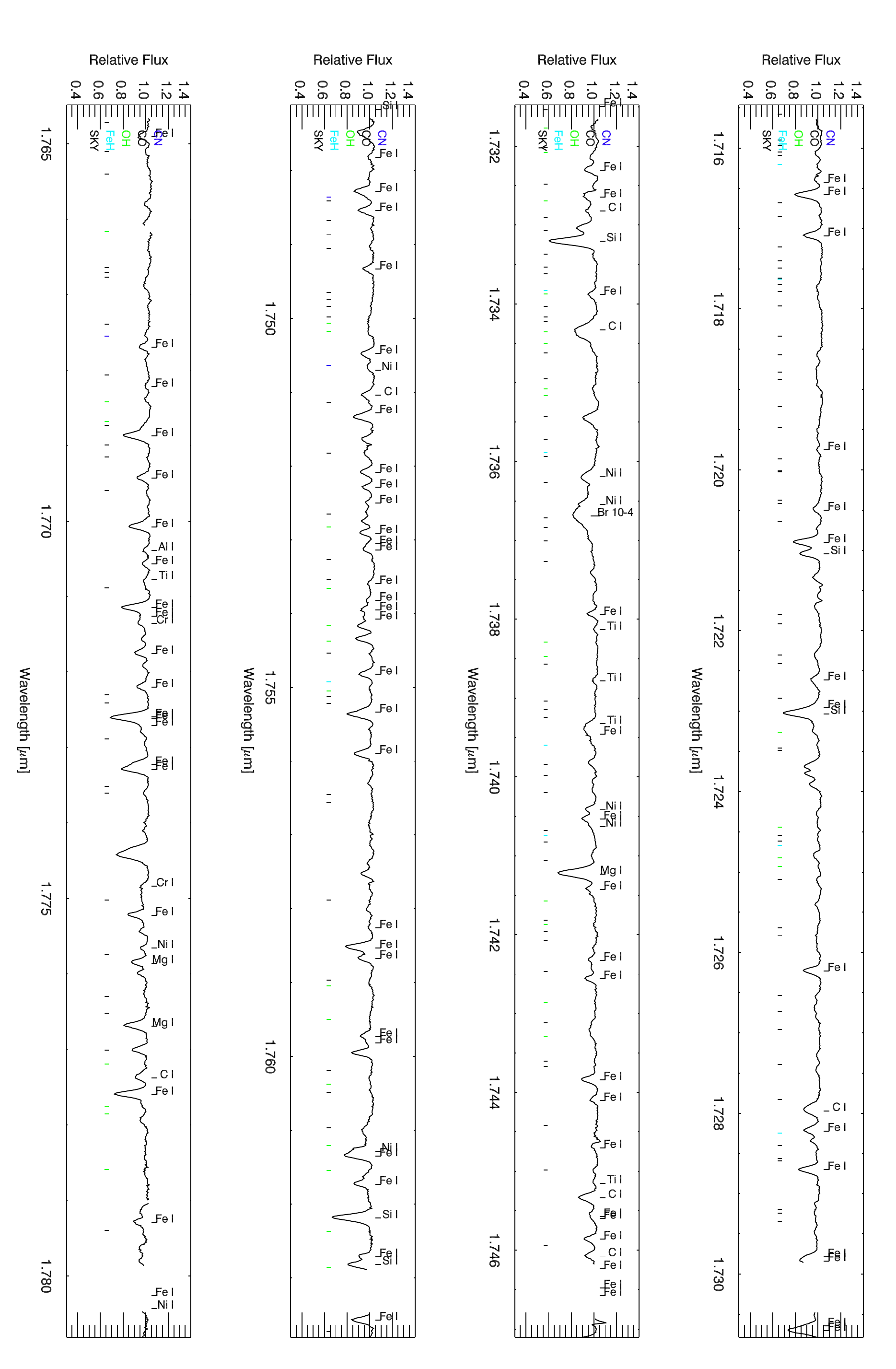}
\end{tabbing}
\center{\textbf{Figure~\ref{hband_appendix}.} Continued.}
\label{hband_appendix}
\end{figure*}


\begin{figure*}
\begin{tabbing}
\epsscale{0.95}
\plotone{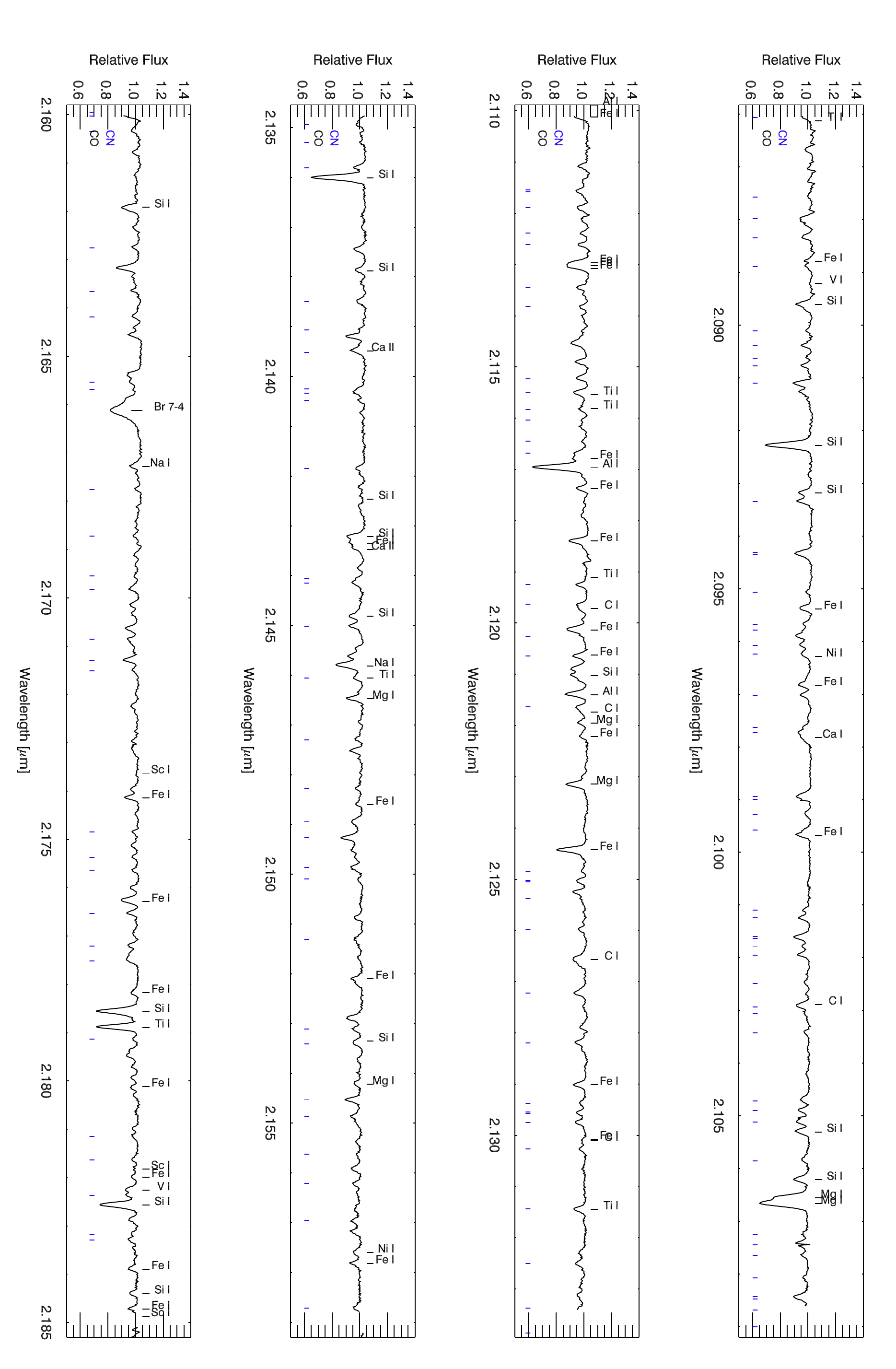}
\end{tabbing}
\caption{The IGRINS K-band spectrum of HD 44391 (K0 Ib). 
Line identification was made using the Arcturus line list from \citet{hinkle95}.
\label{kband_appendix}
}
\end{figure*}

\begin{figure*}
\begin{tabbing}
\epsscale{0.95}
\plotone{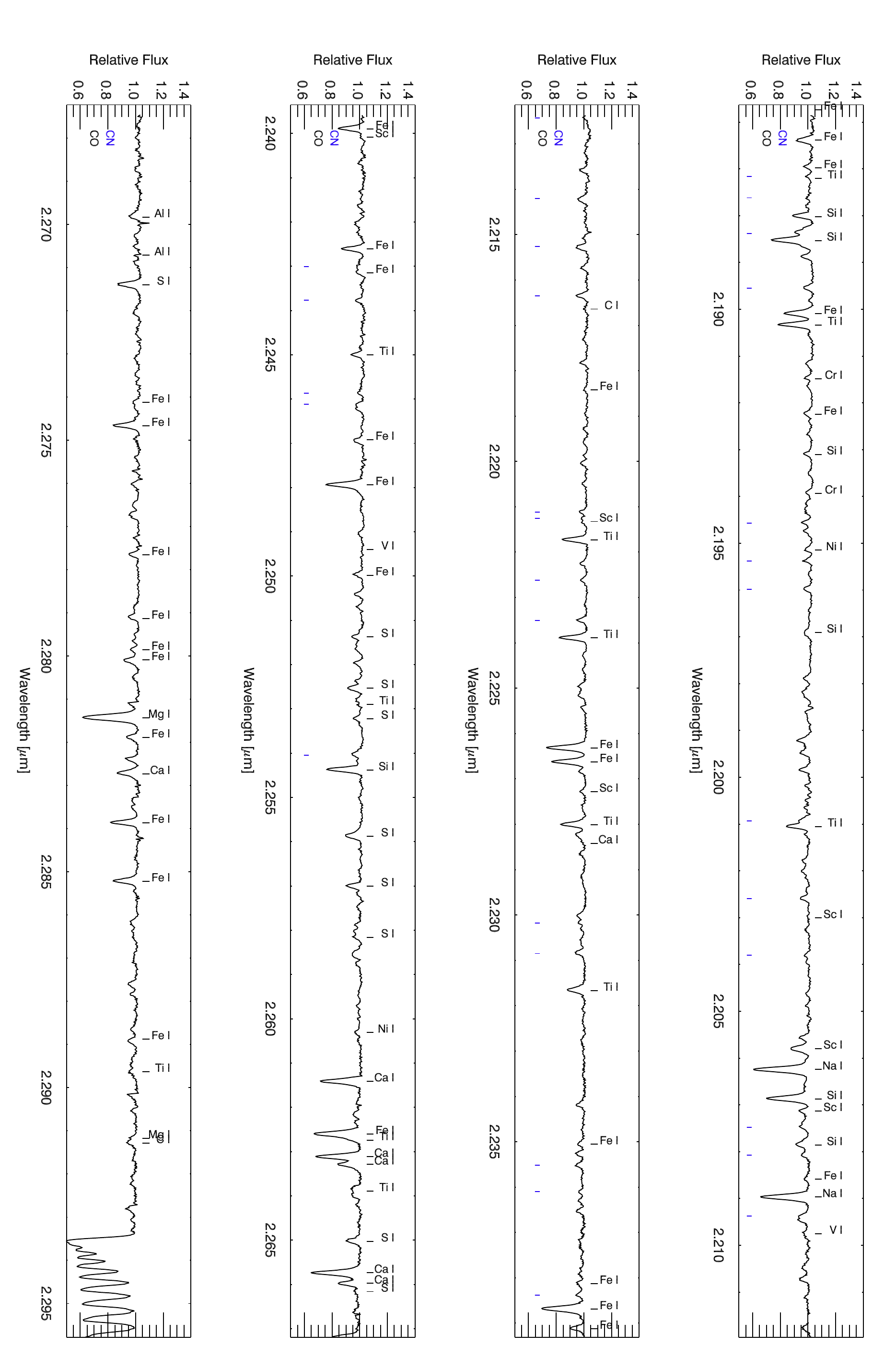}
\end{tabbing}
\center{\textbf{Figure~\ref{kband_appendix}.} Continued.}
\label{kband_appendix}
\end{figure*}

\begin{figure*}
\begin{tabbing}
\epsscale{0.95}
\plotone{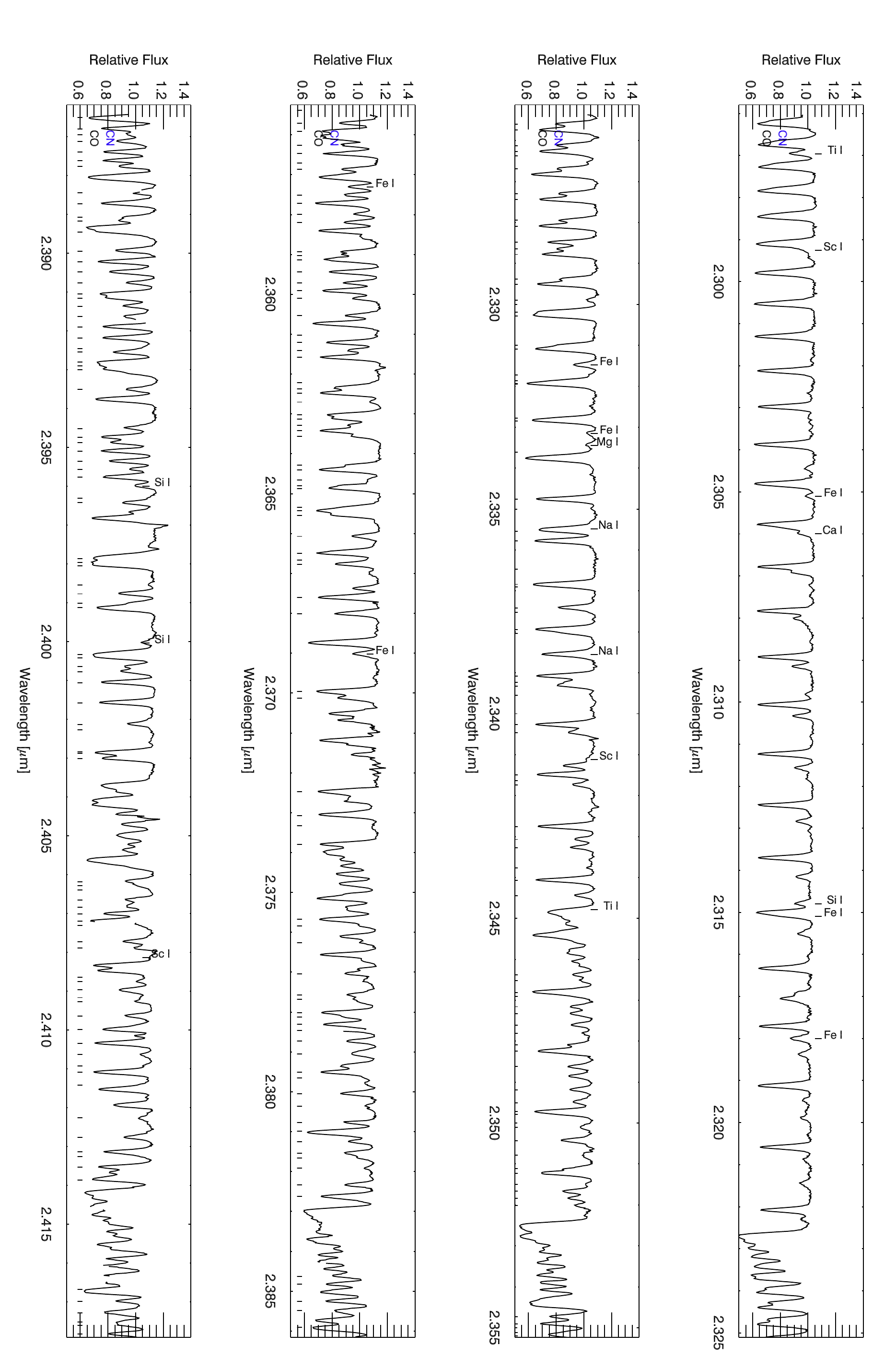}
\end{tabbing}
\center{\textbf{Figure~\ref{kband_appendix}.} Continued.}
\label{kband_appendix}
\end{figure*}

\begin{figure*}
\begin{tabbing}
\epsscale{0.95}
\plotone{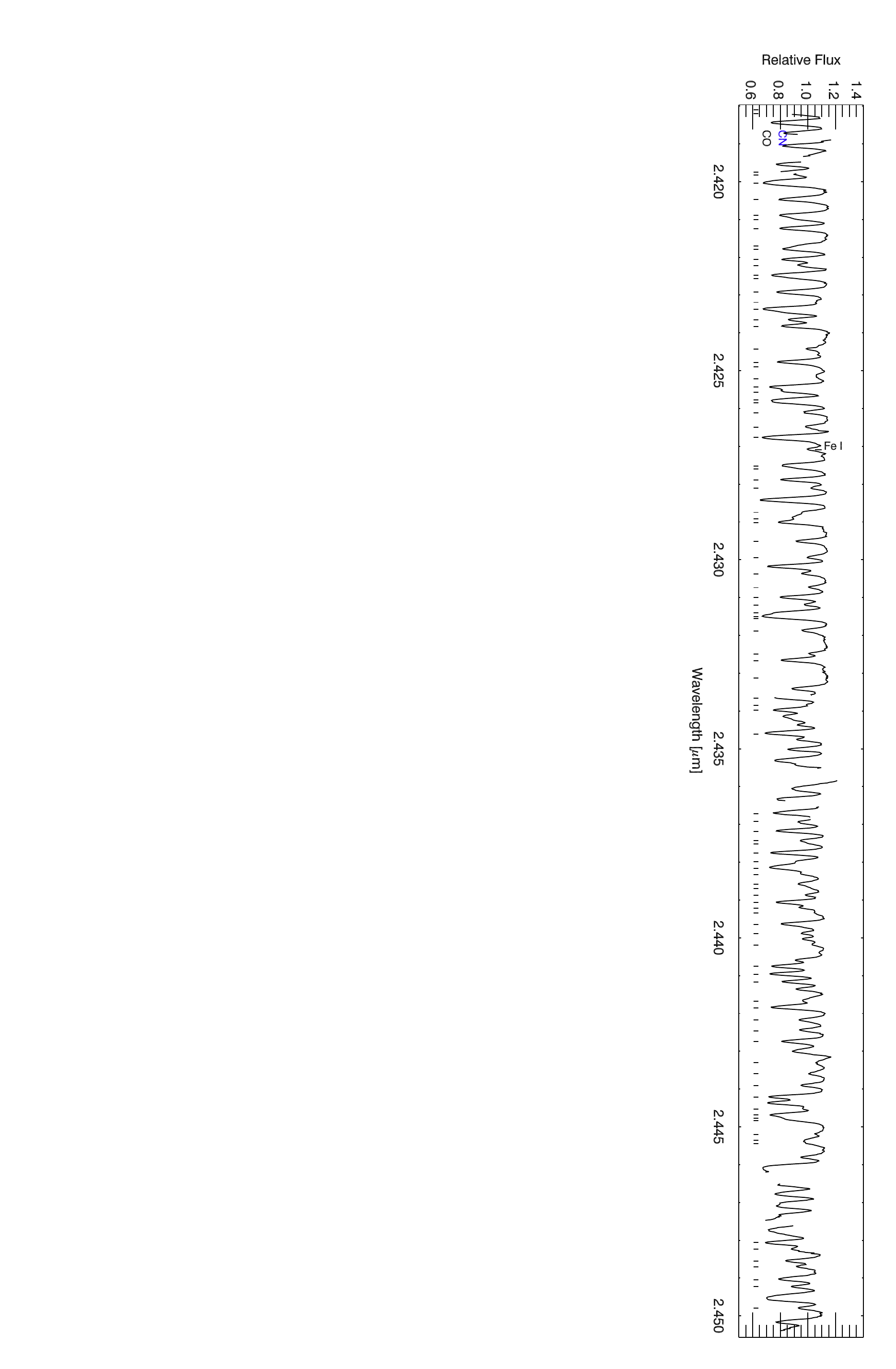}
\end{tabbing}
\center{\textbf{Figure~\ref{kband_appendix}.} Continued.}
\label{kband_appendix}
\end{figure*}

\clearpage

\clearpage

\end{document}